\documentclass[aps,pra,superscriptaddress,singlecolumn,notitlepage,longbibliography]{revtex4-1}
\usepackage{appendix}
\usepackage[table]{xcolor}
\newcommand{\nocontentsline}[3]{}
\newcommand{\tocless}[2]{\bgroup\let\addcontentsline=\nocontentsline#1{#2}\egroup}
\usepackage{graphicx, color, graphpap}
\usepackage{enumitem}
\usepackage{amssymb}
\usepackage{amsthm}
\usepackage[OT1]{fontenc}
\usepackage{multirow}
\usepackage[margin=1.0in]{geometry}
\usepackage[colorlinks=true,citecolor=blue,linkcolor=blue]{hyperref}
\usepackage[T1]{fontenc}

\usepackage{bbm}
\usepackage{thmtools,thm-restate}
\usepackage{verbatim}
\usepackage{mathtools}
\usepackage{titlesec}
\usepackage{amsmath}
\usepackage{tikz}
\usepackage{soul}
\usetikzlibrary{quantikz}
\usepackage[caption = false]{subfig}
\usepackage{float}

\usepackage{bm}
\usepackage{tabularx}
\usepackage{lmodern,adjustbox}
\usepackage[skins,breakable]{tcolorbox}
\tcbuselibrary{listings}
\tcbuselibrary{breakable}
\newtcolorbox{Code}{enhanced,fonttitle=\sffamily\bfseries\large,valign=center
,drop fuzzy shadow,sidebyside,lefthand ratio=0.4,lower separated=false}

\long\def\ca#1\cb{} 

\newcommand{\ketbra}[2]{| \hspace{1pt} #1 \rangle \langle #2 \hspace{1pt} |}

\newcommand{\btheta}{\boldsymbol \theta}

\newcolumntype{s}{>{\columncolor[HTML]{AAACED}} p{3cm}}

\newcommand{\mixHam}{H_{\mathrm{M}}}

{}
{}

\begin{document}

\title{Quantum Optimization for Training Quantum Neural Networks}

\author{Yidong Liao}
\email{yidong.liao@student.uts.edu.au}
\affiliation{Centre for Quantum Software and Information, University of Technology Sydney, Sydney, NSW, Australia}
\affiliation{Sydney Quantum Academy, Sydney, NSW, Australia}

\author{Min-Hsiu Hsieh} \email{minhsiuh@gmail.com}
\affiliation{Hon Hai Quantum Computing Research Center, Taipei, Taiwan}

\author{Chris Ferrie} \email{christopher.ferrie@uts.edu.au}
\affiliation{Centre for Quantum Software and Information, University of Technology Sydney, Sydney, NSW, Australia}

\begin{abstract}

Training quantum neural networks (QNNs) using gradient-based or gradient-free classical optimisation approaches is severely impacted by the presence of barren plateaus in the cost landscapes. In this paper, we devise a framework for leveraging quantum optimisation algorithms to find optimal parameters of QNNs for certain tasks. To achieve this, we coherently encode the cost function of QNNs onto relative phases of a superposition state in the Hilbert space of the network parameters. The parameters are tuned with an iterative quantum optimisation structure using adaptively selected Hamiltonians. The quantum mechanism of this framework exploits hidden structure in the QNN optimisation problem and hence is expected to provide beyond-Grover speed up, mitigating the barren plateau issue.
\end{abstract}
\maketitle

\begin{figure}[h!]
    \centering
    \includegraphics[width=\linewidth]{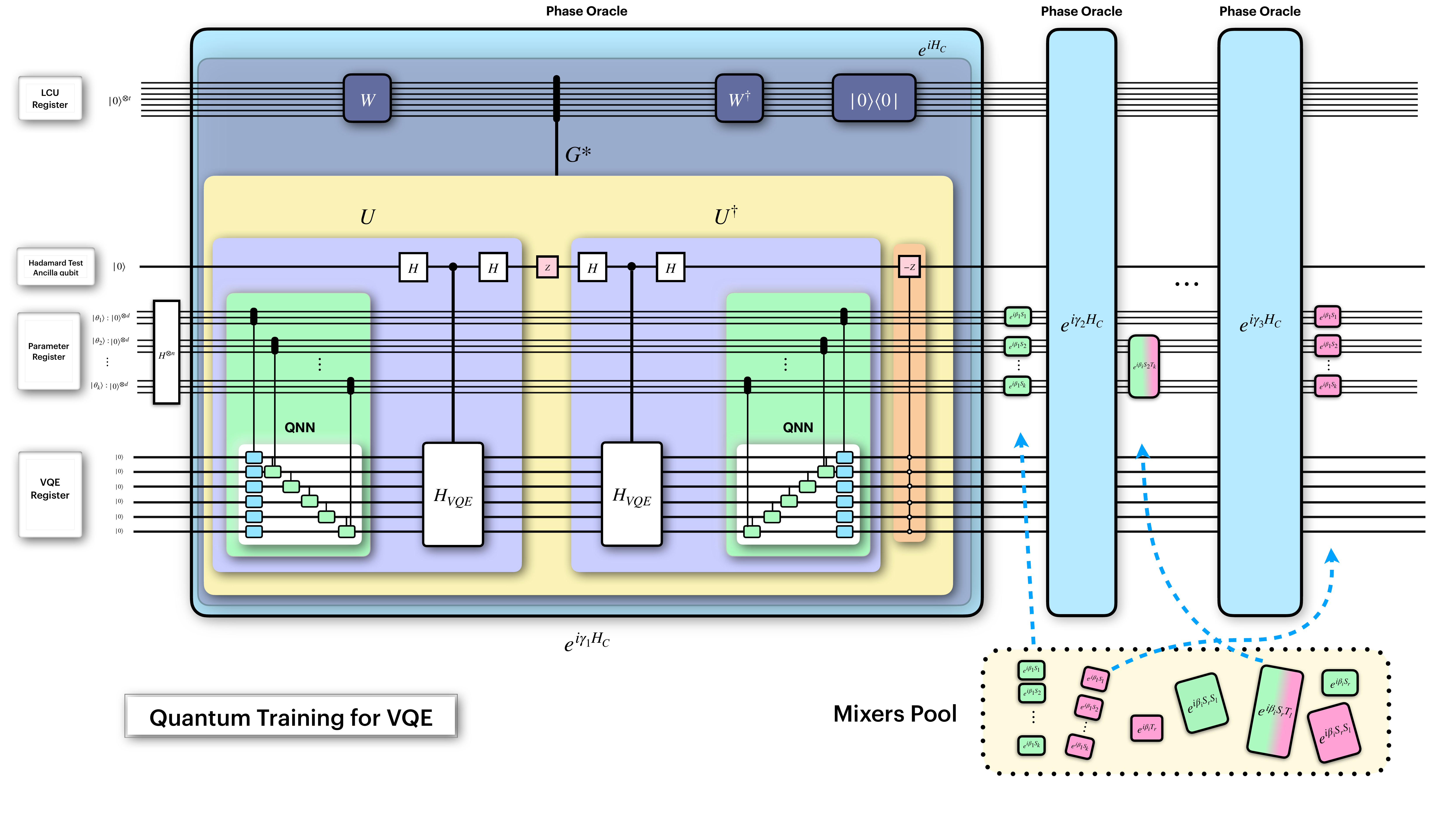}
    \caption{\emph{Schematic of Our quantum training algorithm for VQE.} Here we use the training of VQE as an example, to present the schematic circuit construction of our quantum training algorithm for QNN.
    A video animation of the circuit construction is available at  \href{https://youtu.be/RVWkJZY6GNY}{https://youtu.be/RVWkJZY6GNY}. (This is vector image and best view with the zoom feature in standard PDF viewers.) Note: \textit{1.} In all figures of this Paper, we omit the minus signs in all time-evolution-like terms (i.e. exponential of a Hamiltonian $e^{-iHt}$) for sake of brevity and space. \textit{2.} Some quantum registers are not depicted in this figure due to the limitation of space.}
    \label{fig:vqe}
\end{figure}


\section{Introduction}

\subsection{Quantum Neural Networks}

Quantum Neural Networks (QNNs) are considered to be a leading candidate to achieve a quantum advantage in noisy intermediate-scale quantum (NISQ) devices. A QNN consists of a set of parameterized quantum gates within an predefined circuit ansatz. The design of the ansatz together with the value of the gate parameters determine the outcome of the QNN. In order to successfully perform certain tasks, QNNs must be trained to find optimal parameters for generating desired outcomes. In the majority of QNN research, the training is carried out by employing variational hybrid quantum-classical algorithms \cite{mcclean2016theory}, in which the parameters are optimized by a classical optimizer using gradient-based or gradient-free approaches.
In this paper, we achieve a scalable, maximally quantum pipeline of the applications of QNNs by replacing the classical optimizer by quantum optimizer. In short, we employ quantum optimisation methods for training QNNs. \newline

There are two main avenues for the application of QNNs. The first uses QNNs to generate quantum states that minimize the expectation value of a given Hamiltonian, such as the case in Variational Quantum Eigensolvers (VQE) \cite{VQE} for chemistry problems or Quantum Approximate Optimization Algorithms (QAOA) \cite{qaoa2014} for combinatorial optimization problems. The second path uses QNNs as data-driven machine learning models to perform discriminative \cite{farhi2018classification,Schuld_2020,du2018implementable} and generative \cite{Benedetti_2019,Zeng_2019,Hamilton_2019,mitarai2018quantum,huang2020experimental} tasks for which QNNs could have more expressive power than their classical counterparts \cite{du2018expressive}. Though an ever increasing amount of effort is being put into QNN research, there is evidence that they will be difficult to train due to flat optimisation landscapes called barren plateaus \cite{mcclean2018barren}. \newline

The barren plateau issue has spawned several studies on the strategies to avoid them, including layerwise training \cite{skolik2020layerwise}, using local cost functions \cite{cerezo2020cost}, correlating parameters \cite{volkoff2020large}, and pre-training \cite{verdon2019learning}, among others \cite{du2020learnability,du2020quantum,zhang2020trainability}. Such strategies give hope that the variational quantum-classical algorithms may avoid the exponential scaling due to the barren plateau issue. However, it has been shown that these strategies do not avoid another type of Barren Plateaus induced by hardware noise \cite{wang2020noiseinduced}, and some strategies may lack theoretical grounding \cite{campos2020abrupt}. In addition to noise, there is also other sources of barren plateaus due to entanglement growth \cite{marrero2020entanglement}. {Moreover, it has been shown that gradient-free approaches are also adversely affected by barren plateaus~\cite{arrasmith2020effect}.} \newline 

Our work presents a new alternative to training QNNs with a maximally coherent (i.e., quantum) protocol.

\subsection{Prior work}

The above noted results indicate that training QNNs using classical optimisation methods have unprecedented challenges as the system scales up. Therefore, one seeks to leverage alternative optimisation methods for training QNNs. Indeed, preliminary attempts have been made in this direction. Verdon et al. proposed a QAOA-like training protocol for QNNs \cite{verdon2018universal} and Gilyén et al. developed a quantum algorithm for calculating gradients faster than classical methods \cite{Gily_n_2019}. In these two works, to cast the optimisation problem of training QNNs into the context of quantum optimisation, the network parameters in the QNN are quantized --- moved from being classical to being stored in quantum registers, {which are in addition to those upon which the QNN is performing its computation}. The quantized parameters are used as control registers of the parameterized gates on the QNN registers. The parameters can now be in superposition, which one hopes would allow for a \emph{quantum parallelism}-type computation of the QNN with multiple parameter configurations. \newline

\begin{figure}[h!]
\centering
\includegraphics[width=0.88\linewidth]{ 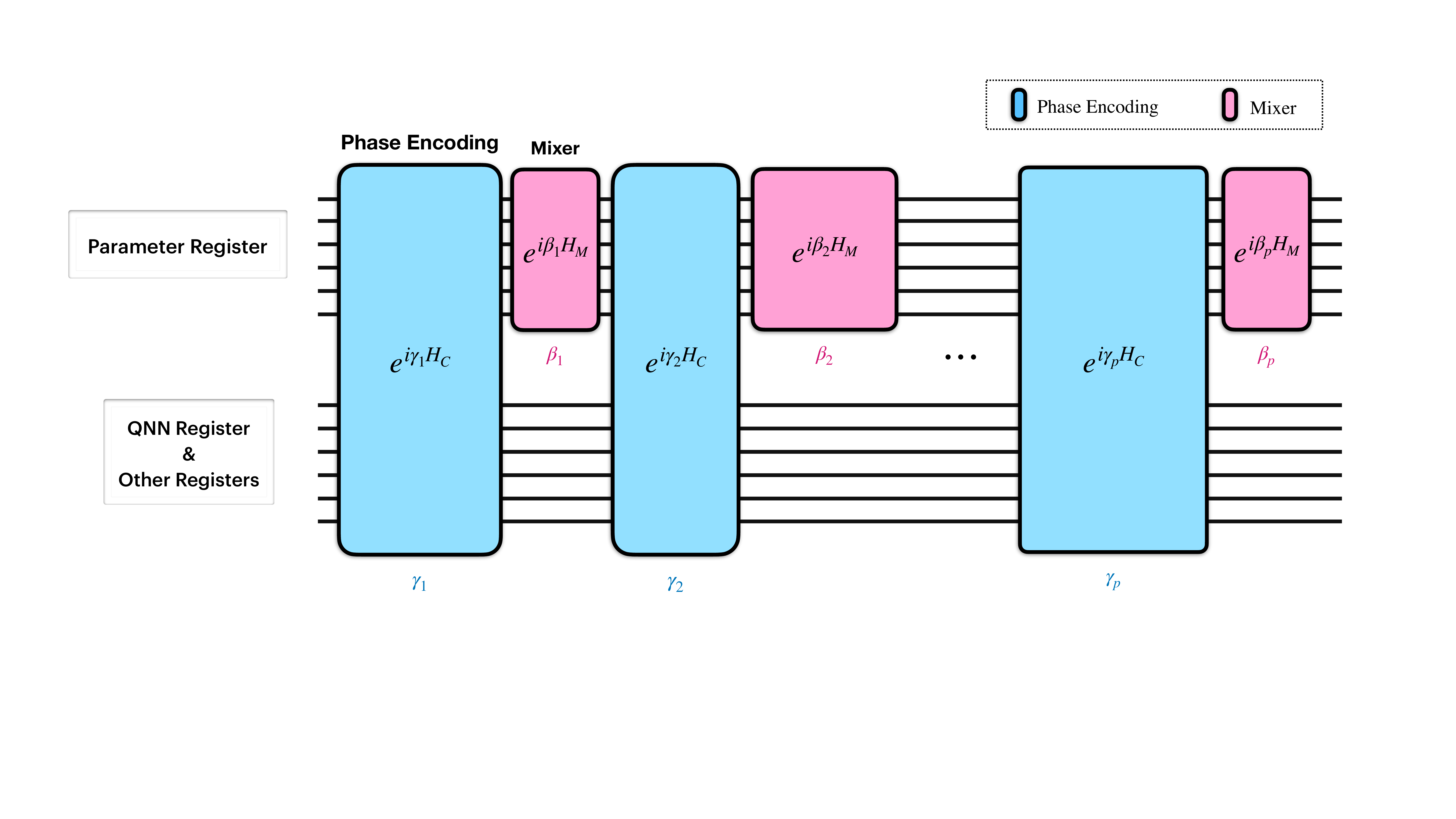}
\caption{\textit{QAOA-like training protocol for QNN, proposed in Ref.~\cite{verdon2018universal}}. The quantum training protocol consists of two alternating operations in a QAOA fashion --- the first operation acts on both the parameter register and QNN register to encode the cost function of QNN onto a relative phase of the parameter state. This operation is represented by the blue blocks in the figure.  The second operation acts only on the parameter register and it is a variant of the original QAOA Mixers, tailored for the case that the parameters in the QNN are continuous variables. This operation is represented by the pink blocks in the figure. These two operation can be mathematically expressed as $e^{-i \gamma_i C(\btheta)}$ and $e^{-i\beta_i H_M}$, where $\btheta$ are the parameters of QNN, $C(\btheta)$ is the cost function of the QNN, and $\gamma_i$ and $\beta_i$ are tunable hyperparameters,$H_M$ is the Mixer Hamiltonian. The width of each block represents the hyperparameters $\gamma_i$ and $\beta_i$ --- the wider the block, the larger the value of the hyperparameters. The phase encoding operation $e^{-i \gamma_i H_C}$ act as $e^{-i \gamma_i C(\btheta)}$. }
\label{verdon}
\end{figure}

In Ref.~\cite{verdon2018universal}, the quantum training process can be described as the state evolution in the joint Hilbert space of the parameter register and the QNN register. Their quantum training protocol consists of two alternating operations in a QAOA fashion --- the first operation acts on both the parameter register and QNN register to encode the cost function of QNN onto a relative phase of the parameter state. The second operation acts only on the parameter register and it is a variant of the original QAOA Mixers, tailored for the case that the parameters in the QNN are continuous variables. These two operation can be mathematically expressed as $e^{-i \gamma_i C(\btheta)}$ and $e^{-i\beta_i H_M}$, where $\btheta$ are the parameters of QNN, $C(\btheta)$ is the cost function of the QNN, and $\gamma_i$ and $\beta_i$ are tunable hyperparameters, $H_M$ is the Mixer Hamiltonian. By heuristically tuning the hyperparameters, the quantum training is expected to home in on the optimal parameters of the QNN after several iterations of the QAOA alternating operations. We illustrate the alternating operations of their quantum training in Fig.~\ref{verdon}. \newline

Despite being the pioneering application of the QAOA method for training QNNs, the protocol in Ref.~\cite{verdon2018universal} has some limitations. In the phase encoding operation, the parameter register and the QNN register are generally always entangled. This will have the effect of causing phase decoherence in the parameter eigenbasis. To minimize the effect of this decoherence, the tuneable hyper-parameter $\gamma_i$ must be sufficiently small --- in other words, the phase encoding is coherent only in the first order of $\gamma_i$. To overcome this limitation --- to enact phase encoding operation with arbitrary hyperparameters --- the phase encoding operation with a small hyperparameter $\Delta \gamma$ should be repeated an excessive amount of times. This simulates the phase encoding operation with a large hyperparameter $\gamma$ via $e^{-i \gamma C(\btheta)}=  e^{-i \Delta \gamma C(\btheta)} e^{-i \Delta \gamma C(\btheta)} e^{-i \Delta \gamma C(\btheta)}...$ These repetitions will yield large overhead in the complexity of the algorithm. In Ref.~\cite{Gily_n_2019}, a \textit{phase oracle} is designed for the phase encoding and can achieve it coherently and efficiently. (Note that throughout this paper, the term \textit{phase oracle} has different meaning than the one in Ref.~\cite{Gily_n_2019}, our \textit{phase oracle} stands for the term \textit{fractional phase oracle} in Ref.~\cite{Gily_n_2019}.) Nevertheless, they did not utilise the phase encoding as a component of QAOA routine to accomplish a fully quantum training algorithm for QNNs. Instead they use the phase oracle as a component of quantum evaluation of the gradient of a QNN, which serves for gradient based classical training of QNNs. However this improvement will not be practically useful due to the barren plateau issue of QNNs.\newline

\begin{figure}[h!]
\centering
\includegraphics[width=0.88\linewidth]{ 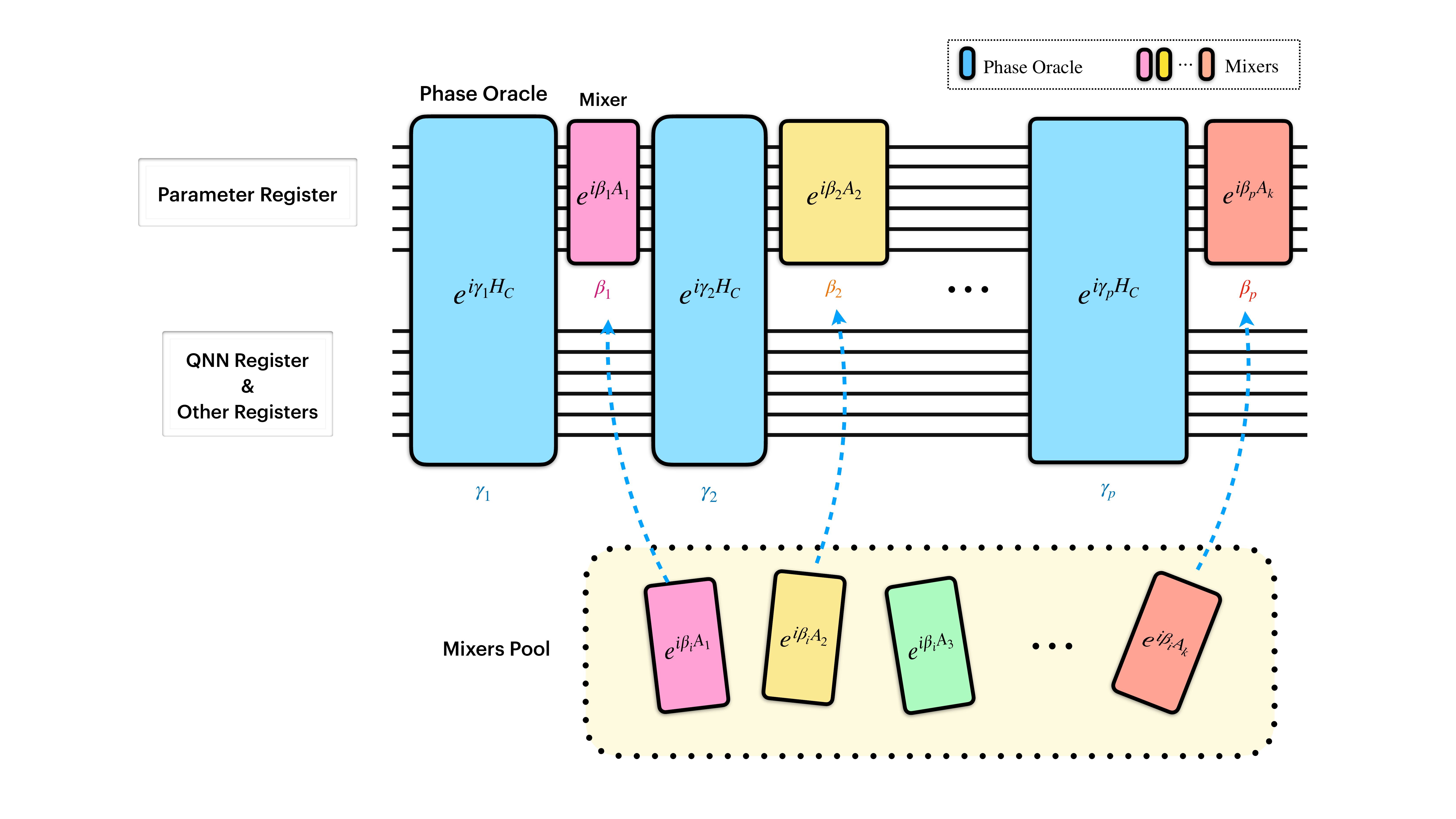}
\caption{\textit{Schematic of our framework for quantum training of QNN}. Our quantum training for QNN taking advantage of the well-established parts in Refs.~\cite{verdon2018universal} and ~\cite{Gily_n_2019}, while eliminating their shortcomings. We replace the phase encoding operations in QAOA-like protocol of Ref.~\cite{verdon2018universal}(as depicted in Fig~\ref{verdon}) by the \textit{phase oracle} in Ref.~\cite{Gily_n_2019}. For the mixers in the QAOA-like routine, we allow different mixers for each layer, similar to Ref.~\cite{zhu2020adaptive}. In this figure, the color of each block represents the nature of the corresponding Hamiltonian: different color corresponds to different Hamiltonian (One can see that the Cost Hamiltonian is the same throughout the training whereas the mixer varies from layer to layer). The mixers pool contains the proper mixers tailored to our QNN training problem. These rules also apply to the other circuit schematic in this paper.}
\label{ours}
\end{figure}

In this paper, we devise an improved framework for training QNNs, taking advantage of the well-established parts in Refs.~\cite{verdon2018universal} and \cite{Gily_n_2019}, while eliminating the shortcomings. A schematic of our quantum training framework for QNNs is  depicted in Fig.~\ref{ours}. More specifically, we achieve the following:
\begin{itemize}
    \item[\textsc{1.}] We replace the phase encoding operations in QAOA-like protocol of Ref.~\cite{verdon2018universal} by the \textit{phase oracle} in Ref.~\cite{Gily_n_2019}. This achieves coherent encoding of the cost function onto a relative phase of parameter state, while avoiding the limitations of the hyper-parameters in the phase encoding.

    \item[\textsc{2.}] For the mixers in the QAOA-like routine we adopt a similar approach to Ref.~\cite{zhu2020adaptive} by making the mixers adaptive --- that is, we allow different mixers for each layer (particularly, to allow entangling mixers that act across different parameters). This potentially leads to a dramatic shortening of the depth of QAOA layers while significantly improving the quality of the solution (the optimal QNN parameters found by the QAOA routine).
\end{itemize}

By making the mixers flexible and adaptive to specific optimisation problems, it is demanding to find an efficient way of determining the best sequence of mixers and the optimized hyperparameters. To address these we adopt machine learning approaches (in particular, Recurrent Neural Networks and Reinforcement Learning) as proposed in Refs.~\cite{PhysRevResearch.2.033446, yao2020reinforcement, Warren2020RNNVQEAM, verdon2019learning}. The quantum mechanism of this framework is the best candidate to exploit hidden structure in the QNN optimisation problem, which would provide beyond-Grover speed up and mitigate the barren plateau issues for training QNNs.

\subsection{Paper Outline}

The remainder of this paper is organized as follows: in Section \ref{Preliminaries} we review some essential preliminaries --- particularly on the details of QAOA and its variants, from which we designed a new variant of QAOA tailored for our QNN training problem. Section \ref{par} introduces a way of quantising parameters of a QNN --- that is, we show how to create superposition of a QNN with multiple parameter configurations.
In Section \ref{GAS} we present quantum training by Grover adaptive search as a baseline prior to our quantum training framework using QAOA. In Section \ref{ourf} we present the details of our framework including how to implement the phase oracle, that can achieve coherent phase encoding of the cost function of a QNN, and which mixers to use for the QAOA routine, as well as the strategy to determine the mixers sequence and the optimize their hyper-parameters. {Section \ref{applications} presents the deployment potential of our quantum training to a variety of application including training VQE, learning a pure state, and training a quantum classifier.}  The final section summarise our work and provides outlook for future work.

\section{Preliminaries}\label{Preliminaries}

\subsection{Quantum Optimisation Algorithms}
\subsubsection{Zoo of Quantum Optimisation Algorithms}

For completeness and context, we list some typical quantum optimisation algorithms in Table.~\ref{table}, including the primitive ones (adiabatic, quantum walks, QAOA, Grover adaptive search), their hybridizations, and their variants. In this paper, for the training of QNNs, we focus on utilising QAOA and its variants as well as Grover adaptive search, which we will review in the following subsections.

\begin{table}[h!]
\centering

\begin{tabularx}{\textwidth}
{ | >{\raggedright\arraybackslash}X | >{\raggedright\arraybackslash}X | >{\raggedright\arraybackslash}X | >{\raggedright\arraybackslash}X| >{\raggedright\arraybackslash}X  |}
\hline
\textbf{Primitives}& {Adiabatic} & {Quantum Walk}& {QAOA}& {Grover adaptive search} \\
\hline
\textbf{Hybridization of Primitives} & \multicolumn{4}{c|}{Hybrid adiabatic–quantum-walk algorithms~\cite{PhysRevA.99.022339}, others~\cite{marsh2018quantum,Jiang_2017,mbeng2019quantum,wang2017quantum}}\\
\hline
 \textbf{Variants of Primitives} & Shortcut to adiabaticity~\cite{Gu_ry_Odelin_2019,hegade2020shortcuts} & Quantum stochastic walk~\cite{Whitfield_2010} & Adaptive QAOAs~\cite{zhu2020adaptive,yao2020reinforcement}, Others~\cite{nasaQAOA2019,verdon2018universal}& Quantum basin hopping~\cite{2005quant.ph..7193B}\\
\hline
\end{tabularx}
\caption{\textit{Zoo of Quantum Optimisation Algorithms}. {The first row contains the four primitive quantum optimisation algorithms by adiabatic quantum evolution, quantum walks, QAOA and Grover adaptive search. The second row contains the hybridization among these four primitives, e.g. hybrid adiabatic–quantum-walk algorithms \cite{PhysRevA.99.022339}. The third row contains the variants of the primitives, e.g. variants of QAOA include  Adaptive QAOAs~\cite{zhu2020adaptive,yao2020reinforcement}, and others~\cite{nasaQAOA2019,verdon2018universal}}.}\label{table}
\end{table}

\begin{figure}
\centering
\includegraphics[width=0.98\linewidth]{ 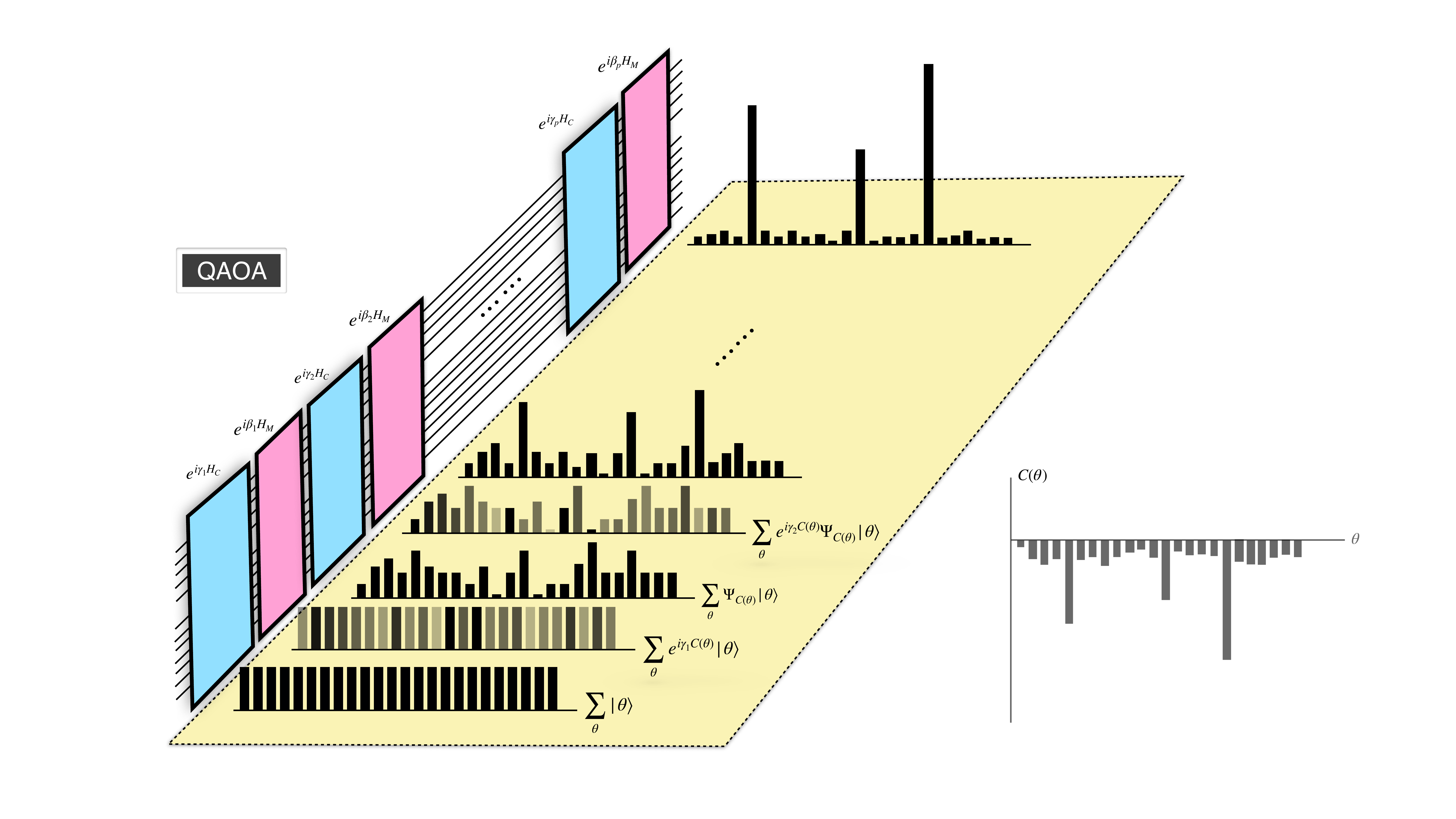}
\caption{\textit{Interference process of QAOA.} QAOA is an interference-based algorithm such that non-target states interfere destructively while the target states interfere constructively. Here we illustrate this interference process by presenting the evolution of the quantum state of the parameters (black bar graphs on the yellow plane) alongside with the QAOA operations (blue and pink boxes on circuit lines, representing the Phase encoding and Mixers respectively). The starting state $\sum_{\theta} \ket{\theta}$ (omitting the normalization factor) is the even superposition state of all possible parameter configurations. After the first Phase encoding operation, the state becomes $\sum_\theta e^{-i{\gamma}_{1}C(\theta)} |{\theta}\rangle$ for which we use opacity of the bars indicate the value of the phase, the magnitudes of the amplitudes in the state remains unchanged. After the first Mixer, the state becomes $\sum_\theta  \Psi_{C(\theta)}|{\theta}\rangle$ in which the  magnitudes of the amplitudes in the state has changed. Similar process happens to the following operations, until the amplitudes of the optimal parameter configurations are amplified significantly (the furthest bar graph). The grey bar graph in the right corner is the cost function being optimized by QAOA.     }
\label{inter}
\end{figure}

Before that, however, some remarks on the fundamental differences of the adiabatic and QAOA protocols are in order. QAOA can be seen as a ``trotterized'' version of adiabatic evolution: the mixer Hamiltonians being the initial Hamiltonian in the analogous adiabatic algorithm, and the cost Hamiltonians being the final Hamiltonian. However short-depth QAOA is not really the digitized version of the adiabatic problem, but rather an ad hoc ansatz. In Ref.~\cite{streif2019comparison} it is shown that QAOA is able to deterministically find the solution of specially constructed optimization problems in cases where quantum annealing fail. We emphasise that QAOA is an interference-based algorithm such that non-target states interfere destructively while the target states interfere constructively. In Fig.\ref{inter} we depict this interference process of QAOA.

\subsubsection{QAOA and its variants}\label{variants}

\begin{figure}[h!]
\centering
\includegraphics[width=\linewidth]{ 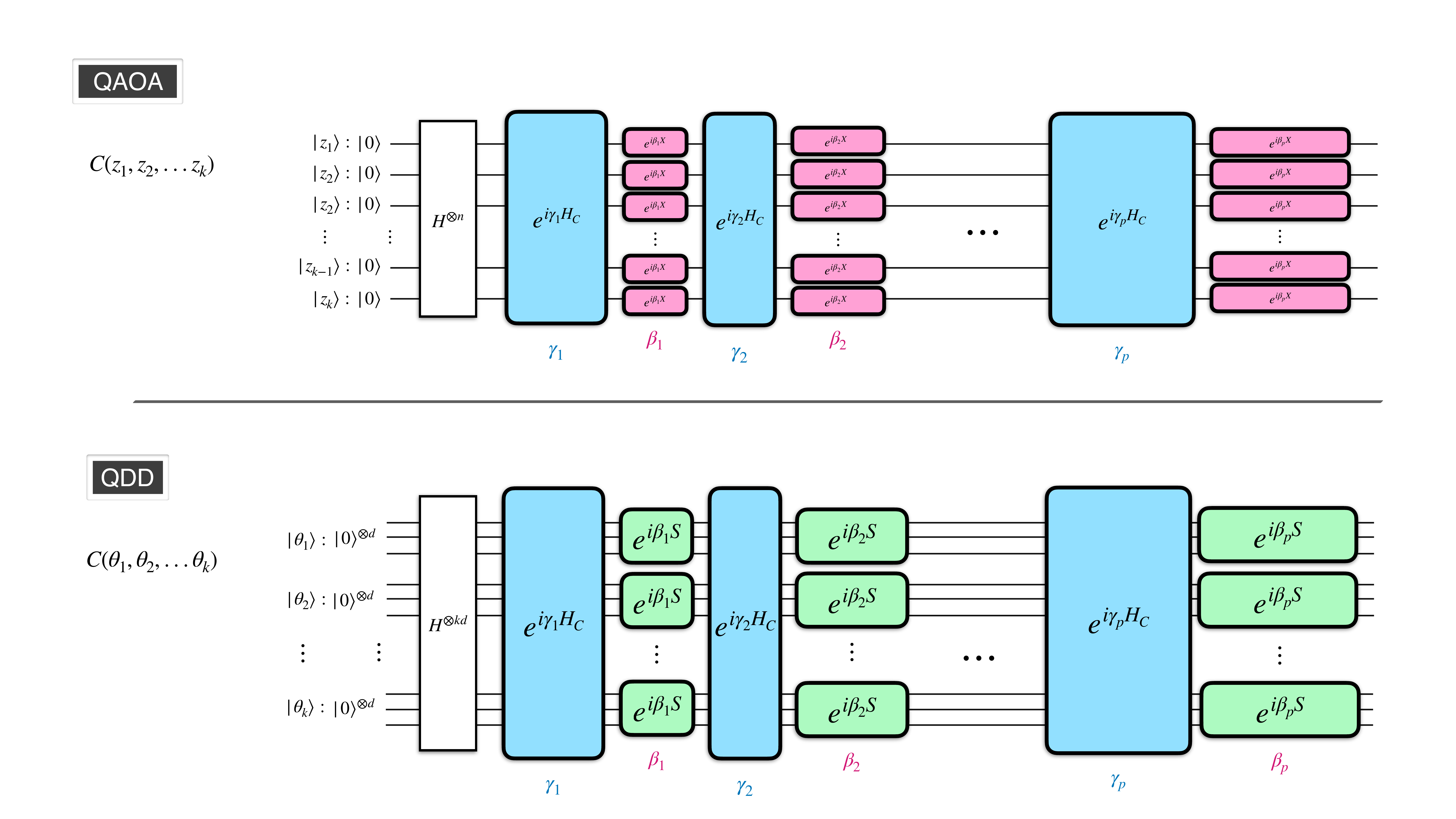}
\caption{\textit{Quantum circuit schematic of the operations in the original QAOA}. The state is initialized by applying Hadamard gates on each qubit, represented as $H^{\otimes n}$. This results in the equal superposition state of all possible solutions. QAOA consists of alternating time evolution under the two Hamiltonians $H_C$ and $\mixHam$ for~$p$ rounds,
where the duration in round $j$ is specified by the parameters $\gamma_j$ and $\beta_j$, respectively. In the original QAOA, the mixing Hamiltonian $\mixHam$ is chosen as to be
$ \mixHam = \sum_{j=1}^n X_j,$ After all $p$ rounds, the state becomes
$\ket{\boldsymbol \beta, \boldsymbol \gamma}
=
e^{-i \beta_p \mixHam}e^{-i \gamma_p  H_C} \dots e^{-i \beta_2 \mixHam}e^{-i \gamma_2  H_C}e^{-i \beta_1 \mixHam}e^{-i \gamma_1  H_C}
\ket{s}.$}
\label{oqaoa}
\end{figure}

In this section, we review the original quantum approximation optimization algorithm (QAOA) proposed
in Ref.~\cite{qaoa2014} and its variants. Consider an unconstrained optimization problem on $n$-bit strings $\mathbf z=(z_1,z_2,z_3,....z_n)$ where $z_i \in \{-1,1 \}$ We seek the optimal bit string $\mathbf z$ that maximizes (or minimizes) a cost function $C(\mathbf z)$ . 
Given the cost function $C(\mathbf z)$ of a problem instance, the algorithm is characterized by
two Hamiltonians: the \textit{cost Hamiltonian} $H_C$ and the \textit{Mixing Hamiltonian} $\mixHam$.
The cost Hamiltonian $H_C$ encodes the cost function~$C(\mathbf z)$ to be optimized,
and acts on $n$-qubit computational basis states as
$$H_C \ket{\mathbf z}= C(\mathbf z) \ket{\mathbf z}.$$
  The mixing Hamiltonian $\mixHam$ is chosen as to be
$$ \mixHam = \sum_{j=1}^n X_j,$$
where $X_j$ is the Pauli $X$ operator acting on the $j$th qubit.
The initial state is the even superposition state of all possible solutions:
$\ket{s} = \frac{1}{\sqrt{2^n}} \sum_{\mathbf{z}} \ket{\mathbf{z}}\;$.
The QAOA algorithm consists of alternating
time evolution under the two Hamiltonians $H_C$ and $\mixHam$ for~$p$ rounds,
where the duration in round $j$ is specified by the parameters $\gamma_j$ and $\beta_j$, respectively. After all $p$ rounds, the state becomes
$$\ket{\boldsymbol \beta, \boldsymbol \gamma}
=
e^{-i \beta_p \mixHam}e^{-i \gamma_p  H_C} \dots e^{-i \beta_2 \mixHam}e^{-i \gamma_2  H_C}e^{-i \beta_1 \mixHam}e^{-i \gamma_1  H_C}
\ket{s}.$$
The alternating operations can be illustrated as in Fig.~\ref{oqaoa}. Finally a measurement in the computational basis is performed on the state. Repeating the above state preparation and measurement, the expected value of the cost function,
$$ \langle C\rangle = \bra{\boldsymbol \beta, \boldsymbol \gamma}  H_C \ket{\boldsymbol \beta, \boldsymbol \gamma},$$
can be estimated from the samples produced from the measurements. \newline

The above steps are then repeated altogether, with updated sets of time parameters $\gamma_1,\dots,\gamma_p,\beta_1, \dots, \beta_p$. Typically a classical optimization loop (such as gradient descent) is used to find the optimal parameters that maximize(or minimize) the the expected value of the cost function $\langle C\rangle$. Then measuring the resulting state of the optimal parameters provide an approximate solution to the optimization problem.\newline

There has been a lot of progress on QAOA recently on both the experimental and theoretical fronts. There is evidence suggesting that QAOA may provide a significant quantum advantage over classical algorithms \cite{niu2019optimizing,Barkoutsos_2020}, and that it is computationally universal \cite{morales2019universality,lloyd2018quantum}. Despite these advances, there are limitations of QAOA. The performance improves with circuit depth, but circuit depth is still limited in near-term quantum processors. Moreover, deeper circuits translate into more variational parameters, which introduces challenges for the classical optimizer in minimizing the objective function. Ref.~\cite{bravyi2019obstacles} show that the locality and symmetry of QAOA can limit its performance. These issues can be attributed to the form of the QAOA ansatz. A short-depth ansatz that is further tailored to a given combinatorial problem could therefore address the issues with the standard QAOA ansatz. However, identifying such an alternative is a highly non-trivial problem given the vast space of possible ansatzes. Farhi et al. \cite{farhi2017quantum} allowed the mixer to rotate each qubit by a different angle about the $x$-axis and modified the cost Hamiltonian based on hardware connectivity. This modification was made primarily out of hardware capability concerns with the hope that superior-than-classical performance can be experimentally verified. \newline

\begin{figure}[htbp]
\centering
\includegraphics[width=\linewidth]{ 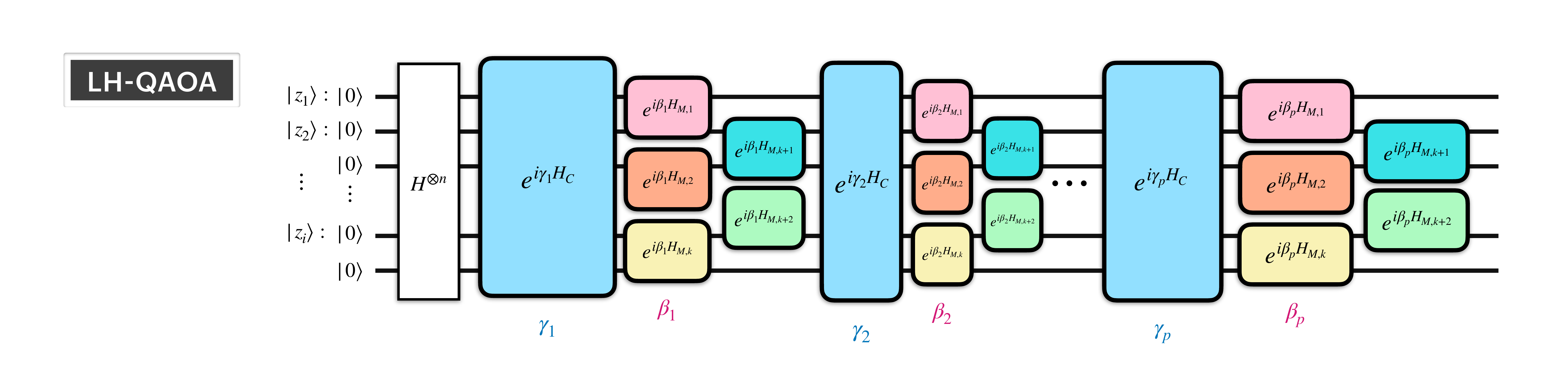}
\caption{\textit{Quantum circuit schematic of the operations in \textit{LH-QAOA}}. The overall process of LH-QAOA is similar to that of the original QAOA in Fig.~\ref{oqaoa}, where the difference is that the mixer of LH-QAOA contains entangling an mixer Hamiltonian on two qubits. These are represented by the $H_{M,i}$ blocks with various colors in the figure. Note that in order to avoid an excessive amount of hyper-parameters, Hadfield et al. \cite{hadfield2019quantum} choose the $\beta_j$ for each  $H_{M,i}$ to be the same in every layer.}
\label{lh}
\end{figure}

\textbf{LH-QAOA}.
 In Ref.~\cite{hadfield2019quantum} Hadfield et al. considered alternative mixers including entangling ones on two qubits. The selection of mixers is based on the criteria of preserving the relevant subspace for the given combinatorial problem, for which they entitled it Local Hamiltonian-QAOA (LH-QAOA). Here we depict the quantum circuit schematic of LH-QAOA in Fig.~\ref{lh}.

\begin{figure}[h!]
\centering
\includegraphics[width=\linewidth]{ 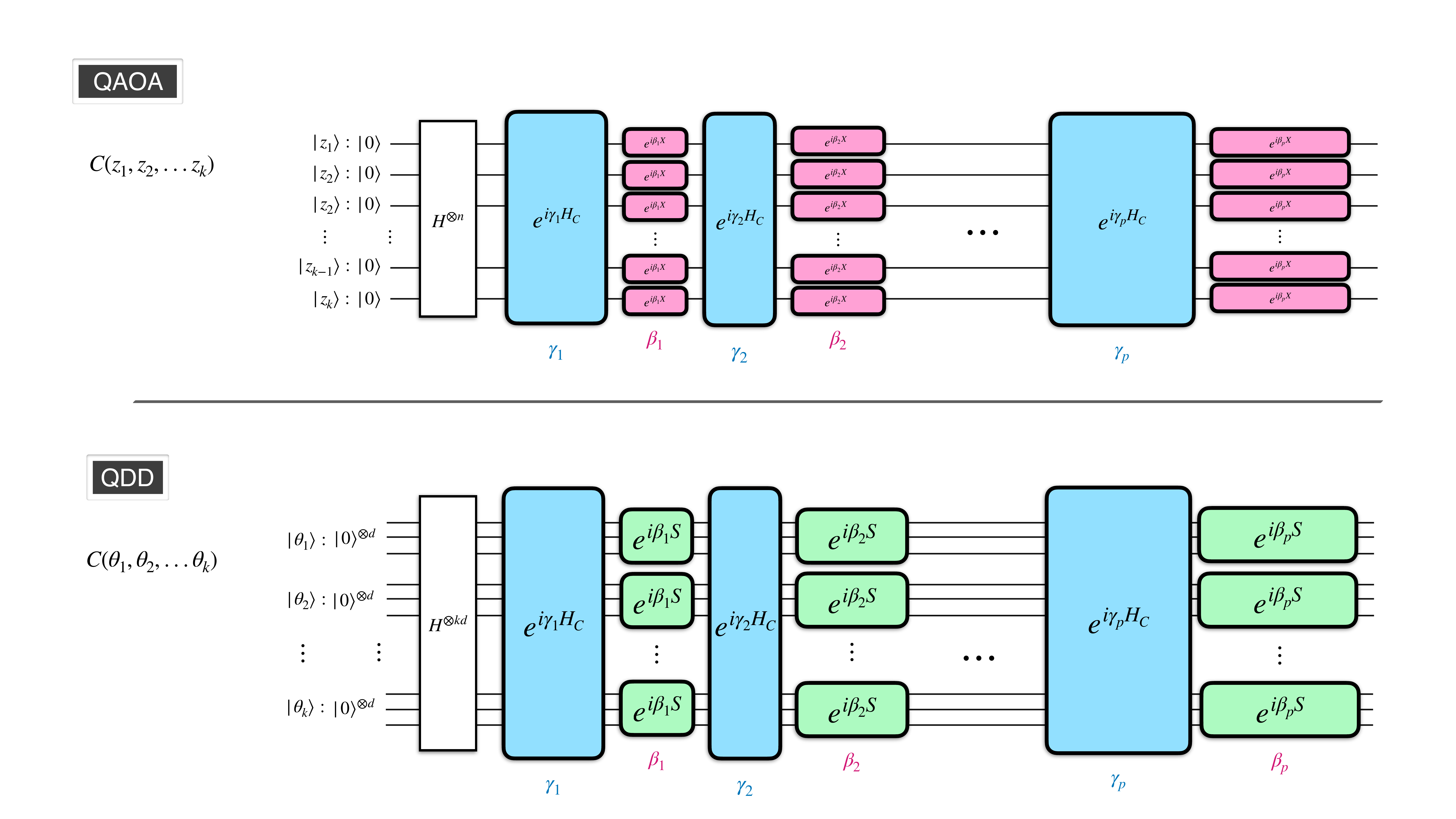}
\caption{\textit{Quantum circuit schematic of \textit{QDD}}. QDD solves optimization problems of continuous variable. In this figure, $\theta_i$ are the continuous variables to be optimized in the training, where each $\theta_i$ is digitized into binary form and stored in an independent register. The overall process of QDD is similar to that of the original QAOA, where the difference is that the mixer of QDD with Hamiltonian $S$ is acting on the registers of $\theta_i$ (rather than single qubits as in the original QAOA). The effect of the mixer in QDD is to shift the value for each $\theta_i$.}
\label{qdd}
\end{figure}

\textbf{QDD}. In Refs.~\cite{verdon2019quantum3,verdon2018universal} Verdon et al. adjusted the mixers for continuous optimization problem in which the parameters to be optimized are continuous variables. In the original QAOA ansatz, the mixer is chosen to be single-qubit $X$ rotations applied on all qubits.  These constitute an uncoupled sum of generators of shifts in the computational basis.
Similarly, the appropriate mixers in the continuous case should shift the value for each digitized continuous variables stored in independent registers. They entitled it Quantum Dynamical Descent (QDD). Here we depict the quantum circuit schematic of QDD in Fig.~\ref{qdd}. \newline

\begin{figure}[h!]
\centering
\includegraphics[width=0.95\linewidth]{ 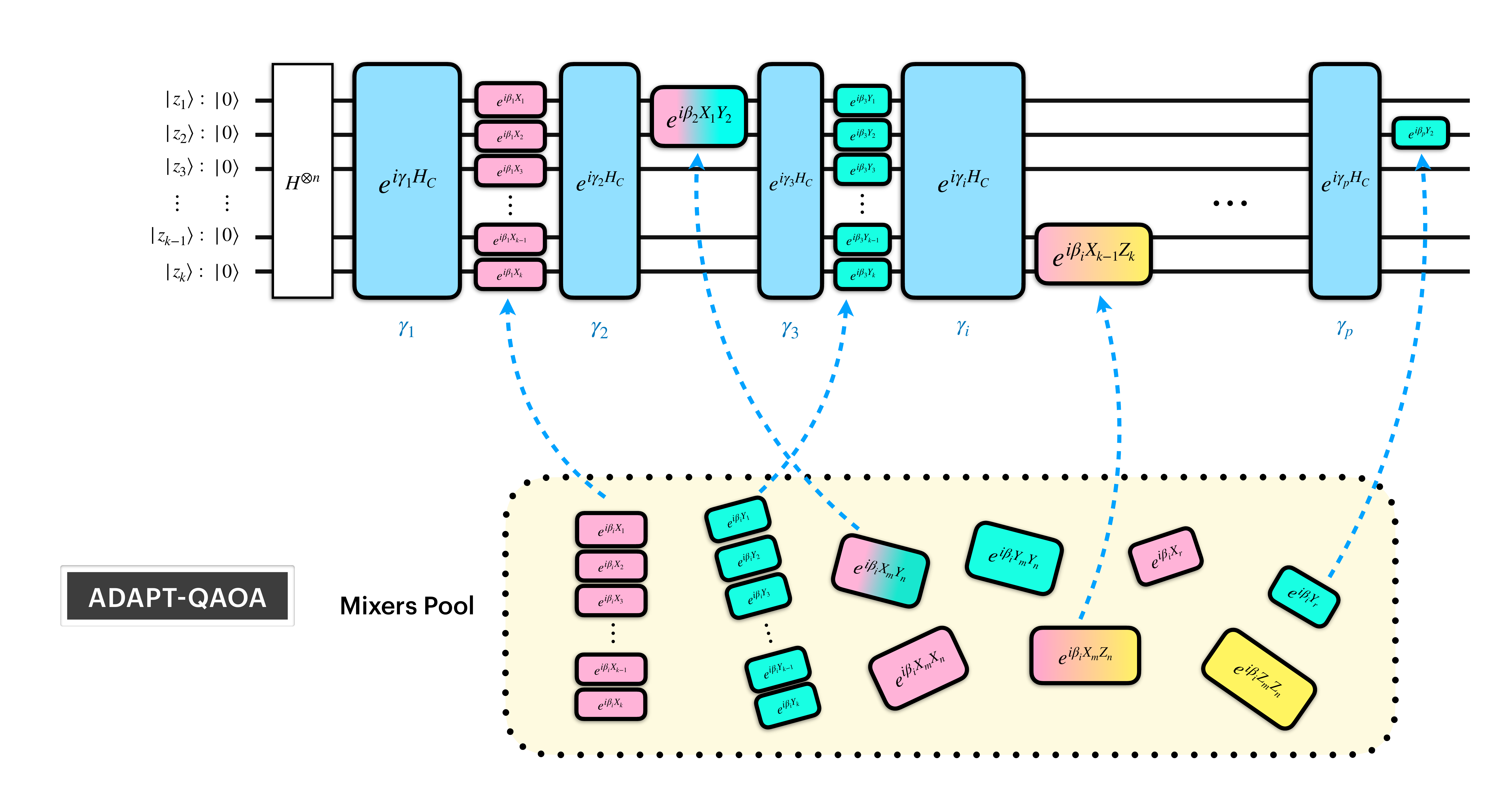}
\caption{\textit{Quantum circuit schematic of ADAPT-QAOA}. The overall process of LH-QAOA is similar to that of the original QAOA in Fig.~\ref{oqaoa}, where the difference is that the mixer of LH-QAOA contains variable mixers taken from a \emph{mixers pool}. Define $Q$ to be the set of qubits. The mixer pool of ADAPT-QAOA is  $P_\text{ADAPT-QAOA} = \cup_{i \in Q} \left\{ X_i, Y_i, \sum_{i \in Q}X_i , \sum_{i \in Q}Y_i \right\} \cup_{i,j \in Q \times Q} \left\{ B_i C_j | B_i, C_j \in \left\{ X, Y, Z \right\} \right\} $.}
\label{adapt}
\end{figure}

\textbf{ADAPT-QAOA}. LH-QAOA and QDD showcase the potential of problem-tailored mixers, but do not provide a general strategy for choosing mixers for different optimization problems. In Ref.~\cite{zhu2020adaptive} Zhu et al. replaced the fixed mixer $H_M$ by a set of different mixers $A_k$ that change from layer to layer. They entitled this variation of QAOA as ADAPT-QAOA. This adaptive approach would dramatically shorten the depth of QAOA layers while significantly improving the quality of the solution. Here we depict the quantum circuit schematic of ADAPT-QAOA in Fig.~\ref{adapt}. \newline

\begin{figure}[h!]
\centering
\includegraphics[width=0.9\linewidth]{ 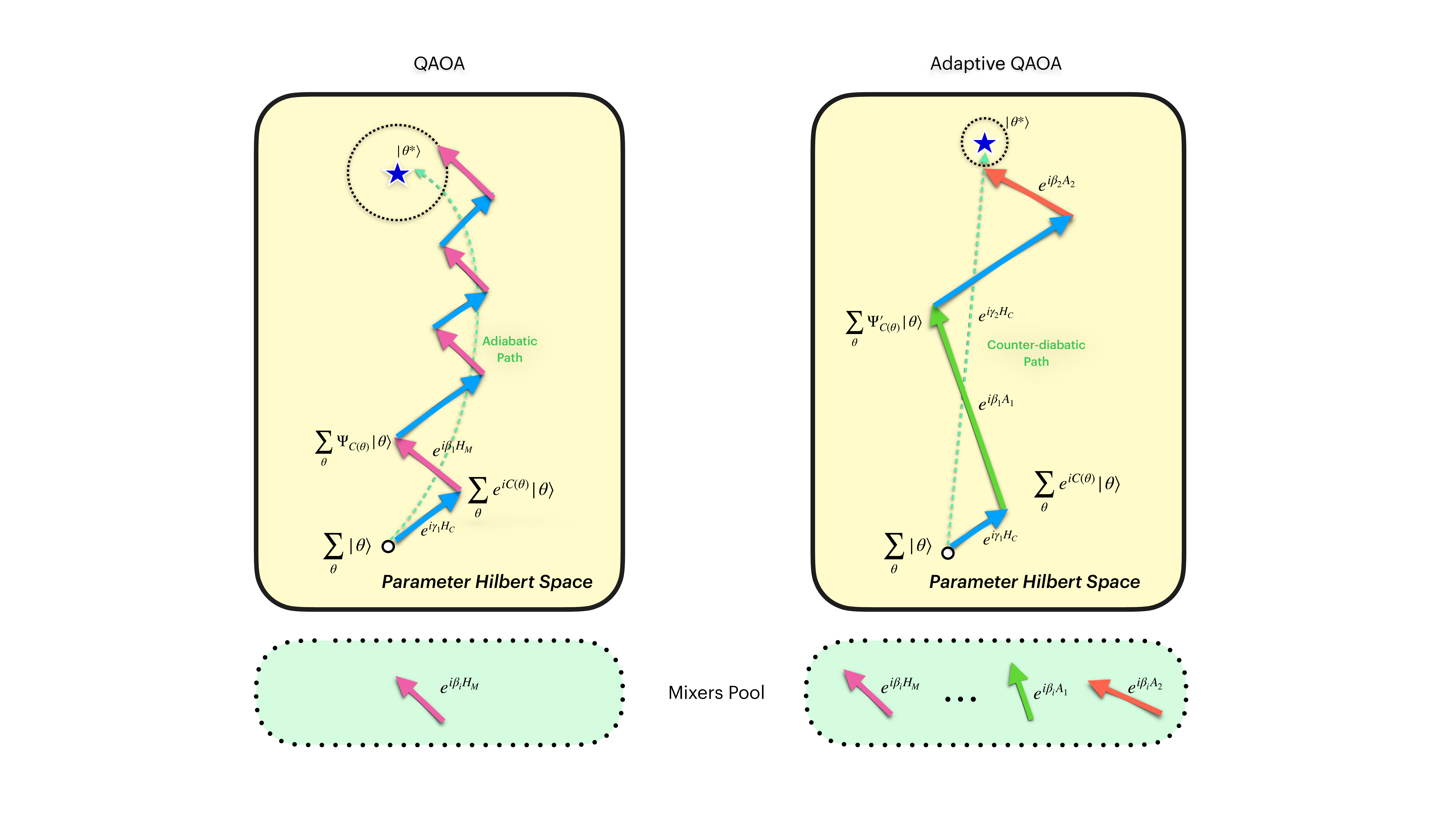}
\caption{{\textit{Comparison of original QAOA and ADAPT-QAOA}}. In the left and right panels of this figure, we depict the state change in the Hilbert space of the parameters to be optimized, for the original QAOA and ADAPT-QAOA respectively. The starting state $\sum_{\theta} \ket{\theta}$ (omitting the normalization factor), represented by the rounded dot at the bottom of each space, is the even superposition state of all possible solutions. The arrows represent the state evolution generated by the cost Hamiltonian and mixer Hamiltonian, and the color and direction of the arrows indicate the nature of the evolution. Blue arrows represent the evolution by the cost Hamiltonian. Arrows of other colors represent the evolution by different mixer Hamiltonians. In the original QAOA, there is only one mixer (shown in pink) available. Whereas, in ADAPT-QAOA there are more alternative mixers to chose from the mixers pool. The two algorithm try to reach the target state $\ket{\theta^{*}}$ (represented by the blue star) by stacking these arrows, which represent the alternating operations of two QAOAs. For reference we sketched the relevant paths --- adiabatic path for the original QAOA and counter-diabatic path for ADAPT-QAOA --- along the state evolution of the two QAOAs. As can be seen, the ADAPT-QAOA takes much fewer iterations to reach a closer point to the target state. This illustrates that compared to the original QAOA, allowing alternative mixers enables ADAPT-QAOA to dramatically improve algorithmic performance while achieving rapid convergence.}
\label{adaptspace}
\end{figure}

Compared to the original QAOA, allowing $Y$ mixers and entangling mixers enables ADAPT-QAOA to dramatically improve algorithmic performance while achieving rapid convergence for problems with complex structures. This effect of the adaptive mechanism can be illustrated in Fig.~\ref{adaptspace}.\newline

The advantage of this adaptive ansatz may come from the counter-diabatic (CD) driving mechanism. Numerical evidence shows that the adaptive mixer sequence chosen by the algorithm coincides with that of ``shortcut to adiabaticity'' by CD driving \cite{zhu2020adaptive}. Inspired by the CD driving procedure, another variant of QAOA, CD-QAOA \cite{yao2020reinforcement}, also uses an adaptive ansatz to achieve similar advantages. CD-QAOA is designed for preparing the ground state of quantum-chaotic many-body spin chains. By using terms occurring in the adiabatic gauge potential as additional control unitaries, CD-QAOA can achieve fast high-fidelity many-body control.

\begin{figure}
\centering
\includegraphics[width=\linewidth]{ 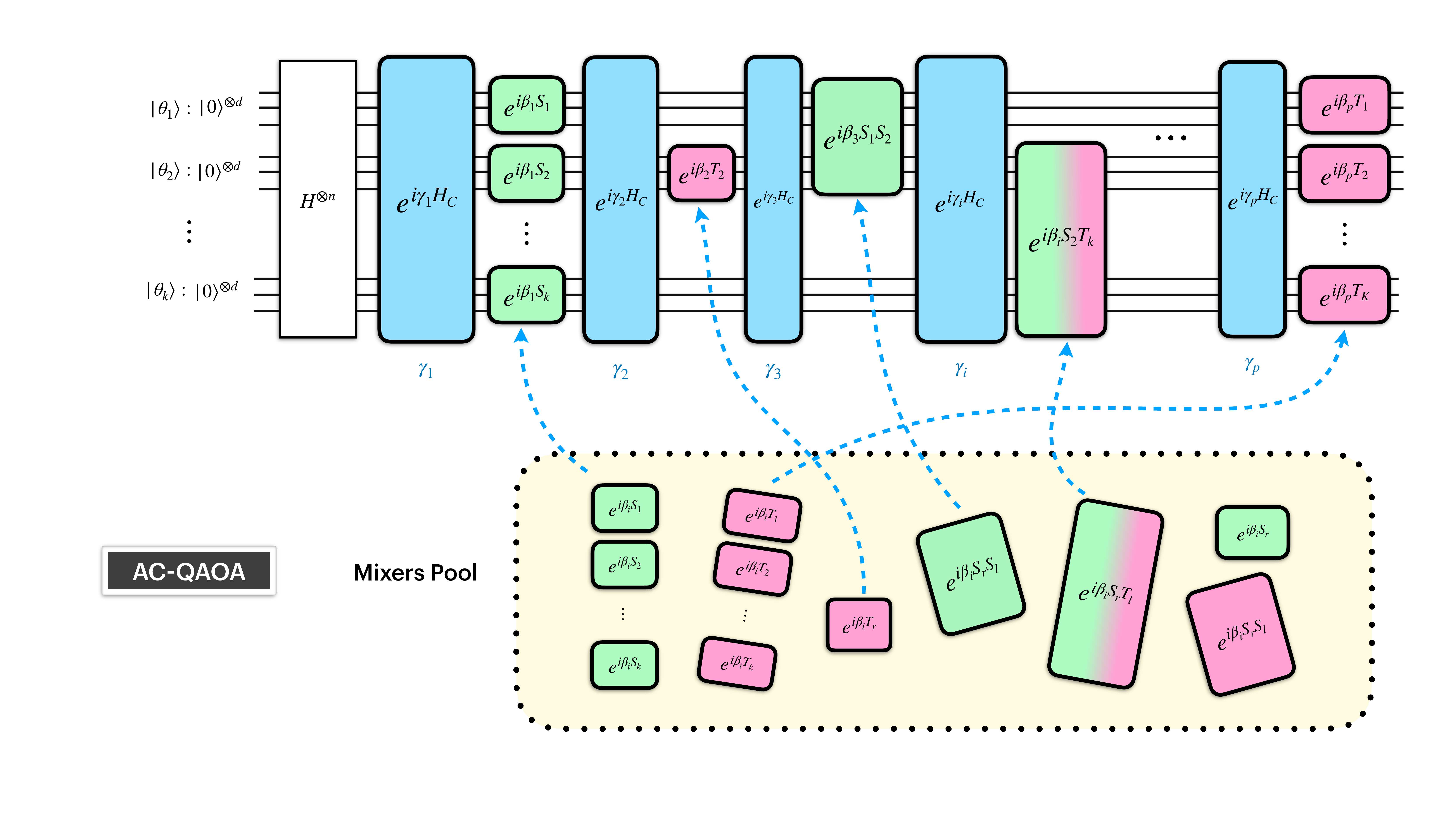}
\caption{\emph{Quantum circuit schematic of AC-QAOA}. AC-QAOA is a variant of QAOA we designed for solving optimisation of continuous variables with the short-depth advantage of QAOA layers.  In this figure, $\theta_i$ are the continuous variables to be optimized in the training. Each $\theta_i$ is digitized into binary form and stored in an independent register. The overall process of AC-QAOA is similar to that of the original QAOA, with the difference being as follows. \textit{1.} The mixers of AC-QAOA  with Hamiltonians $S_i$ and $T_i$ are acting on the registers of $\theta_i$ (rather than single qubits as in the original QAOA). \textit{2.} The mixers of AC-QAOA contain alternative mixers taken from a \emph{mixers pool} and can vary from layer to layer.}
\label{acqaoa}
\end{figure}

Inspired by above variants of QAOA, we design a new variant of QAOA tailored for our QNN training problem. In our case, for QNN training, the parameters we are optimizing (the angles of rotation gates) are continuous (real) values. Therefore, the choice of mixer Hamiltonian has to be adapted (as in QDD). We also want take advantage of including alternative mixers and allowing adaptive mixers for different layers (as in ADAPT-QAOA). Thus, the proper QAOA ansatz for our QNN training problem should be an adaptive continuous version of QAOA, which we call we call \emph{AC-QAOA}. Here we depict the the quantum circuit schematic of AC-QAOA in Fig.~\ref{acqaoa}.

\subsubsection{Grover Adaptive Search}\label{grovers}

Grover’s algorithm is generally used as a search method to find a set of desired
solutions from a set of possible solutions. Dürr and Høyer presented an
algorithm based on Grover’s method that finds
an element of minimum value inside an array of $N$ elements using on the order of $O( \sqrt{N} ) $ queries to the oracle \cite{1996quant.ph..7014D}.
Baritompa et al. \cite{Baritompa2005} applied Grover’s algorithm for global optimization, which they call Grover Adaptative Search (GAS). GAS has been applied in training classical neural networks \cite{10.1088/1367-2630/abc9ef} and polynomial binary optimization \cite{gilliam2020grover}. In the following we outline GAS.\newline

Consider a function $f: X \rightarrow \mathbb{R}$, where for ease of presentation assume $X = \{0, 1\}^n$. we are interested in solving $\min_{x \in X} f(x)$. The main idea of GAS is to construct an “adaptive” oracle for a given threshold $y$ such that it flags all states $x \in X$ satisfying $f(x) < y$, namely the oracle marks a solution $x$ if and only if another boolean function $g_y$ satisfies $g_y(x) = 1$, where
\begin{equation}
  g_{y}(x) =
    \begin{cases}
      1 & \text{if} \quad f(x)<y\\
      0 & \text{otherwise}
    \end{cases}       ,
\end{equation}
The oracle $\mathcal{O}_{\textsf{\textit{Grover}}}$ then act as
\begin{equation}
  \mathcal{O}_{\textsf{\textit{Grover}}}\ket{x} =(-1)^{g_{y}(x)}\ket{x}
  \label{groveroracle}.
\end{equation}

We use Grover search to find a solution $\tilde{x}$ with a function value better than $y$. Then we set $y = f(\tilde{x})$ and repeat until some formal termination criteria is met --- for example, based on the number of iterations, time, or progress in $y$.\newline

\subsection{Swap test, Hadamard test, and the Grover operator}

This section introduces the Swap test, Hadamard test, and their corresponding Grover operators, which will be used in the phase encoding of the cost function of QNNs.

\tocless\subsubsection{Swap Test and its Grover operator}

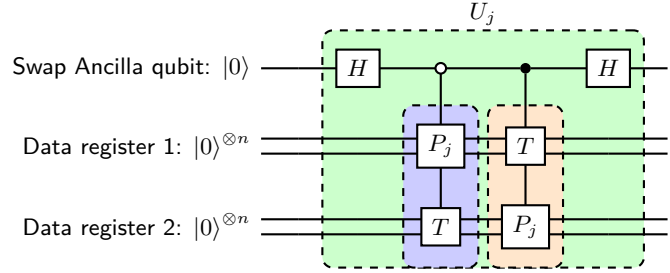
\begin{figure}[htbp]
\centering
\begin{quantikz}[row sep=0.5cm]
  \lstick{\textsf{Swap Ancilla qubit:} \ket{0}}& \qw &\gate{H}\gategroup[3,steps=4,style={dashed,rounded corners,fill=green!20, inner xsep=2pt},background]{{$U_j$}} &\octrl{2} &\ctrl{2} &\gate{H} &\qw\\
 \lstick{\textsf{Data register 1:} $\ket{0}^{\otimes n}$}& \qwbundle[alternate=2]{} & \qwbundle[alternate=2]{} & \gate[nwires=1]{P_j}\qwbundle[alternate=2]{}\gategroup[2,steps=1,style={dashed,rounded corners,fill=blue!20, inner xsep=2pt},background]{{}} & \gate[nwires=1]{T}\qwbundle[alternate=2]{}\gategroup[2,steps=1,style={dashed,rounded corners,fill=orange!20, inner xsep=2pt},background]{{}} &\qwbundle[alternate=2]{}&\qwbundle[alternate=2]{}\\
 \lstick{\textsf{Data register 2:} $\ket{0}^{\otimes n}$}& \qwbundle[alternate=2]{} & \qwbundle[alternate=2]{} & \gate[nwires=1]{T}\qwbundle[alternate=2]{} & \gate[nwires=1]{P_j}\qwbundle[alternate=2]{} &\qwbundle[alternate=2]{}&\qwbundle[alternate=2]{}
\end{quantikz}
\caption{\textit{Circuit diagram of swap test}. Here we present an alternative form of swap test: instead of applying the swap operation on two quantum states, the circuit in this figure simulate the ``swap'' effect by applying two unitaries $P_j,T$ on two registers in different order controlled by an ancilla qubit. The ``anti-control'' symbol is defined as: when the control qubit is in state $\ket{0}$, the unitary being controlled is executed; when the control qubit is in state $\ket{1}$, the unitary being controlled is not executed.}
\label{swaptest}
\end{figure}

Let $\ket{p_j}, \ket{t}$ be the resulting quantum states of unitary operators $P_j$ and $T$, respectively --- that is,  $\ket{p_j}=P_j|0\rangle^{\otimes n}$ and $\ket{t}=T|0\rangle^{\otimes n}$.
The swap test is a technique that can be used to estimate $|\langle p_j|t\rangle|^2$ \cite{Buhrman_2001}.
The circuit of swap test is shown in Fig.~\ref{swaptest}.

We denote the unitary of the Swap test circuit (dotted green box in Fig.~\ref{swaptest}) as $U_{j}$, which can be written as
\begin{align}
U_{j}\coloneqq [H\otimes I\otimes I]\cdot[\ket{0}\bra{0}\otimes P_{j}\otimes T+\ket{1}\bra{1}\otimes T\otimes P_{j}]\cdot[H\otimes I\otimes I]
\label{defineUj}.
\end{align}

The output state from $U_{j}$ is denoted as $|\phi_j\rangle$:
\begin{equation}
  |\phi_j\rangle=\frac{1}{\sqrt{2}}(\ket{+}\ket{p_j}\ket{t}+\ket{-}\ket{t}\ket{p_j}).
\end{equation}

Rearranging the terms we have
\begin{equation}
  |\phi_j\rangle=\frac{1}{2}\left(|0\rangle\otimes(|p_j\rangle\ket{t}+|t\rangle\ket{p_j})+|1\rangle\otimes(|p_j\rangle\ket{t}-|t\rangle\ket{p_j})\right).
\label{sw}
\end{equation}

Denote $|u_j\rangle$ and $|v_j\rangle$ as the normalized states of $|p_j\rangle\ket{t}+|t\rangle\ket{p_j}$ and $|p_j\rangle\ket{t}-|t\rangle\ket{p_j}$ respectively.
Then there is a real number $\theta_j\in[{\pi}/{4},{\pi}/{2}]$ such that
\begin{align}\label{amplitudeencoding}
|\phi_j\rangle=\sin\theta_j|0\rangle|u_j\rangle+\cos\theta_j|1\rangle|v_j\rangle.
\end{align}
 $\theta_j$ satisfies $\cos\theta_j=\sqrt{1- |\langle p_j|t\rangle|^2}/\sqrt{2}$, $\sin\theta_j=\sqrt{1+ |\langle p_j|t\rangle|^2}/\sqrt{2}$, therefore we have:
\begin{equation}
 |\langle p_j|t\rangle|^2=-\cos{2\theta_j}.
\label{relation}
\end{equation}

From Eq.~\ref{relation} and Eq.~\ref{amplitudeencoding} we can see that the value of $|\langle p_j|t\rangle|^2$ is encoded in the amplitude of the output state $|\phi_j\rangle$ of swap test. This will be used in the amplitude encoding of QNN cost function which is a crucial component of the quantum training.\newline

Applying the Schmidt decomposition method to state $|\phi_j\rangle$ we arrive at
\begin{align}\label{eq_phi_schmidt}
|\phi_j\rangle=\frac{-  i}{\sqrt{2}}(e^{  i \theta_j}|w_+\rangle_j-e^{-  i\theta_j}|w_-\rangle_j),
\end{align}
where
$
|w_{\pm}\rangle_j=\frac{1}{\sqrt{2}}(|0\rangle|u_j\rangle\pm  i|1\rangle|v_j\rangle).
$
\newline

\begin{figure}[htbp]
\begin{quantikz}
 \lstick{\textsf{Swap Ancilla qubit:} \ket{0}}& \qw &\gate{Z}\gategroup[3,steps=5,style={dashed,rounded corners, inner xsep=2pt},background]{{$G_j$}} & \gate[wires=3,nwires={2,3},style={rounded corners,fill=green!20}]{U_{j}^{\dag}} & \gate[wires=3,nwires={2,3},style={fill=blue!12, inner xsep=2pt}]{C} & \gate[wires=3,nwires={2,3},style={rounded corners,fill=green!20}]{U_{j}} &\qw \\
 \lstick{\textsf{Data register 1:} $\ket{0}^{\otimes n}$}&\qwbundle[alternate=2]{}&\qwbundle[alternate=2]{}&\qwbundle[alternate=2]{}&\qwbundle[alternate=2]{}&\qwbundle[alternate=2]{}&\qwbundle[alternate=2]{}\\
 \lstick{\textsf{Data register 2:} $\ket{0}^{\otimes n}$}&\qwbundle[alternate=2]{}&\qwbundle[alternate=2]{}&\qwbundle[alternate=2]{}&\qwbundle[alternate=2]{}&\qwbundle[alternate=2]{}&\qwbundle[alternate=2]{}
\end{quantikz}
\caption{\emph{Quantum circuit to implement $G_j$}. $G_j$ is defined as $G_j:= U_{j}C U_{j}^\dag (Z\otimes I^{\otimes 2n})$.}
\label{gjf}
\end{figure}
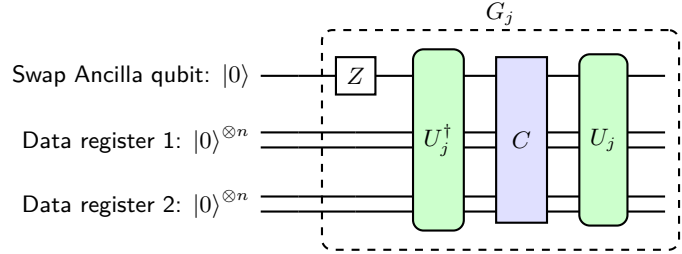

One can construct a Grover operator using $U_j$ as follows:

\begin{align}
G_j &:= (I^{\otimes (2n+1)}-2|\phi_j\rangle\langle\phi_j|)(Z\otimes I^{\otimes 2n}), \\
&= U_{j}C U_{j}^\dag (Z\otimes I^{\otimes 2n}),
\label{gj}
\end{align}
where $Z=|0\rangle\langle 0| - |1\rangle\langle 1|$ is the Pauli-$Z$ operator,
 $C=I^{\otimes (2n+1)}-2|0\rangle^{\otimes (2n+1)}\langle0|^{\otimes (2n+1)}$ can be implemented as the circuit shown in Fig.~\ref{uc}. The circuit representation of $G_j$ is shown in Fig.~\ref{gjf}.\newline

\begin{figure}[htbp]
\begin{quantikz}
&\octrl{1}\gategroup[5,steps=1,style={dashed,fill=blue!12, inner xsep=2pt},background]{{$C$}}&\qw\\
&\octrl{1}&\qw\\
&\vdots&\\
&\octrl{1}&\qw\\
&\gate{-Z}\qw&\qw
\end{quantikz}
\caption{Quantum circuit to implement unitary $C=I^{\otimes (2n+1)}-2|0\rangle^{\otimes (2n+1)}\langle0|^{\otimes (2n+1)}$.}
\label{uc}
\end{figure}
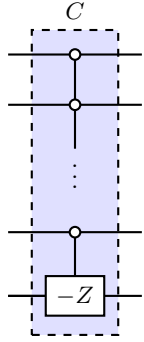

It is easy to check that $|w_{\pm}\rangle_j$ are the eigenstates of $G_j$. --- that is,
\begin{align}\label{eq_Gw}
G_j|w_{\pm}\rangle=e^{\pm  i  2\theta_j}|w_{\pm}\rangle_j.
\end{align}
Recall from Eq.~\ref{relation} the value of $|\langle p_j|t\rangle|^2$ is encoded in the phase of the eigenvalue of $G_j$. This will be used in the phase encoding of QNN cost function which is a crucial component of the quantum training.\\

\tocless\subsubsection{Hadamard Test and its ``Grover operator''}

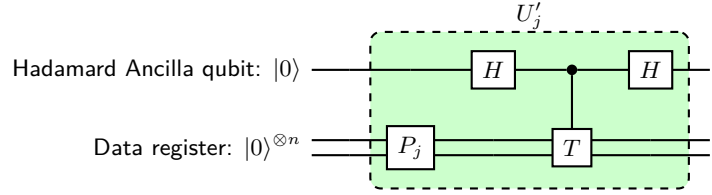
\begin{figure}[h!]
  \centering
  \begin{quantikz}
  \lstick{\textsf{Hadamard Ancilla qubit:} \ket{0}} & \qw&\qw\gategroup[wires=2,steps=4,style={dashed,rounded corners,fill=green!20, inner xsep=3pt},background]{$U'_{j}$} &\gate{H}  & \ctrl{1} &\gate{H}& \qw \\
  \lstick{\textsf{Data register:}  $\ket{0}^{\otimes n}$}\qwbundle[alternate=2]{}& \qwbundle[alternate=2]{} & \gate[nwires=1]{P_j}\qwbundle[alternate=2]{}& \qwbundle[alternate=2]{}& \gate[nwires=1]{T}\qwbundle[alternate=2]{}& \qwbundle[alternate=2]{}& \qwbundle[alternate=2]{}
  \end{quantikz}
  \caption{\emph{Circuit diagram of Hadamard Test}. The circuit is used to estimate  $\bra{0}P_j^{\dagger}TP_j\ket{0}$, for two unitary $P_j$ and $T$. The Hadamard test will be used the phase encoding of QNN cost function which is a crucial component of the quantum training.  }
  \label{Hadamard}
\end{figure}

Similar to the swap test, the Hadamard test is a technique that can be used to estimate  $\bra{0}P_j^{\dagger}TP_j\ket{0}$, for two unitary operators $P_j$ and $T$ (assuming $T$ is Hermitian).
The circuit of Hadamard test is shown in Fig.~\ref{Hadamard}.

We denote the unitary of the Hadamard test circuit (the dotted green box in Fig.~\ref{Hadamard}) as  $U'_{j}$ and the output state from $U'_{j}$ as

\begin{equation}
  |\phi'_j\rangle=\frac{1}{\sqrt{2}}(\ket{+}P_j\ket{0}+\ket{-}TP_j\ket{0}).
\end{equation}

Rearranging the terms we have
\begin{equation}
  |\phi'_j\rangle=\frac{1}{2}(|0\rangle\otimes(P_j\ket{0}+TP_j\ket{0})+|1\rangle\otimes(P_j\ket{0}-TP_j\ket{0}).
\end{equation}

Denote $|u'_j\rangle$ and $|v'_j\rangle$ as the normalized states of $P_j\ket{0}+TP_j\ket{0}$ and $P_j\ket{0}-TP_j\ket{0}$ respectively.
Then there is a real number $\theta'_j\in[0, {\pi}/{2}]$ such that
\begin{align}\label{hadard}
|\phi'_j\rangle=\sin\theta'_j|0\rangle|u'_j\rangle+\cos\theta'_j|1\rangle|v'_j\rangle.
\end{align}
$\theta'_j$ satisfies $\cos\theta'_j=\sqrt{1- \bra{0}P_j^{\dagger}TP_j\ket{0}}/\sqrt{2}$, $\sin\theta'_j=\sqrt{1+ \bra{0}P_j^{\dagger}TP_j\ket{0}}/\sqrt{2}$, (assuming $T$ is Hermitian). Therefore we have
\begin{equation}
  \label{hadamardrelation}
 \bra{0}P_j^{\dagger}TP_j\ket{0}=-\cos{2\theta'_j}.
\end{equation}

We can define the Grover operator $G'_j$ from $U'_j$ in the same way as in last subsection for the swap test and obtain similar eigen-relation. The value of $\bra{0}P_j^{\dagger}TP_j\ket{0}$ is encoded in the phase of the eigenvalue of $G'_j$. This will be used in the phase encoding of QNN cost function which is a crucial component of the quantum training.\newline

\subsection{Creating Superpositions of QNNs}\label{par}

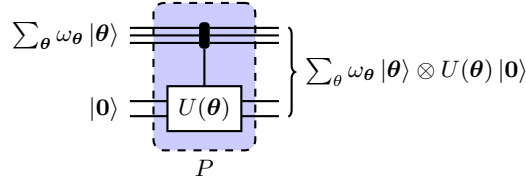
\begin{figure}[h!]
    \centering
\begin{quantikz}
\lstick{$\sum_{\btheta}\omega_{\btheta}\ket{\btheta}$} & \ctrlbundle{1} \gategroup[2,steps=1,style={dashed,rounded corners,fill=blue!20, inner xsep=2pt},background,label style={label position=below,anchor=north,yshift=-0.2cm}]{$P$}& \rstick[wires=2]{$\sum_{\theta}\omega_{\btheta}\ket{\btheta}\otimes U({\btheta})\ket{\mathbf{0}}$}\qwbundle[alternate]{} \\
\lstick{\ket{\mathbf{0}}}& \gate[nwires=1]{U({\btheta})}\qwbundle[alternate=2]{} & \qwbundle[alternate=2]{}
\end{quantikz}
    \caption{\emph{Action of the controlled unitary $P$}. In this figure, the upper register is parameter register and the lower register is the QNN register. $\boldsymbol \theta = (\theta_1,\dots, \theta_M)$ is the set of trainable parameters in the QNN and $U({\btheta})$ is the unitary of the QNN with corresponding parameters. The qubits in the parameter register act as control qubits on the rotation gates in the QNN. The controlled operations (in the dotted blue box) is denoted as $P$. When $P$ acts on a superposition state of parameters $\sum_{\btheta}\omega_{\btheta}\ket{\btheta}$, the output state is $\sum_{\btheta}\omega_{\btheta}\ket{\btheta}\otimes U({\btheta})\ket{\mathbf{0}}.$ in which the parameter register and QNN register are entangled.
  }
    \label{controlp}
\end{figure}

 As an essential building block for our quantum training protocol, we present a way to create superpositions of QNNs entangled with corresponding parameters. That is, we construct a controlled unitary $P$ such that
\begin{equation}\label{Aoperator}
    P \ket{\btheta} \ket{\mathbf{0}} \to  \ket{\btheta}\otimes U({\btheta})\ket{\mathbf{0}} \quad \textrm{for every $\btheta$}
\end{equation}
in which $\boldsymbol \theta = (\theta_1,\dots, \theta_M)$ is the set of trainable parameters in the QNN and $U({\btheta})$ is the unitary of the QNN with corresponding parameters. When $P$ acts on a superposition state of parameters $\sum_{\btheta}\omega_{\btheta}\ket{\btheta}$, we have
\begin{equation}\label{cqnn}
    P \sum_{\btheta}\omega_{\btheta}\ket{\btheta} \ket{\mathbf{0}} \to  \sum_{\btheta}\omega_{\btheta}\ket{\btheta}\otimes U({\btheta})\ket{\mathbf{0}}.
\end{equation}
The action of the controlled unitary $P$ is depicted in Fig.~\ref{controlp}.\newline

\begin{figure}[h!]
\centering
\begin{adjustbox}{width=0.7\textwidth}
\begin{quantikz}[row sep=0.5cm]
\lstick[wires=3]{\textsf{Parameter register}\\for $\theta$ in $R_z(\theta)$} & \lstick{$\ket{0}$} & \gate{H} & \qw\gategroup[wires=5,steps=5,style={dashed,rounded corners,fill=blue!20, inner xsep=3pt},background]{$P$} & \ctrl{3} & \qw  & \qw & \qw& \qw\\
&\lstick{$\ket{0}$} & \gate{H} & \qw& \qw      & \ctrl{2} & \qw & \qw& \qw\\
&\lstick{$\ket{0}$} & \gate{H} & \qw& \qw      & \qw       & \ctrl{1} & \qw\\
&\lstick{$\ket{0}$} & \qw & \qw     & \gate{R_z(\bar{\theta}/2)}\hphantom{very wide}\gategroup[1,steps=3,style={dashed,rounded corners,fill=red!20, inner xsep=2pt},background,label style={label position=below,anchor=north,yshift=-0.2cm}]{{\sc $R_z(\theta)$}}  &\gate{R_z(\bar{\theta}/4)}\hphantom{wide}   &\gate{R_z(\bar{\theta}/8)} & \qw \\
&  & &&&&& &
\end{quantikz}
\end{adjustbox}
\caption{\emph{An example of the construction of $P$ for one rotation gate $R_z({\theta})$}. In this example, the parameter register consist of three qubits, each qubit controls a ``partial'' rotation on the fourth qubit. The ``partial'' rotation are the binary segments $R_z({\bar{\theta}/2})$,$R_z({\bar{\theta}/4})$,$R_z({\bar{\theta}/8})$ in which $\bar{\theta}$ is the maximum value that angle $\theta$ can take. }
\label{control1}
\end{figure}
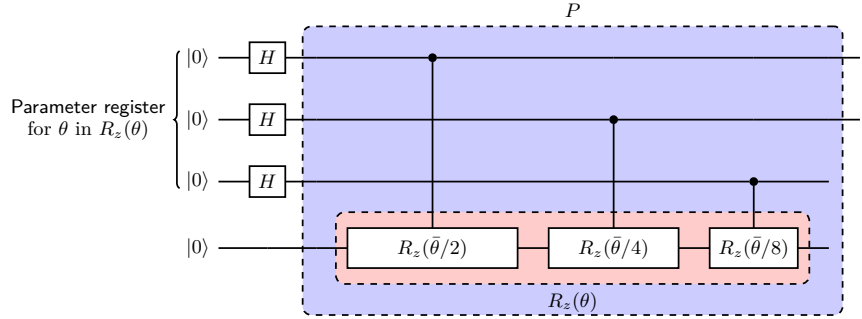

This controlled unitary can be realized by dividing each rotation gate in QNN into a sequence of binary segments, followed by applying controlled operations on them. A simple example of one rotation gate, for example $U({\btheta})=R_z(\theta)$, is illustrated in Fig.~\ref{control1}. \newline

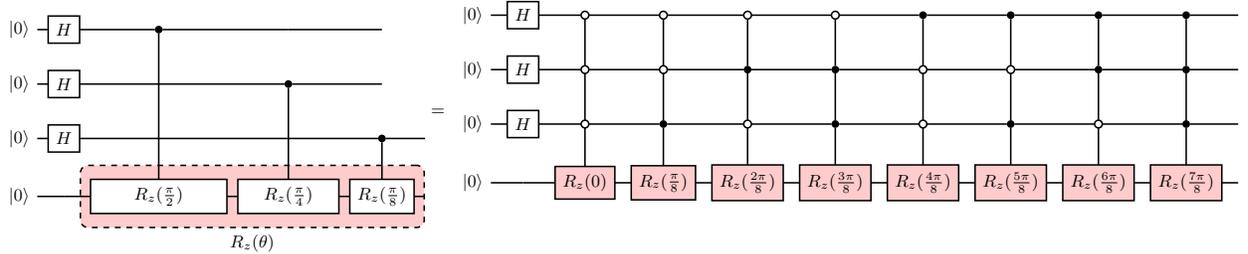
\begin{figure}[h!]
\centering
\begin{adjustbox}{width=\textwidth}
\begin{quantikz}[row sep=0.5cm,column sep=0.2cm]
\lstick{$\ket{0}$} & \gate{H} & \ctrl{3} & \qw  & \qw \\
\lstick{$\ket{0}$} & \gate{H} & \qw      & \ctrl{2} & \qw \\
\lstick{$\ket{0}$} & \gate{H} & \qw      & \qw       & \ctrl{1} & \qw\\
\lstick{$\ket{0}$} & \qw      & \gate{R_z(\frac{\pi}{2})}\hphantom{very wide}\gategroup[wires=1,steps=3,style={dashed,rounded corners,fill=red!20, inner xsep=2pt},background,label style={label position=below,anchor=north,yshift=-0.2cm}]{{\sc $R_z(\theta)$}}  &\gate{R_z(\frac{\pi}{4})}\hphantom{wide}   &\gate{R_z(\frac{\pi}{8})} & \qw
\end{quantikz}
=
\begin{quantikz}[column sep=0.3cm]
\lstick{$\ket{0}$} 	& 	\gate{H} 	& 	\octrl{1} 	& 	\octrl{1} 	& 	\octrl{1} 	& 	\octrl{1} 	& 	\ctrl{1} 	& 	\ctrl{1} 	& 	\ctrl{1} 	& 	\ctrl{1} 	&	\qw	\\
\lstick{$\ket{0}$} 	& 	\gate{H} 	& 	\octrl{1} 	& 	\octrl{1} 	& 	\ctrl{1} 	& 	\ctrl{1} 	& 	\octrl{1} 	& 	\octrl{1} 	& 	\ctrl{1} 	& 	\ctrl{1} 	&	\qw	\\
\lstick{$\ket{0}$} 	& 	\gate{H} 	& 	\octrl{1} 	& 	\ctrl{1} 	& 	\octrl{1} 	& 	\ctrl{1} 	& 	\octrl{1} 	& 	\ctrl{1} 	& 	\octrl{1} 	& 	\ctrl{1} 	&	\qw	\\
\lstick{$\ket{0}$} 	& 	\qw	& 	\gate[wires=1,style={fill=red!20}]{R_z(0)}	& 	\gate[wires=1,style={fill=red!20}]{R_z(\frac{\pi}{8})}	& 	\gate[wires=1,style={fill=red!20}]{R_z(\frac{2\pi}{8})}	& 	\gate[wires=1,style={fill=red!20}]{R_z(\frac{3\pi}{8})}	& 	\gate[wires=1,style={fill=red!20}]{R_z(\frac{4\pi}{8})}	& 	\gate[wires=1,style={fill=red!20}]{R_z(\frac{5\pi}{8})}	& 	\gate[wires=1,style={fill=red!20}]{R_z(\frac{6\pi}{8})}	& 	\gate[wires=1,style={fill=red!20}]{R_z(\frac{7\pi}{8})}	&	\qw	\\
\end{quantikz}
\end{adjustbox}
\caption{\emph{An example of the effect of $P$ defined in Fig.~\ref{control1}}. Each bit string of the parameter register can be seen as a binary representation of the rotation angle and the associated basis state of the register is entangled with the rotation gate of the corresponding angle. For instance, in the example above, the bit string {111} corresponds to the angle $7\bar{\theta}/8$ and \ket{111} is associated with $R_z(7\bar{\theta}/8)$. }
\label{equal}
\end{figure}

Each bit string of the parameter register can be seen as a binary representation of the rotation angle and the associated basis state of the register is entangled with the rotation gate of the corresponding angle. For instance, in the example above, the bit string {111} corresponds to the angle $7\bar{\theta}/8$ and \ket{111} is associated with $R_z(7\bar{\theta}/8)$, where $\bar{\theta}$ is the maximum value that angle $\theta$ can take. This relation can be fully illustrated in Fig.~\ref{equal}, in which we take  $\bar{\theta}=\pi$. \newline

\begin{figure}[h]
\centering
\begin{adjustbox}{width=0.68\textwidth}
\begin{quantikz}[row sep=0.5cm]
\lstick[wires=3]{$\ket{0}^{\otimes 3}$\\\textit{Parameter register} \\for $\theta_1$ in $R_z(\theta_1)$} 	&	\gate{H}&	\qw	\gategroup[wires=8,steps=8,style={dashed,rounded corners,fill=blue!20, inner xsep=3pt},background]{$P$} &	\ctrl{6}	&	\qw	&	\qw	&	\qw	&	\qw	&	\qw	&	\qw		&	\qw	\\
	&	\gate{H}&	\qw	&	\qw	&	\ctrl{5}	&	\qw	&	\qw	&	\qw	&	\qw	&	\qw		&	\qw	\\
	&	\gate{H}&	\qw	&	\qw	&	\qw	&	\ctrl{4}	&	\qw	&	\qw	&	\qw	&	\qw		&	\qw	\\
\lstick[wires=3]{$\ket{0}^{\otimes 3}$\\\textit{Parameter register} \\for $\theta_2$ in $R_z(\theta_2)$} 	&	\gate{H}&	\qw	&	\qw	&	\qw	&	\qw	&	\ctrl{4}	&	\qw	&	\qw	&	\qw		&	\qw	\\
	&	\gate{H}&	\qw	&	\qw	&	\qw	&	\qw	&	\qw	&	\ctrl{3}	&	\qw	&	\qw		&	\qw	\\
	&	\gate{H}&	\qw	&	\qw	&	\qw	&	\qw	&	\qw	&	\qw	&	\ctrl{2}	&	\qw		&	\qw	\\
\lstick{$\ket{0}$}&	\qw	&	\qw	&	\gate{R_z(\bar{\theta_1}/2)}\gategroup[1,steps=3,style={dashed,rounded corners,fill=red!20, inner xsep=2pt},background,label style={label position=below,anchor=north,yshift=-0.2cm}]{{\sc $R_z(\theta_1)$}}	&	\gate{R_z(\bar{\theta_1}/4)}	&	\gate{R_z(\bar{\theta_1}/8)}	&	\qw	&	\qw	&	\qw	&	\qw		&	\qw	\\
\lstick{$\ket{0}$}&	\qw	&	\qw	&	\qw	&	\qw	&	\qw	&	\gate{R_z(\bar{\theta_2}/2)}\gategroup[1,steps=3,style={dashed,rounded corners,fill=red!20, inner xsep=2pt},background,label style={label position=below,anchor=north,yshift=-0.2cm}]{{\sc $R_z(\theta_2)$}}	&	\gate{R_z(\bar{\theta_2}/4)}	&	\gate{R_z(\bar{\theta_2}/8)}	&	\qw	&	\qw
\end{quantikz}
\end{adjustbox}
\caption{\emph{Example of the construction of $P$ for QNN consisting of two rotation gates}. In this example, the QNN consist of two rotation gates $R_z(\theta_1)$, $R_z(\theta_2)$ on the lower two qubits. The upper 6 qubits are divided into two parameter register for the two rotation angles $\theta_1$, $\theta_2$ respectively. Each qubit controls a "partial" rotation. For instance, the "partial" rotations of $R_z(\theta_1)$ are the binary segments $R_z({\bar{\theta_1}/2})$,$R_z({\bar{\theta_1}/4})$,$R_z({\bar{\theta_1}/8})$ in which $\bar{\theta_1}$ is the maximum value that angle $\theta_1$ can take. }
\label{control}
\end{figure}
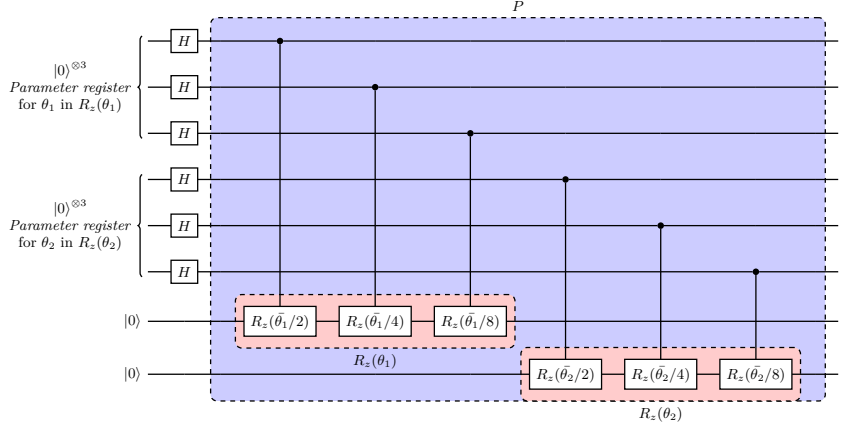

The unitary operator of $P$ can be written as:
\begin{equation}
\label{QNNc}
P =\sum_{j}\left|j \left>\right<j\right|\otimes P_{j},
\end{equation}
in which $P_{j}$ is a specific configuration of the QNN defined by its control bit string $j$. This representation does not only apply to a single rotation gate, but also to the case where there are multiple parameterized rotation gates in the QNN. An example of two rotation gates is depicted in Fig.~\ref{control}.  \newline

In order to achieve precision $\epsilon_0$ for each rotation angle, the number of control qubits needed is $d=\lceil \log_2 (1/\epsilon_0) \rceil $ . Let $r$ be the number of rotation gates in a QNN, then the total number of control qubits needed is $dr$.

\section{QNN training by Grover Adaptive Search}\label{GAS}

In this section we discuss using Grover adaptive search to perform global optimisation of QNNs. As presented in Section \ref{grovers}, the core of the Grover adaptive search is the adaptive oracle defined in Eq.~\ref{groveroracle}. Next we detail how to construct such oracle for QNN training.

\subsection{Construction of the Grover Oracle}
The adaptive Grover Oracle $\mathcal{O}_{\textsf{\textit{Grover}}}$ in the context of QNN training acts as
\begin{equation}
    \mathcal{O}_{\textsf{\textit{Grover}}}\ket{\btheta}\otimes\ket{\mathbf{0}}_{\textsf{\textit{QNN}+\textsf{\textit{ancillas}}}} = (-1)^{g(C(\btheta)-C^*)}\ket{\btheta}\otimes\ket{\mathbf{0}}_{\textsf{\textit{QNN}+\textit{ancillas}}},  \quad   \textrm{for every $\btheta$,}
    \label{groverqnn}
\end{equation}
in which $C^*$ is the adaptive threshold for the cost function and the function $g$ is defined as
\begin{equation}
  g(x) =
    \begin{cases}
      1 & x<0\\
      0 & \text{otherwise}
    \end{cases}.
\end{equation}

When $\mathcal{O}_{\textsf{\textit{Grover}}}$ is acting on a superposition state of parameters $\sum_{\btheta}\omega_{\btheta}\ket{\btheta}$, we have
\begin{equation}
    \mathcal{O}_{\textsf{\textit{Grover}}}\sum_{\btheta}\omega_{\btheta}\ket{\btheta}\otimes\ket{\mathbf{0}}_{\textsf{\textit{QNN}}+\textsf{\textit{ancillas}}} = \sum_{\btheta}(-1)^{g(C(\btheta)-C^*)}\omega_{\btheta}\ket{\btheta}\otimes\ket{\mathbf{0}}_{\textsf{\textit{QNN}}+\textsf{\textit{ancillas}}}.
\end{equation}

The QNN Grover oracle $\mathcal{O}_{\textsf{\textit{Grover}}}$ can be constructed by the following steps.

\tocless\subsubsection{ \textbf{Amplitude Encoding}} \label{amplitude}

The first step is to encode the cost function of QNN into amplitude. Depending on the form of the cost function of the QNN, the amplitude encoding can be achieved by the swap test or Hadamard test. The correspondences are summarized in Table \ref{taskstable}.\newline

\begin{table}[htbp]
\begin{center}
\begin{tabular}{ |p{4cm}||p{2cm}|p{4.3cm}|p{3cm}|}
\hline
\textbf{Task}& \multicolumn{2}{c|}{\textbf{Cost function}}&\textbf{Amplitude encoding Method}\\
\hline
\textsf{Generating Ground state of $T$}  & \textsf{Expectation value }   &$C({\boldsymbol{\theta}})=\bra{0}U^{\dagger}(\boldsymbol{\theta}) T U(\boldsymbol{\theta})\ket{0}$&   \textit{\textbf{Hadamard test}}\\
\hline
\textsf{Generating a pure state $\ket{\psi}=T\ket{0}$ (  $T$ is a given unitary)} & \textsf{Fidelity}  & $C({\boldsymbol{\theta}})=|\bra{0}U^{\dagger}(\boldsymbol{\theta}) T\ket{0}|^2$   &\textit{\textbf{Swap test}}\\
\hline
\end{tabular}
\end{center}
\caption{\emph{QNN Cost functions for two type of tasks}. Here we present the Cost functions for two tasks respectively: For the task of Generating Ground state of some given Hamiltonian $T$ (we use $T$ instead of $H$ here, and assume $T$ is Hermitian), the cost function is chosen to be the Expectation value of $T$. For the task of Generating a pure state $\ket{\psi}=T\ket{0}$ ($T$ is a given unitary), the cost function is chosen to be the Fidelity between the generated state from QNN and the state $\ket{\psi}=T\ket{0}$.}\label{taskstable}
\end{table}

\paragraph{\textbf{Amplitude Encoding by Swap test.}}\label{swapsection}

\begin{figure}[h!]
  \centering
  \begin{quantikz}
  \lstick{\textsf{Swap Ancilla qubit:} \ket{0}} & \qw\slice{$\left|\Psi_{0}\right\rangle$}& \qw &\gate{H}\gategroup[wires=4,steps=6,style={dashed,rounded corners,fill=green!20, inner xsep=3pt},background]{$U$} & \octrl{1} & \octrl{3}& \ctrl{1} & \ctrl{2} &\gate{H}& \qw\slice{$\left|\Psi_{1}\right\rangle$}& \qw \\
  \lstick{\textsf{Parameter register:} $\ket{0}^{\otimes dr}$}&\gate{H^{\otimes dr}}\qwbundle[alternate]{} &\qwbundle[alternate]{} &\qwbundle[alternate]{}  & \ctrlbundle{1}\gategroup[3,steps=2,style={dashed,rounded corners,fill=blue!20, inner xsep=2pt},background,label style={label position=below,anchor=north,yshift=-0.2cm}]{{}}&\qwbundle[alternate]{} &\ctrlbundle{2}\gategroup[3,steps=2,style={dashed,rounded corners,fill=orange!20, inner xsep=2pt},background,label style={label position=below,anchor=north,yshift=-0.2cm}]{{}}&\qwbundle[alternate]{}&\qwbundle[alternate]{}&\qwbundle[alternate]{} &\qwbundle[alternate]{} \qwbundle[alternate]{} \\
  \lstick{\textsf{QNN register:}  $\ket{0}^{\otimes n}$}\qwbundle[alternate=2]{}& \qwbundle[alternate=2]{}& \qwbundle[alternate=2]{}& \qwbundle[alternate=2]{} & \gate[nwires=1]{P_j} \qwbundle[alternate=2]{}& \qwbundle[alternate=2]{}\qwbundle[alternate=2]{}& \qwbundle[alternate=2]{}& \gate[nwires=1]{T} \qwbundle[alternate=2]{}&\qwbundle[alternate=2]{}& \qwbundle[alternate=2]{}& \qwbundle[alternate=2]{}\\
  \lstick{\textsf{QNN register2:}  $\ket{0}^{\otimes n}$}\qwbundle[alternate=2]{}& \qwbundle[alternate=2]{}& \qwbundle[alternate=2]{}& \qwbundle[alternate=2]{}& \qwbundle[alternate=2]{} & \gate[nwires=1]{T}\qwbundle[alternate=2]{}& \gate[nwires=1]{P_j}\qwbundle[alternate=2]{}& \qwbundle[alternate=2]{}& \qwbundle[alternate=2]{}& \qwbundle[alternate=2]{}& \qwbundle[alternate=2]{}
  \end{quantikz}
  \caption{\emph{Amplitude Encoding by Swap test}. This circuit can perform the swap test depicted in Fig.~\ref{swaptest} in parallel for multiple $P_j$. Here, $P_j$ represents QNN with specific (the ``$j$th'') parameter configuration. To achieve swap test in parallel, we add an extra register ---the parameter register--- as the control of $P_j$: each computational basis $j$ of the parameter register corresponds to a specific parameter configuration in $P_j$. As illustrated in Fig.~\ref{controlp}, once the parameter register is in superposition state(by the Hadamard gates $H^{\otimes dr}$), the corresponding $P_j$ are in superposition. We refer the control operation on QNN as ``controlled-QNN''. Comparing with the normal swap test depicted in Fig.~\ref{swaptest}, the difference here is that the Swap ancilla qubit is anti-controlling /controlling the "controlled-QNN" together with the Unitary $T$(as gathered together in the dotted blue/orange box). It can be proven that the entire circuit in dotted the green box (denoted as $U$) can be expressed as $ U=\sum_{j}\left|j \left>\right<j\right|\otimes U_{j}$ where $U_{j}$ is the swap test unitary for $P_{j}$ defined in Fig.~\ref{swaptest}. This indicates that $U$ effectively perform the swap test in parallel for multiple $P_j$. Recall the fact that the normal swap test $U_{j}$ encode $|\langle p_j|t\rangle|^2$ in the amplitude of the output state (Eq.~\ref{relation} and Eq.~\ref{amplitudeencoding}), here the "parallel swap test" $U$ encodes the QNN cost function $|\langle p_j|t\rangle|^2$ in the amplitude of a superposition of $P_{j}$(QNN) with different parameters.   }

  \label{swappa}
\end{figure}
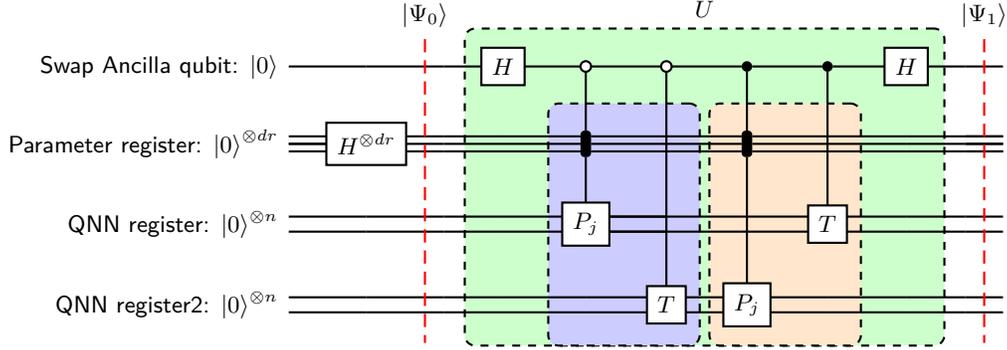

For the task of learning a pure state $\ket{\psi}=T\ket{0}$ ($T$ is a given unitary), the cost function is the fidelity between the generated state from the QNN and the state $\ket{\psi}=T\ket{0}$. In this case the amplitude encoding can be achieved by swap test, as shown in the circuit in Fig.~\ref{swappa}. \newline

We denote the unitary for the swap test circuit (in dotted green box) as $U$, and the input and output state of $U$ as $\ket{\Psi_{0}}$ and $\ket{\Psi_{1}}$, respectively.
The input to $U$, $\ket{\Psi_{0}}$, can be written as (note here and throughout the paper, we omit the normalization factor):
\begin{equation}
\left|\Psi_{0}\right\rangle=\ket{0}\otimes(\sum_{j}\left|j\right\rangle) \otimes \left|0\right\rangle^{\otimes n}_{\textsf{\textit{QNN1}}} \left|0\right\rangle^{\otimes n}_{\textsf{\textit{QNN2}}}
\end{equation}

Then $U$ can be written explicitly as
\begin{align}
U\coloneqq[H\otimes I\otimes I\otimes I]\cdot
[\ket{0}\bra{0}\otimes (\sum_{j}\left|j \left>\right<j\right|\otimes P_{j}\otimes T )+\ket{1}\bra{1}\otimes (\sum_{j}\left|j \left>\right<j\right|\otimes T\otimes P_{j} )]\cdot[H\otimes I\otimes I\otimes I],
\end{align}

Here, $P_j$ represents QNN with specific  parameter configuration defined by its control bit string $j $, as defined in Eq.~\ref{QNNc}. It can be proven (see Appendix \ref{app1}) that $U$ can be rewritten as
\begin{align}
 U=\sum_{j}\left|j \left>\right<j\right|\otimes U_{j},
\end{align}
where $U_{j}$ is the \textit{individual swap test unitary} on unitary $P_j$ and target unitary $T$, defined as in Eq.~\ref{defineUj}:
\begin{align}
\label{uj}
U_{j}\coloneqq [H\otimes I\otimes I]\cdot[\ket{0}\bra{0}\otimes P_{j}\otimes T+\ket{1}\bra{1}\otimes T\otimes P_{j}]\cdot[H\otimes I\otimes I].
\end{align}

As in Eq.~\ref{amplitudeencoding}, the resulting state of $U_{j}$ acting on $\ket{\Psi_{0}}$ is $\ket{\phi_{j}} \coloneqq U_{j}\ket{0}\left|0\right\rangle^{n}_{\textsf{\textit{QNN1}}} \left|0\right\rangle^{n}_{\textsf{\textit{QNN2}}}$ and has the following form:
\begin{equation}
\left|\phi_{j}\right\rangle =\sin \theta_{j}\ket{u_{j}}\ket{0}+\cos \theta_{j}\ket{v_{j}}\ket{1}.
\end{equation}
The final output state from $U$,  $\left|\Psi_{1}\right\rangle=U\left|\Psi_{0}\right\rangle $,  is therefore
\begin{equation}
\left|\Psi_{1}\right\rangle=\sum_{j}|j\rangle (\underbrace{\left.\left.\sin \theta_{j}\left|u_{j}\right\rangle\ket{0}+\cos \theta_{j}\left|v_{j}\right\rangle|1\right\rangle\right)}_{\mid \phi_{j}\rangle} = \sum_{j}\left|j\right\rangle  \left|\phi_{j}\right\rangle
\label{amplitudeencoding1}
\end{equation}

From Eqs.~\ref{amplitudeencoding1} and \ref{relation} we can see that the cost function (fidelity $|\langle p_j|t\rangle|^2$) for different parameters has been encoded into the amplitudes of the state $\left|\Psi_{1}\right\rangle$. \newline

\paragraph{\textbf{Amplitude encoding by Hadamard Test.}} \label{hadamardsection} For the task of generating ground states of given Hamiltonian $T$, the cost function is the expectation value of $T$ with respect to the generated state from the QNN. In this case the amplitude encoding can be achieved by the Hadamard test, as shown in the circuit in Fig.~\ref{ampha}.\newline

\begin{figure}
  \centering
  \begin{quantikz}
  \lstick{\textsf{Hadamard test Ancilla qubit:} \ket{0}} & \qw&\qw\gategroup[wires=3,steps=4,style={dashed,rounded corners,fill=green!20, inner xsep=3pt},background]{$U'$} &\gate{H}  & \ctrl{2} &\gate{H}& \qw \\
  \lstick{\textsf{Parameter register:} $\ket{0}^{\otimes dr}$}&\gate{H^{\otimes dr}}\qwbundle[alternate]{}    & \ctrlbundle{1}\gategroup[2,steps=1,style={dashed,rounded corners,fill=blue!20, inner xsep=2pt},background,label style={label position=below,anchor=north,yshift=-0.2cm}]{{\sc $P$}}&\qwbundle[alternate]{} &\qwbundle[alternate]{}&\qwbundle[alternate]{} &\qwbundle[alternate]{} \qwbundle[alternate]{} \\
  \lstick{\textsf{QNN register:}  $\ket{0}^{\otimes n}$}\qwbundle[alternate=2]{}& \qwbundle[alternate=2]{} & \gate[nwires=1]{P_j}\qwbundle[alternate=2]{}& \qwbundle[alternate=2]{}& \gate[nwires=1]{T}\qwbundle[alternate=2]{}& \qwbundle[alternate=2]{}& \qwbundle[alternate=2]{}
  \end{quantikz}
  \caption{\textit{Amplitude encoding by Hadamard Test} This circuit can perform the Hadamard test depicted in Fig.~\ref{Hadamard} in parallel for multiple $P_j$. Here, $P_j$ represents QNN with specific (the ``$j$th'') parameter configuration. To achieve Hadamard test in parallel, we add an extra register ---the parameter register--- as the control of $P_j$: each computational basis $j$ of the parameter register corresponds to a specific parameter configuration in $P_j$. As illustrated in Fig.~\ref{controlp}, once the parameter register is in superposition state (by the Hadamard gates $H^{\otimes dr}$), the corresponding $P_j$ are in superposition. It can be proven that the entire circuit in dotted the green box (denoted as $U$) can be expressed as $ U'=\sum_{j}\left|j \left>\right<j\right|\otimes U'_{j}$ where $U'_{j}$ is the Hadamard test unitary for $P_{j}$ defined in Fig.~\ref{Hadamard}. This indicates that $U'$ effectively perform the Hadamard test in parallel for multiple $P_j$. Recall the fact that the normal Hadamard test $U'_{j}$ encode $\bra{0}P_j^{\dagger}TP_j\ket{0}$ in the amplitude of the output state (Eq.~\ref{hadard} and Eq.~\ref{hadamardrelation}), here the ``parallel Hadamard test'' $U'$ encodes the QNN cost function $\bra{0}P_j^{\dagger}TP_j\ket{0}$ in the amplitude of a superposition of $P_{j}$(QNN) with different parameters. }
  \label{ampha}
\end{figure}

Since the analysis for the case of the Hadamard test is very similar to that of the swap test, we omit the details here. For the same reason, we only present the case using the swap test also in the next section when discussing phase encoding.

\tocless\subsubsection{\textbf{Amplitude estimation}}

The second step following the amplitude encoding is to use amplitude estimation \cite{2000quant.ph..5055B} to extract and store the cost function into an additional register which we call the ``amplitude register''. In the following we present the details of amplitude estimation.\newline

After the amplitude encoding by the swap test, we introduce an extra register $\ket{0}^{\otimes t}_{\textsf{\textit{amplitude}}}$ and
the output state $\ket{\Psi_{1}}$ (using the same notation) becomes
\begin{align}
  \left|\Psi_{1}\right\rangle=\sum_{j}\left|j\right\rangle  \ket{\phi_{j}}  \ket{0}^{\otimes t}_{\textsf{\textit{amplitude}}},
\end{align}
where $\ket{\phi_{j}}$ can be decomposed as
\begin{equation}
\ket{\phi_{j}}=\frac{-i}{\sqrt{2}}\left(e^{i \theta_{j}}\ket{\omega_{+}}_j-e^{i(-\theta_{j})}\ket{\omega_{-}}_j\right).
\end{equation}
Hence, we have
\begin{equation}
\left|\Psi_{1}\right\rangle=\sum_{j}\frac{-i}{\sqrt{2}}\left( e^{i \theta_{j}}\left|j\right\rangle\ket{\omega_{+}}_j-e^{i(-\theta_{j})}\left|j\right\rangle\ket{\omega_{-}}_j\right) \ket{0}^{\otimes t}_{\textsf{\textit{amplitude}}}.
\label{phi1}
\end{equation}

The overall Grover operator $G$ is defined as
\begin{equation}
G \coloneqq UC_{2}U^{-1}C_{1},
\label{grovero}
\end{equation}
where $C_{1}$ is the $Z$ gate on the swap ancilla qubit, and $C_{2}$ is ``flip zero state'' unitary which is similar to that defined in Fig.~\ref{uc}.
It can be shown (see Appendix \ref{app1}) that $G$ can be expressed as
\begin{align}
    G=\sum_{j}\left|j \left>\right<j\right|\otimes G_{j},
\end{align}
where $G_{j}$ is the individual Grover operator as defined in Eq.~\ref{gj}. The overall Grover operator $G$ possess the following eigen-relation:
\begin{align}
 G\left|j\right\rangle\ket{\omega_{\pm}}_j=   e^{i (\pm 2\theta_{j})}\left|j\right\rangle\ket{\omega_{\pm}}_j.
\end{align}

Next we apply phase estimation of the overall Grover operator $G$ on the input state
$\left|\Psi_{1}\right\rangle$. The resulting state $\left|\Psi_{2}\right\rangle$ can be written as
\begin{equation}\label{statephase}
\left|\Psi_{2}\right\rangle=\sum_{j}\frac{-i}{\sqrt{2}}\left( e^{i \theta_{j}}\left|j\right\rangle\ket{\omega_{+}}_{j}\ket{2\theta_{j}}-e^{i(-\theta_{j})}\left|j\right\rangle\ket{\omega_{-}}_{j}\ket{-2\theta_{j}}\right).
\end{equation}

Note here in Eq.~\ref{statephase}, $\ket{\pm 2\theta_{j}}$ denotes the eigenvalues $\pm 2\theta_{j}$ being stored in the amplitude register with some finite precision.\newline

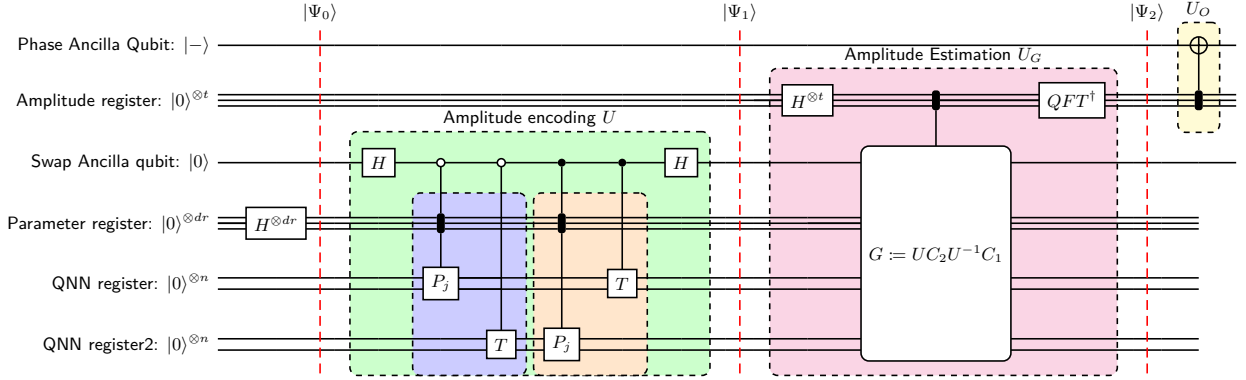
\begin{figure}
  \centering
  \begin{adjustbox}{width=\textwidth}
  \begin{quantikz}
  \lstick{\textsf{Phase Ancilla Qubit:} \ket{-}} &\qw&\qw&\qw&\qw&\qw&\qw&\qw&\qw&\qw&\qw&\qw &\qw&\qw&\qw\slice{$\left|\Psi_{2}\right\rangle$}&\qw&\targ{} \gategroup[wires=2,steps=1,style={dashed,rounded corners,fill=yellow!20, inner xsep=3pt},background]{\textit{{$U_O$}}}&\qw\\
  \lstick{\textsf{Amplitude register:}  $\ket{0}^{\otimes t}$} &\qwbundle[alternate]{} &\qwbundle[alternate]{} &\qwbundle[alternate]{} &\qwbundle[alternate]{}&\qwbundle[alternate]{} &\qwbundle[alternate]{} &\qwbundle[alternate]{} &\qwbundle[alternate]{}&\qwbundle[alternate]{} &\qwbundle[alternate]{}&\gate{H^{\otimes t}}\gategroup[wires=5,steps=3,style={dashed,rounded corners,fill=magenta!20, inner xsep=3pt},background]{\textsf{Amplitude Estimation} \textit{\textbf{$U_G$}}}\qwbundle[alternate]{} & \ctrlbundle{1}  &\gate{QFT^{\dagger}}\qwbundle[alternate]{} &\qwbundle[alternate]{}&\qwbundle[alternate]{}& \ctrlbundle{-1}&\qwbundle[alternate]{} \\
  \lstick{\textsf{Swap Ancilla qubit:} \ket{0}} & \qw\slice{$\left|\Psi_{0}\right\rangle$}& \qw &\gate{H}\gategroup[wires=4,steps=6,style={dashed,rounded corners,fill=green!20, inner xsep=3pt},background]{\textsf{Amplitude encoding} $U$} & \octrl{1} & \octrl{3}& \ctrl{1} & \ctrl{2} &\gate{H}& \qw\slice{$\left|\Psi_{1}\right\rangle$}& \qw& \qw & \gate[wires=4,nwires={3,4},style={rounded corners}]{G\coloneqq UC_{2}U^{-1}C_{1}}& \qw& \qw& \qw& \qw& \qw\\
  \lstick{\textsf{Parameter register:} $\ket{0}^{\otimes dr}$}&\gate{H^{\otimes dr}}\qwbundle[alternate]{} &\qwbundle[alternate]{} &\qwbundle[alternate]{}  & \ctrlbundle{1}\gategroup[3,steps=2,style={dashed,rounded corners,fill=blue!20, inner xsep=2pt},background,label style={label position=below,anchor=north,yshift=-0.2cm}]{{}}&\qwbundle[alternate]{} &\ctrlbundle{2}\gategroup[3,steps=2,style={dashed,rounded corners,fill=orange!20, inner xsep=2pt},background,label style={label position=below,anchor=north,yshift=-0.2cm}]{{}}&\qwbundle[alternate]{}&\qwbundle[alternate]{}&\qwbundle[alternate]{} &\qwbundle[alternate]{} &\qwbundle[alternate]{}&\qwbundle[alternate]{}&\qwbundle[alternate]{}&\qwbundle[alternate]{}&\qwbundle[alternate]{}&\qwbundle[alternate]{}  \\
  \lstick{\textsf{QNN register:}  $\ket{0}^{\otimes n}$}\qwbundle[alternate=2]{}& \qwbundle[alternate=2]{}& \qwbundle[alternate=2]{}& \qwbundle[alternate=2]{} & \gate[nwires=1]{P_j} \qwbundle[alternate=2]{}& \qwbundle[alternate=2]{}\qwbundle[alternate=2]{}& \qwbundle[alternate=2]{}& \gate[nwires=1]{T} \qwbundle[alternate=2]{}&\qwbundle[alternate=2]{}& \qwbundle[alternate=2]{}& \qwbundle[alternate=2]{}& \qwbundle[alternate=2]{}& \qwbundle[alternate=2]{}& \qwbundle[alternate=2]{}& \qwbundle[alternate=2]{}& \qwbundle[alternate=2]{}& \qwbundle[alternate=2]{}\\
  \lstick{\textsf{QNN register2:}  $\ket{0}^{\otimes n}$}\qwbundle[alternate=2]{}& \qwbundle[alternate=2]{}& \qwbundle[alternate=2]{}& \qwbundle[alternate=2]{}& \qwbundle[alternate=2]{} & \gate[nwires=1]{T}\qwbundle[alternate=2]{}& \gate[nwires=1]{P_j}\qwbundle[alternate=2]{}& \qwbundle[alternate=2]{}& \qwbundle[alternate=2]{}& \qwbundle[alternate=2]{}& \qwbundle[alternate=2]{}& \qwbundle[alternate=2]{}& \qwbundle[alternate=2]{}& \qwbundle[alternate=2]{}& \qwbundle[alternate=2]{}& \qwbundle[alternate=2]{}& \qwbundle[alternate=2]{}
  \end{quantikz}
  \end{adjustbox}
  \caption{\emph{Major steps in the Construction of the Grover Oracle.} \textsf{Step 0:} We initialize the system by applying Hadamard gates on the parameter register, leading to the state $\left|\Psi_{0}\right\rangle=\ket{0}_{s}\otimes(\sum_{j}\left|j\right\rangle) \otimes \left|0\right\rangle^{n}_{\textsf{\textit{QNN1}}} \left|0\right\rangle^{n}_{\textsf{\textit{QNN2}}}$. \textsf{Step 1(dotted green box)}: Amplitude encoding of the cost function, as illustrated in Fig.~\ref{swappa} (refer the caption of Fig.~\ref{swappa} for the meaning of each symbol), resulting in the state $\left|\Psi_{1}\right\rangle=\sum_{j}|j\rangle (\left.\left.\sin \theta_{j}\left|u_{j}\right\rangle\ket{0}+\cos \theta_{j}\left|v_{j}\right\rangle|1\right\rangle\right)$, in which $\theta_i$ contains the cost function. \textsf{Step 2(dotted pink box)}: Amplitude estimation to extract and store the cost function into an additional register which we call the ``amplitude register'', resulting in the state $\left|\Psi_{2}\right\rangle=\sum_{j}\frac{-i}{2}\left( e^{i \theta_{j}}\left|j\right\rangle\ket{\omega_{+}}_{j}\ket{2\theta_{j}}-e^{i(-\theta_{j})}\left|j\right\rangle\ket{\omega_{-}}_{j}\ket{-2\theta_{j}}\right)$. \textsf{Step 3(dotted yellow box)}: Threshold Oracle to encode the cost function into relative phase by using a Phase ancilla qubit, resulting in the state $ \left|\Psi_{3}\right\rangle=\sum_{j}\frac{-i}{2}(-1)^{ g(\theta_{j}-\theta^*)}\left( e^{i \theta_{j}}\left|j\right\rangle\ket{\omega_{+}}_{j}\ket{2\theta_{j}}-e^{i(-\theta_{j})}\left|j\right\rangle\ket{\omega_{-}}_{j}\ket{-2\theta_{j}}\right)$. }
  \label{pepe}
\end{figure}

\tocless\subsubsection{\textbf{Threshold Oracle and Uncomputations}}

Next we apply a threshold oracle $U_{O}$ on the amplitude register and an extra phase ancilla qubit, which acts as
\begin{equation}
 U_{O}\ket{\pm 2\theta_{j}}=(-1)^{ g(\theta_{j}-\theta^*)}\ket{\pm2\theta_{j}},
 \label{threshold}
\end{equation}
where $\theta^*$ is implicitly defined as
\begin{equation}
C^*=-\cos{2\theta^*},\quad  \theta^*\in[{\pi}/{4}, {\pi}/{2}].
\label{thresoracle}
\end{equation}

Note that in Eq.~\ref{threshold} we omit the state of the phase ancilla qubit.\newline

The state after the oracle $\left|\Psi_{3}\right\rangle$ can be written as
\begin{equation}
 \left|\Psi_{3}\right\rangle=\sum_{j}\frac{-i}{\sqrt{2}}(-1)^{ g(\theta_{j}-\theta^*)}\left( e^{i \theta_{j}}\left|j\right\rangle\ket{\omega_{+}}_{j}\ket{2\theta_{j}}-e^{i(-\theta_{j})}\left|j\right\rangle\ket{\omega_{-}}_{j}\ket{-2\theta_{j}}\right),
\end{equation}

The procedure thus far can be illustrated in a circuit as in Fig.~\ref{pepe}. \newline

After we perform the uncomputation of Phase estimation, the resulting state is
\begin{align}
  \left|\Psi_{4}\right\rangle&=\sum_{j}\frac{-i}{\sqrt{2}} (-1)^{ g(\theta_{j}-\theta^*)}\left(e^{i \theta_{j}}\left|j\right\rangle\ket{\omega_{+}}_{j}\ket{0}^{\otimes t}_{\textsf{\textit{amplitude}}}-e^{i(-\theta_{j})}\left|j\right\rangle\ket{\omega_{-}}_{j}\ket{0}^{\otimes t}_{\textsf{\textit{amplitude}}}\right), \\
  &=\sum_{j}(-1)^{ g(\theta_{j}-\theta^*)}\left|j\right\rangle  \ket{\phi_{j}}  \ket{0}^{\otimes t}_{\textsf{\textit{amplitude}}}.
\end{align}
Finally, we perform the uncomputation of the swap test and the resulting state is
\begin{equation}
\left|\Psi_{5}\right\rangle=\sum_{j}(-1)^{ g(\theta_{j}-\theta^*)}\left|j\right\rangle \left|0\right\rangle^{\otimes n}_{\textsf{\textit{QNN1}}}\left|0\right\rangle^{\otimes n}_{\textsf{\textit{QNN2}}}  \ket{0}  \ket{0}^{\otimes t}_{\textsf{\textit{amplitude}}}.
\label{dec}
\end{equation}

As can be seen from Eqs.~\ref{dec} and \ref{relation}, the above steps implemented the Grover oracle $  \mathcal{O}_{\textsf{\textit{Grover}}}$ (defined in Eq.~\ref{groverqnn}) After the above procedure a relative phase, which depends on the cost function of the QNN $ |\langle p_j|t\rangle|^2 $ and the threshold, have been coherently added to the parameter state. Importantly, uncomputation allows the parameter register to be decoupled from the QNN and other registers.\newline

\subsection{Performance of the Quantum training by Grover Adaptive Search }

Taking training VQE as an example, in Table.~\ref{complexity} we present the result for the number of “controlled-QNN” runs, the number of QNN runs and the number of measurements needed in the quantum training by Grover Adaptive Search. The derivation are included in Appendix \ref{app2}.

\begin{table}[h]
\centering
\begin{tabular}{ |p{5cm}|p{3.6cm}|p{3.6cm}|   }
\hline
 \textsf{Number of “controlled-QNN” runs} & \textsf{Number of QNN runs} & \textsf{Number of measurements}\\
\hline
$O\left(\frac{1}{\epsilon_2\epsilon_1}{\left(\frac{1}{{\epsilon_0}}\right)}^{r/2}s^{-0.5}\right)$&$ O\left(\frac{1}{\epsilon_2} \left(r\log\left(\frac{1}{\epsilon_0}\right)\right)^{1.5}\right)$ & $ O\left(\left(r\log\left(\frac{1}{\epsilon_0}\right)\right)^{1.5}\right)$ \\
\hline
\end{tabular}
    \caption{\emph{Performance of the Quantum training by Grover Adaptive Search.} Here we present the result for the number of “controlled-QNN” runs, the number of QNN runs and the number of measurements needed in the quantum training by Grover Adaptive Search. In this table, $r$ is the number of parameters (rotation angles) in QNN, $1-\epsilon_1$ is the probability of success of the phase estimation, $\epsilon_2$ is the precision we set up for the evaluation of the cost function using amplitude estimation, $\epsilon_0$ is the precision of each angle value, $s$ is the number of global optima of the QNN cost function.}
    \label{complexity}
\end{table}

\subsection{Advantages and disadvantages of training by Grover Adaptive Search}\label{groverlim}

In the presence of a noise-free barren plateau, the Grover Adaptive Search mechanism can find global optima without an exponential number of measurements. However, it has the following disadvantages:
\begin{itemize}

    \item It can be seen from Table.~\ref{complexity} that in Quantum training by Grover adaptive search, the number of  “controlled-QNN” runs is exponential in the number of parameters in QNN. Even in the case where the number of parameters scales only linearly with the number of qubits  in a QNN, the quantum training by Grover takes excessive runtime. Moreover, it invokes very deep circuit.

    \item Training by Grover adaptive search does not circumvent the noise-induced barren plateau. When the entire cost landscape is flatten in the case of noise-induced barren plateau \cite{wang2020noiseinduced}, it requires exponential precision of the amplitude estimation. That is, $\epsilon_2$ should be exponentially small. According to Table.~\ref{complexity}, this adds another exponentially large factor to the number of  “controlled-QNN” runs and QNN runs.

\end{itemize}

While these disadvantages most probably rule out Grover adaptive search for NISQ-era devices, it still represents a maximally quantum solution. For fault-tolerant devices, this method is the provably optimal approach for QNN cost function with no structure, it enjoys a quadratic speed-up which is a significant improvement compare to the exponential "slow-down" of the classical training methods due to the barren plateau issue.

\section{QNN training by Adaptive QAOA}\label{ourf}

As depicted in Fig.~\ref{ours}, our framework for quantum training of QNNs consists of two major components.
\begin{itemize}
    \item \textbf{Phase oracle.} This coherently encodes the cost function of QNNs onto a relative phase of a superposition state in the Hilbert space of the parameters \cite{Gily_n_2019}.
    \item \textbf{Adaptive Mixers.} These exploit hidden structure in QNN optimisation problems, hence can achieve short-depth circuit \cite{mcclean2020low}.
\end{itemize}
Iterations of the phase oracle and the adaptive mixers constitute a QAOA routine which quantumly homing in on optimal network parameters of QNNs. This section presents the details of our framework.

\subsection{Phase Oracle}\label{phaseo}

We aim to coherently achieve the phase encoding for the cost function of the QNN by a phase oracle $O_{\textsf{\textit{Phase}}}$, which acts as
\begin{equation}
    O_{\textsf{\textit{Phase}}}\ket{\btheta}\ket{\mathbf{0}}_{\textsf{\textit{QNN}+\textsf{\textit{ancillas}}}} \to e^{-i\gamma C(\btheta)}\ket{\mathbf{0}}_{\textsf{\textit{QNN}+\textsf{\textit{ancillas}}}}
\end{equation}
in which $\gamma$ is a free parameter to be optimized. When $O_{\textsf{\textit{Phase}}}$ is acting on a superposition state of parameters $\sum_{\btheta}\omega_{\btheta}\ket{\btheta}$, we have
\begin{equation}
O_{\textsf{\textit{Phase}}}\sum_{\btheta}\omega_{\btheta}\ket{\btheta}\ket{\mathbf{0}}_{\textsf{\textit{QNN}+\textsf{\textit{ancillas}}}}\to \sum_{\btheta}e^{-i\gamma C(\btheta)}\omega_{\btheta}\ket{\btheta}\ket{\mathbf{0}}_{\textsf{\textit{QNN}+\textsf{\textit{ancillas}}}}
    \label{phaseoracle}
\end{equation}

As detailed in Ref.~\cite{Gily_n_2019}, this phase oracle can be constructed based on the amplitude encoding which we have implemented in Section \ref{amplitude}. Next we present the details of how to contruct the phase oracle from the amplitude encoding by amplitude estimation or Linear Combination of Unitaries (LCU) \cite{PhysRevLett.114.090502}.
\newline

\paragraph{\textbf{Phase oracle by amplitude estimation}}

The procedure to achieve $O_{\textsf{\textit{Phase}}}$ by  amplitude estimation is very similar to that of $\mathcal{O}_{\textsf{\textit{Grover}}}$, the only difference is that the threshold  $U_{O}$ (defined in Eq.~\ref{threshold}) needs to be replaced by $U'_{O}$ which acts as
\begin{equation}
 U'_{O}\ket{\pm 2\theta_{j}}=e^{-i\gamma C(\btheta)}\ket{\pm2\theta_{j}}.
 \label{th}
\end{equation}
Recall Eq.~\ref{relation}, Eq.~\ref{hadamardrelation} and the form of the cost function in Table~\ref{taskstable}, the cost function $C(\btheta)$ is encoded in $\theta_j$ as $C(\btheta)=-\cos{2\theta_j}$, therefore $U'_{O}$ acts as
\begin{equation}
 U'_{O}\ket{\pm 2\theta_{j}}=e^{i\gamma \cos{2\theta_j}}\ket{\pm2\theta_{j}}.
 \label{thresholdoracle2}
\end{equation}

Once we have chosen the specific value of $\gamma$, $U'_{O}$ can be constructed according to Eq.~\ref{thresholdoracle2}.  \newline

\paragraph{\textbf{Phase oracle by LCU}}

For this approach, we start with constructing an operator $G^*$ defined similar as in Eq.~\ref{grovero}: \footnote{Note that here our definition of $G^*$ is slightly different from the $G_U$ in Ref.~\cite{Gily_n_2019}: $C_1$ and $C_2$ being negative to their counterpart in the definition of $G_U$. However the two negative sign cancel, therefore we have $G^*=G_U$.}
\begin{equation}
G^* \coloneqq C_{2}U^{-1}C_{1}U,
\label{grovero2}
\end{equation}
It has been shown in Ref.~\cite{Gily_n_2019} that
\begin{equation}
    e^{-i \frac{1}{2}\left(C(\btheta)-1\right)}\cdot I \approx \sum_{m=-M}^{M}\beta_{m}  {G^*}^{m} ,
\end{equation}
where $\beta_{m} = \sum_{k=|m|}^{M}\left(\begin{array}{c}
2 k \\
k-m
\end{array}\right) \frac{(-1)^{m} i^{k}}{k ! 2^{2 k}}$, $M \in \mathbb{N}_{+}$.

Define a new cost function $C'(\btheta):= \frac{1}{2}\left(C(\btheta)-1\right)$ (optimizing $C'(\btheta)$ is equivalent to optimizing $C(\btheta)$), we have
\begin{equation}
    e^{-i C'(\btheta)}\cdot I \approx \sum_{m=-M}^{M}\beta_{m}  {G^*}^{m} ,
\end{equation}
This series of $G^*$ can be implemented using the LCU technique (together with the subsequent "Oblivious Amplitude Amplification") \cite{PhysRevLett.114.090502} in which the number of calls to $G^*$ needed is only logarithmic of the inverse of the desired precision \cite{Gily_n_2019}. Using the techniques in Ref.~\cite{Gily_n_20192}, we can convert phase oracle with $e^{-i C'(\btheta)}$ into phase oracle with $e^{-i\gamma C'(\btheta)}$ for arbitrary $\gamma$ bounded from $[-1, 1]$), by only logarithmic (of the inverse of the desired precision) number of queries of phase oracle with $e^{-i C'(\btheta)}$. \newline

\begin{figure}[h!]
\centering
\includegraphics[width=\linewidth]{ 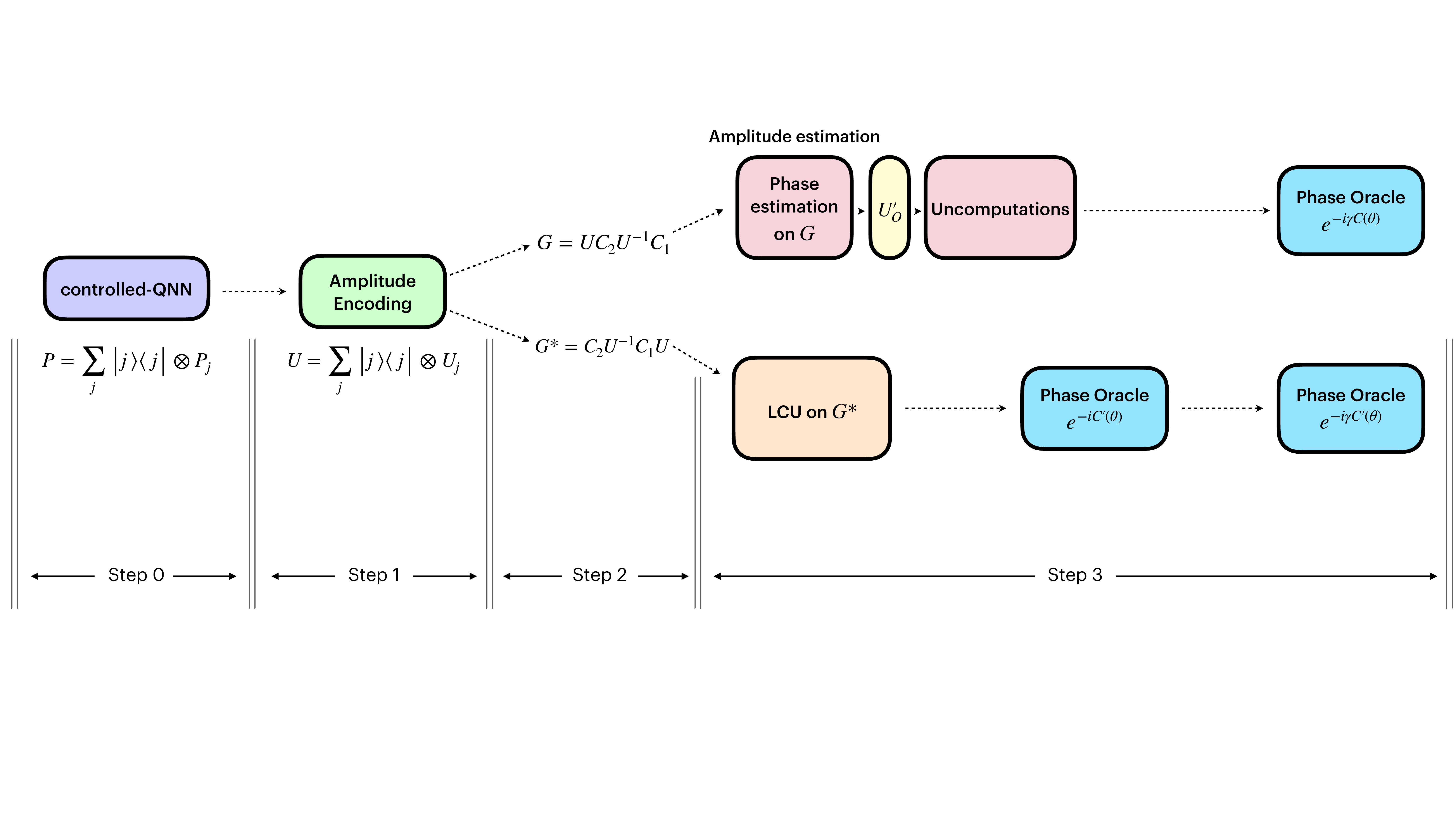}
\caption{\emph{Pipeline of the construction of the phase oracle.} Here we summarise the two approaches by amplitude estimation and by LCU for the Phase encoding of the cost function. \textsf{Step 0:} Creating superposition for QNN with different parameters, which is implemented by "controlled-QNN" (see Fig.~\ref{swaptest}), denoted by $P =\sum_{j}\left|j \left>\right<j\right| \otimes P_{j}$. \textsf{Step 1:} Amplitude encoding of the cost function, by the unitary operation $U =\sum_{j}\left|j \left>\right<j\right| \otimes U_{j}$. \textsf{Step 2:} Constructing the "Grover Operator" upon the amplitude encoding unitary. In the approach using amplitude estimation, the Grover Operator $G$ is constructed as $G = UC_{2}U^{-1}C_{1}$. In the approach using LCU, the Grover Operator $G^*$ is constructed as $G^* = C_{2}U^{-1}C_{1}U$.  \textsf{Step 3:} Phase encoding of the cost function, by amplitude estimation(upper path) or by LCU(lower path). In the upper path, the Phase Oracle is achieved by phase estimation on $G$, threshold oracle $U'_O$, and uncomputation. In the lower path, LCU on $G^*$ (together with the subsequent "Oblivious Amplitude Amplification") \cite{PhysRevLett.114.090502} realizes $e^{-i C'(\btheta)}$ which is then converted to the Phase Oracle with arbitrary $\gamma$ ---$e^{i{\gamma}C'(\btheta)}$ using the method in Ref.~\cite{Gily_n_20192}. $C'(\btheta):= \frac{1}{2}\left(C(\btheta)-1\right)$ is a new cost function, optimizing $C'(\btheta)$ is equivalent to optimizing $C(\btheta)$.  }
\label{pip}
\end{figure}

In Fig. \ref{pip} we summarise the two approaches for the Phase encoding of the cost function.

\subsection{Adaptive Mixers}\label{mixers}

As in Section \ref{variants}, we designed a new variant of QAOA --- ``Adaptive-Continuous(AC-QAOA)'' --- to be the ansatz of our quantum training for QNN. We summarise the reason of this choice as follow:
\begin{itemize}
    \item[1.] \textbf{[Why "Continuous"]} In our optimisation problem of QNN training, the parameters we are optimizing (the angles of rotation gates) are continuous variables (real values), hence the choice of mixer Hamiltonian has to be designed for continuous variables. For example, the mixer Hamiltonian of the original QAOA (X rotations) generate shifts in the computational basis, here in continuous case, the corresponding mixer should shift the value for each digitized continuous variables stored in independent registers.

    \item[2.] \textbf{[Why "Adaptive"]} The Cost function of QNNs is complicated and task-specific (given by the learning objectives). Hence it is hard to analytically determine good mixers for our optimisation problem of QNN training. Therefore we would want to take advantage of including alternative mixers and allowing adaptive mixers for different layer (as in ADAPT-QAOA).

\end{itemize}
  Adopting "AC-QAOA" could exploit hidden structure in QNN optimisation problem and dramatically shorten the depth of QAOA layers while significantly improving the quality of the solution \cite{mcclean2020low}.
\newline

Generally, the mixer pool of AC-QAOA should include two types of Mixer Hamiltonians for continuous variables:

\begin{itemize}
\item[1.]
Quadratic functions of the position operator and the momentum operator for single continuous variables, e.g. the squeezing operator \cite{RevModPhys.84.621}.
 \item[2.]
Entangling mixers that acts on two continuous variables, e.g. the two-mode squeezing operator \cite{RevModPhys.84.621}.

\end{itemize}

These operators could be carried out in continuous variable quantum systems. However, we will focus on the circuit implementation of these mixers when using a collection of qubits to approximate the behavior of continuous variables. \newline

When using a qudit of dimension $d$ to digitally simulate a continuous variable, the position operator can be written as
\begin{equation}
  J_d := \sum_{j=0}^{d-1} j \ket{j}\!\bra{j},
\end{equation}
in which $j$ is the digitized value of the continuous variable.\newline

We can use $N$ qubits to simulate the qudit and construct $J_d$ for $d = 2^N$ as \cite{verdon2018universal},
\begin{equation}
\begin{split}
  J_{2^N} &= \sum_{n=1}^N 2^{n-2} ( I^{(n)} - Z^{(n)} )
,
\end{split}
\label{xop}
\end{equation}
where $I^{(n)}$ and $Z^{(n)}$ are the identity and the Pauli-Z operator (respectively) for the $n^\text{th}$ qubit.\newline

The momentum operator, which act as generator of shifts in the value of a continuous variable (denote as $S$) can be written as the discrete Fourier transform of $J_d$ \cite{verdon2018universal},
\begin{equation}
  S := F_d^\dagger J_d F_d,
  \label{si}
\end{equation}
in which the discrete Fourier transform $F_d$ is defined by
\begin{equation}
  F_d \ket{j} := \frac{1}{\sqrt{d}} \sum_{k=0}^{d-1} \omega_d^{-jk} \ket{k},
\end{equation}
where $\omega_d := e^{2\pi i/d}$.
\newline

\begin{figure}[h!]
\centering
\includegraphics[width=\linewidth]{ 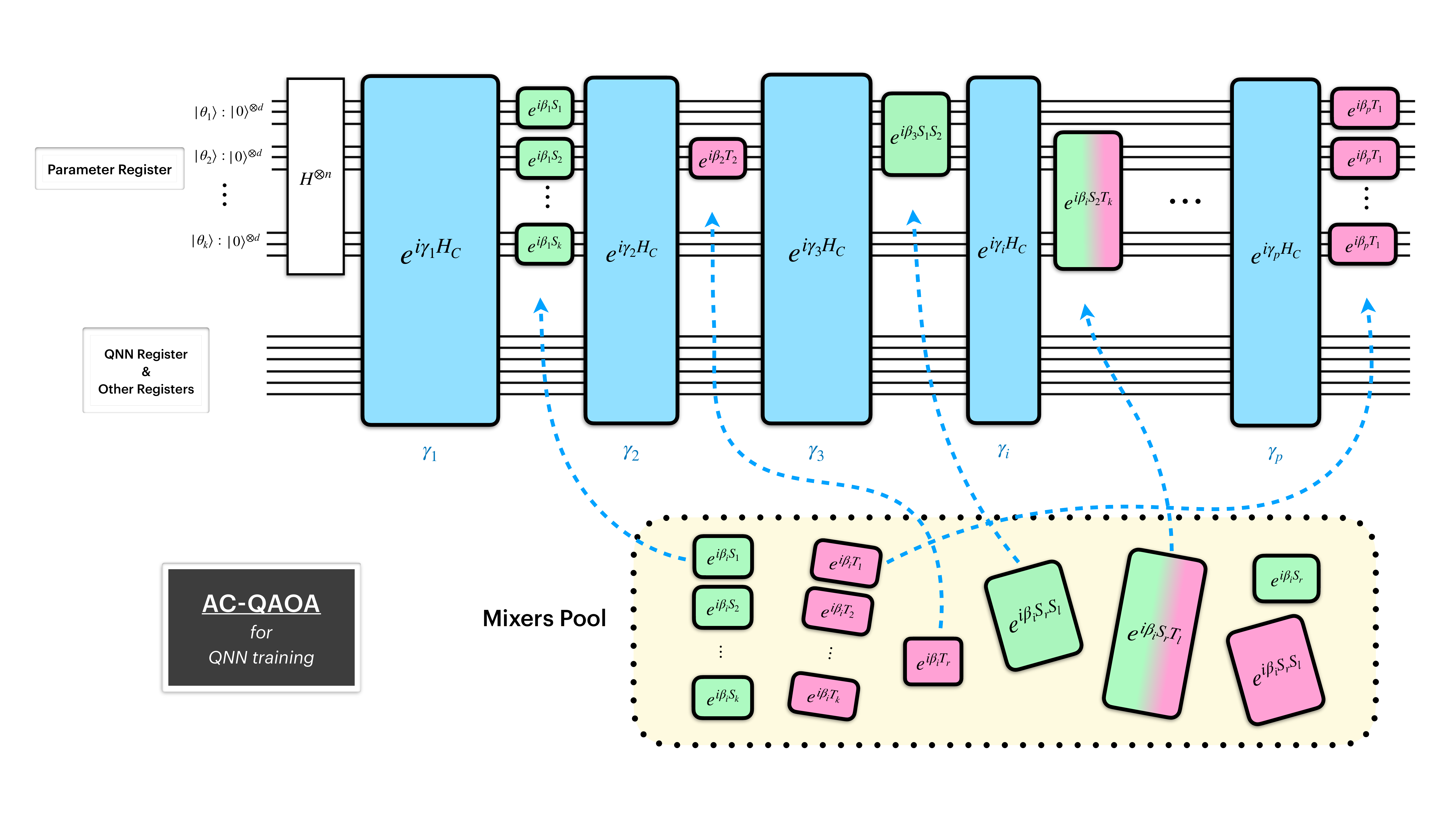}
\caption{\textit{Schematic diagram of applying AC-QAOA to QNN training.} AC-QAOA is a variant of QAOA we designed for solving optimisation of continuous variables with the short-depth advantage of QAOA layers, see Fig.~\ref{acqaoa}. This figure illustrate applying AC-QAOA to QNN training, following the scheme in Fig.~\ref{ours}. The quantum training protocol consists of alternating operations in a QAOA fashion --- the first operation acts on both the parameter register and QNN register to encode the cost function of QNN onto a relative phase of the parameter state. This operation is represented by the blue blocks in the figure. The other operations are the Mixers (green and pink boxes) which act only on the parameter register. In the parameter register, $\theta_i$ are the continuous variables to be optimized in the training, each $\theta_i$ is digitized into binary form and stored in an independent register. The overall process of AC-QAOA is similar to that of the original QAOA, with the difference being as follows. \textit{1.} The mixers of AC-QAOA with Hamiltonians $S_i$ and $T_i$ are acting on the registers of $\theta_i$ (rather than single qubits as in the original QAOA). \textit{2.} The mixers of AC-QAOA contain alternative mixers taken from a \emph{mixers pool} and can vary from layer to layer.}
\label{ourac}
\end{figure}

As mentioned above, a general mixer Hamiltonian is the quadratic functions of the position operator  $J_d$ and the momentum operator $S$, therefore using Eq.~\ref{xop} and Eq.~\ref{si} (set $d=2^N$) we can rewrite a mixer Hamiltonian as a summation of simple unitaries. Hence utilising the Hamiltonian simulation technique in \cite{PhysRevLett.114.090502}, the Mixer operator can be efficiently implemented. For instance, the digitized version of the generator of squeezing operator
(denote as $T$) is defined as:
\begin{equation}
  T:= J_d S + S J_d .
\label{ti}
\end{equation}
Plugging Eq.~\ref{si} into Eq.~\ref{ti}, together with Eq.~\ref{xop} (set $d=2^N$), we can see that $T=J_d F_d^\dagger J_d F_d + F_d^\dagger J_d F_d J_d$ can be expressed as the summation of simple unitaries. Therefore the corresponding Mixer with Hamiltonian $T$ can be efficiently implemented using the Hamiltonian simulation technique in \cite{PhysRevLett.114.090502}. Similarly, the entangling Mixers on two continuous variables with Hamiltonian $S_iS_j,S_iT_j,T_iT_j$ (The subscript $i,j$ indicate that they are for specific variables) can be implemented in the same manner. In Fig.~\ref{ourac}, we depict the schematic diagram of applying AC-QAOA to QNN training. \newline

Due to the fact that the non-Gaussian operators are costly to implement, we only consider up-to-second-order polynomial functions of the position operator  $J_d$ and the momentum operator $S$ for the Mixer Hamiltonian. The Mixer pool can generally include mixers with Hamiltonians: $J_d$, $S$, $J_dS$, $SJ_d$, ${J_d}^2$, ${S}^2$, ${J_d}^2+{S}^2$ for one continuous variable and the entangling Mixers for two continuous variables. Comparing to the Mixer pool of ADAPT-QAOA for discrete variable, we can have the following the analogy:
\begin{itemize}
    \item[1.]
The momentum operator $S$ is the (digitized) continuous version of $X$ mixers that shift the value for each digitized continuous variables stored in independent registers.
 \item[2.]
$J_dS$ is the (digitized) continuous version of $Y$ mixers which ‘unlock’ geodesics in parameter space, allowing the QAOA iterations reaching the target state faster. \cite{yao2020reinforcement}

\end{itemize}

We note that quadratic Hamiltonians are efficiently simulatable (classically), but only when the initial state is from a special class of Gaussian states (e.g. the vacuum state) \cite{PhysRevLett.88.097904}. Here, the initial state in the qubit encoding is far from Gaussian and a continuous variable analog of our technique would use an equivalent encoding. \newline

By making the mixers flexible and adaptive to specific optimisation problems, it is demanding to find an efficient way of determining the mixers sequence and optimizing the hyper-parameters.
There are several research works on using machine learning approaches (Recurrent Neural Networks (RNN) and Reinforcement Learning(RL)) to determine the mixers sequence and optimize the hyper-parameters. These works achieved significantly less measurements than conventional approach(e.g. gradient based methods). We list the papers in the following table:

\begin{center}
\begin{tabular}{ |p{5cm}|p{3cm}|p{3.5cm}|  }
\hline
& \textbf{RNN} & \textbf{RL} \\
\hline
\textsf{Determining     mixers sequence} & Ref. \cite{Warren2020RNNVQEAM} & Ref. \cite{yao2020reinforcement} \\
\hline
 \textsf{Optimizing     hyper-parameter} & Ref. \cite{verdon2019learning} &Refs. \cite{PhysRevResearch.2.033446,yao2020reinforcement,khairy2019reinforcementlearningbased,yao2020policy} \\

\hline
\end{tabular}
\end{center}
We adopt the approaches developed in these works to our quantum training of QNNs for efficiently determining the mixers sequence and optimizing the hyper-parameters.

\subsection{Advantages of training by QAOA}

As we have discussed in \ref{groverlim}, due to the global-search nature of Grover's algorithm, the quantum training using Grover Adaptive Search can circumvent the noise-free barren plateau, however it has certain limitations and disadvantages such as: 1. cannot handle the noise-induced barren plateau; 2. requires an exponential number of calls to the “controlled-QNN” with excessive lengths of circuit and run time.\newline

In contrast, our quantum training using adaptive continuous QAOA could eliminate the limitations of that using Grover Adaptive Search and the advantages come in the following two folds:
\begin{itemize}


 \item[1.] The phase oracle by LCU approach does not explicitly evaluate/store the value of the cost function at any stage of the algorithm and the number of calls to the “controlled-QNN” scales only logarithmic with respect to the inverse of the desired precision~\cite{Gily_n_2019}. Therefore the phase encoding is not affected by the noise-induced barren plateau for which the precision required is exponentially small. This is better than the case using Grover Adaptive Search.
    \item[2.]  The adaptive mixers can dramatically reduce the number of QAOA iterations while significantly increasing the quality of the output solution. This will enable our quantum training to achieve high performance within relatively shallow circuit and short run time. Thanks to the phase encoding faithfully conserving all the information and structure in the cost function, our adaptive QAOA protocol can exploit hidden structures in the QNN training problem. (Whereas, the Grover Oracle ‘cuts off’ the cost function with the threshold effectively losing some information and structure in the cost function.) Therefore, adaptive QAOA can offer beyond-Grover speed-up. Moreover, numerical experiments in \cite{yao2020reinforcement} show that when using the adaptive approach, the depth of the QAOA steps can be independent of the problem size (number of qubits), this would yield even more advantage when system size scales up.

\end{itemize}

\section{Applications}\label{applications}
In this section we discuss several applications of QNN which our quantum training algorithm can apply to. For each application, we first briefly illustrate the usage of QNN and the corresponding cost function for the task, then we present the way of amplitude encoding tailored for this application. Based on the amplitude encoding, the construction of the full quantum training algorithm is similar for every application.

\subsection{Training VQE}

\begin{figure}[h!]
\centering
\includegraphics[width=0.9\linewidth]{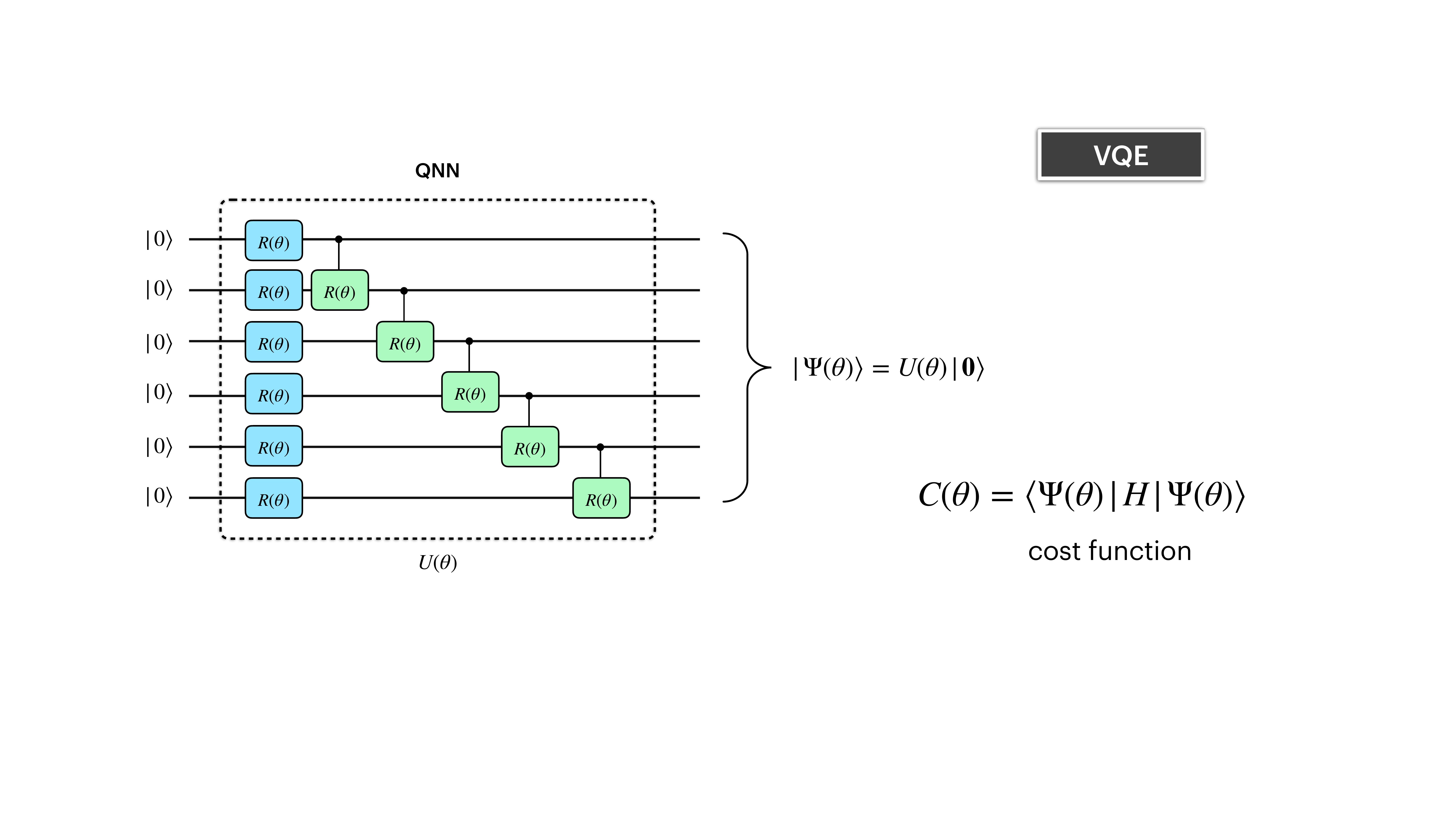}
\caption{\emph{Schematic of VQE for ground state estimation.} The QNN (a parameterized circuit ansatz) is applied to an initial state (e.g. the zero state) over multiple qubits to generate the ground state of a given Hamiltonian $H$. The parameters in the QNN, i.e. the rotation angles of the parametrized gates (here for simplicity we use the same symbol $\theta$ for all the angles of different gates), are optimized so that the generated state of the QNN possess the lowest expectation value of the given Hamiltonian.
 }
\label{vqe}
\end{figure}

 Variational quanutm eigensolvers (VQEs) utilize QNN to estimate the eigenvalue corresponding to some eigenstate of a Hamiltonian. The most common instance is ground state estimation in which the QNN (a parameterized circuit ansatz) is applied to an initial state (e.g.the zero state) over multiple qubits to generate the ground state. The parameters in the QNN are optimized so that the generated state of the QNN possess the lowest expectation value of the given Hamiltonian. A schematic of VQE for groud state estimation is presented in Fig.~\ref{vqe}.

Consider a Hamiltonian$$H=\sum_i a_i U_i$$ where $U_i$ is a unitary, $a_i>0$ and $\sum_{j}a_i=1$. (This assumption can be made without loss of generality by renormalizing the Hamiltonian and absorbing signs into the unitary matrix.)  Let the state $\ket{\psi({\btheta})}$ for $\btheta\in \mathbb{R}^m$ be the variational state prepared by the QNN. ($m$ is the number of parameters in the QNN.)
The cost function of the QNN is:
\begin{equation}
C(\btheta)=\bra{\psi(\btheta)}\sum_i a_i U_i\ket{\psi(\btheta)}.
\end{equation}

Our goal is then to estimate
\begin{equation}
\btheta^{*}= \underset{\btheta}{{\rm argmin}}\left(\bra{\psi(\btheta)}\sum_i a_i U_i\ket{\psi(\btheta)}\right).
\end{equation}

Here we use a technique "Linear Combinations of Unitaries"(LCU) \cite{childs2012hamiltonian} to implement the Hamiltonian. Define new unitary oracles ${W},H_{\textsf{\textit{LCU}}}$ such that
\begin{align}
{W}\ket{0}&= \sum_{i}\sqrt{a_i}\ket{i},\label{eq:prepareW}\\
{H_{\textsf{\textit{LCU}}}}&= \sum_i \ketbra{i}{i}\otimes U_i.\label{eq:selectH}
\end{align}

 The amplitude encoding of the cost function of the QNN can be implemented using the following circuit in Fig.~\ref{vqecircuit} \cite{Gily_n_2019}:
\begin{figure}[h!]
  \centering
  \begin{quantikz}  \lstick{\textsf{Hadamard test Ancilla qubit:} \ket{0}} & \qw&\qw\gategroup[wires=4,steps=4,style={dashed,rounded corners,fill=green!20, inner xsep=3pt},background]{$U'$} &\gate{H}  & \ctrl{2} &\gate{H}& \qw \\
  \lstick{\textsf{LCU Ancilla qubits:} \ket{0}} &\qwbundle[alternate]{}&\qwbundle[alternate]{}   &\gate{W}\qwbundle[alternate]{}  & \ctrlbundle{2}\gategroup[wires=3,steps=1,style={dashed,rounded corners,fill=orange!20, inner xsep=3pt},background,label style={label position=below,anchor=north,yshift=-0.2cm}]{${H_{\textsf{\textit{LCU}}}}$} &\gate{W^{\dagger}}\qwbundle[alternate]{}& \qwbundle[alternate]{}\\
  \lstick{\textsf{Parameter register:} $\ket{0}^{\otimes dr}$}&\gate{H}\qwbundle[alternate]{}    & \ctrlbundle{1}\gategroup[2,steps=1,style={dashed,rounded corners,fill=blue!20, inner xsep=2pt},background,label style={label position=below,anchor=north,yshift=-0.2cm}]{{\sc $P$}}&\qwbundle[alternate]{} &\qwbundle[alternate]{}&\qwbundle[alternate]{} &\qwbundle[alternate]{} \qwbundle[alternate]{} \\
  \lstick{\textsf{QNN register:}  $\ket{0}^{\otimes n}$}\qwbundle[alternate=2]{}& \qwbundle[alternate=2]{} & \gate[nwires=1]{P_j}\qwbundle[alternate=2]{}& \qwbundle[alternate=2]{}& \gate[nwires=1]{U_i}\qwbundle[alternate=2]{}& \qwbundle[alternate=2]{}& \qwbundle[alternate=2]{}
  \end{quantikz}
  \caption{\emph{Circuit for the amplitude encoding of the cost function for VQE.} Here we use the Hadamard Test Circuit for the amplitude encoding of the cost function, as detailed in \ref{hadamardsection}. We use a technique "Linear Combinations of Unitaries"(LCU) \cite{childs2012hamiltonian} to implement the given Hamiltonian $H=\sum_i a_i U_i$. The unitary oracles ${W},H_{\textsf{\textit{LCU}}}$ are defined as
${W}\ket{0}= \sum_{i}\sqrt{a_i}\ket{i},{H_{\textsf{\textit{LCU}}}}= \sum_i \ketbra{i}{i}\otimes U_i.
$}
\label{vqecircuit}
\end{figure}
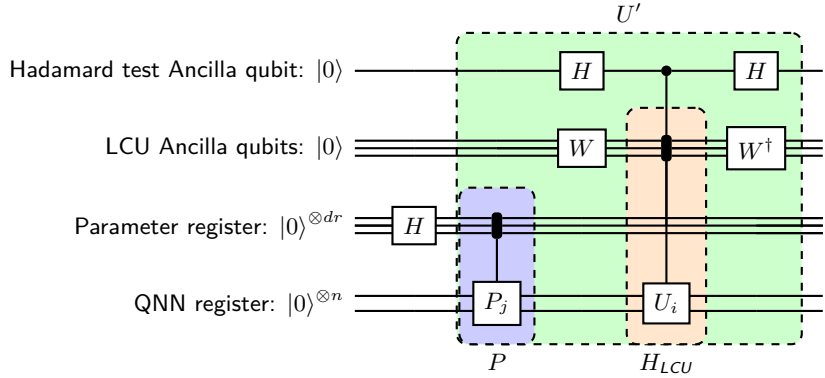

\subsection{Learning to generate a pure state}

\begin{figure}[h!]
\centering
\includegraphics[width=0.9\linewidth]{ 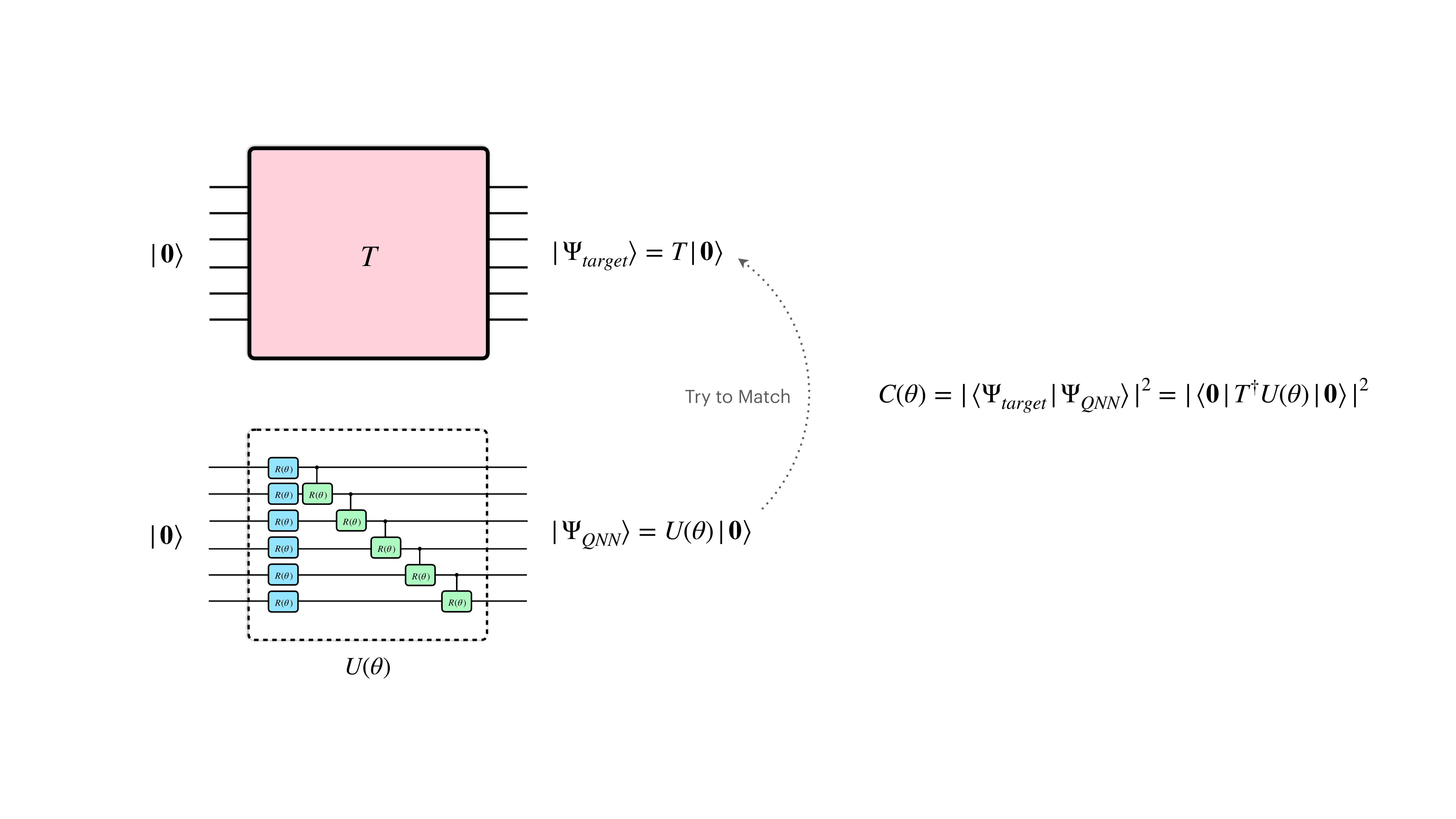}
\caption{\emph{Schematic of using QNN to generate a pure state.} In our scenario, the target state is generated by a given Unitary $T$, i.e. $\ket{\Psi_{target}}=T\ket{0}$), the QNN (denoted as $U(\theta)$) serves as another generator circuit for the target state. The parameters in QNN are optimized such that the generated state of QNN $\ket{\Psi_{QNN}}$ matches the target state. The cost function is the fidelity between the target state and the generated state by QNN.}
\label{learnu}
\end{figure}

Another application of our quantum training is when QNN is served as a generative model to learn a pure state. In our scenario, the target state is generated by a given unitary (e.g. a given sequence of gates), the QNN serves as another generator circuit for the target state. The parameters in QNN are optimized such that the generated state of QNN matches the target state. This approach can be used to transform a given sequence of gates to a different/simpler sequence (e.g. translating circuits from superconducting gate sets to ion trap gate sets) A schematic of this application is presented in Fig.~\ref{learnu}. \newline

The amplitude encoding for this application has been given in section \ref{swapsection}.

\subsection{Training a Quantum Classifier}

Finally, we discuss the application of QNN as a quantum classifier that performs supervised learning which is a standard problem in machine learning. \newline

To formalise the learning task, let $\mathcal{X}$ be a set of inputs and $\mathcal{Y}$ a set of outputs. Given a dataset $\mathcal{D} = \{(x_1,y_1),...,(x_M,y_M)\}$ of pairs of so called \textit{training inputs} $x_m \in \mathcal{X}$ and \textit{target outputs} $y_m \in \mathcal{Y}$ for $m=1,...,M$, the task of the model is to predict the output $y \in \mathcal{Y}$ of a new input $x \in \mathcal{X}$. For simplicity we will assume in the following that $\mathcal{X} = \mathbb{R}^N$ and $\mathcal{Y} =\{0,1\}$, which is a binary classification task on a $N$-dimensional real input space. In summary the quantum classifier aim to learn an effective labeling function $\ell: \mathcal{X} \to \{0,1\}$.  \newline

\begin{figure}[h!]
\centering
\includegraphics[width=0.9\linewidth]{ 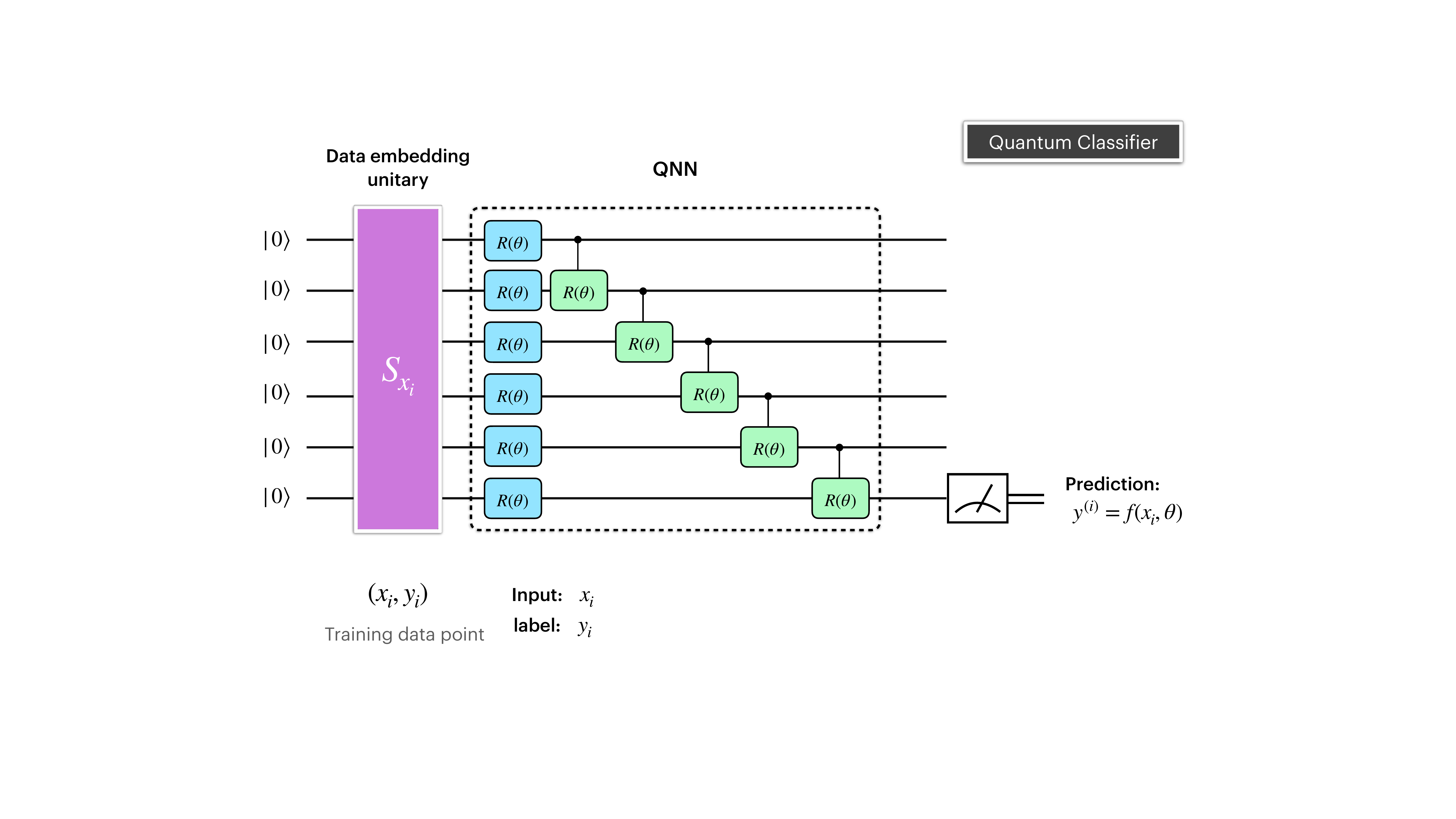}
\caption{\textit{Schematic of a quantum classifier.} For a training data point $(x_i,y_i)$, the quantum classifier first embeds $x_i$ into the state of a $n$-qubit quantum system via a data embedding circuit $S_{x_i}$ (purple box) such that $S_{x_i}\ket{0}=\ket{\varphi({x_i})}$, and subsequently uses a learnable quantum circuit $U(\btheta)$ (QNN) as a predictive model to make inference (here for simplicity we use the same symbol $\theta$ for all the angles of different gates). The predicted class label $ y^{(i)} = f (x_i, \btheta)$ is retrieved by measuring a designated qubit in the state $U(\btheta)\ket{\varphi(x)}$. }
\label{classifier}
\end{figure}

 Given an input $x_i$ and a set of parameters $\btheta$, the quantum classifier first embeds $x_i$ into the state of a $n$-qubit quantum system via a state preparation circuit $S_{x_i}$ such that $S_{x_i}\ket{0}=\ket{\varphi({x_i})}$, and subsequently uses a learnable quantum circuit $U(\btheta)$ (QNN) as a predictive model to make inference. The predicted class label $ y^{(i)} = f (x_i, \btheta)$ is retrieved by measuring a designated qubit in the state $U(\btheta)\ket{\varphi(x)}$. A schematic of the quantum classifier is presented in Fig.~\ref{classifier}. Note although the variational quantum classifier could be operated as a multiclass classifier, here we limit ourselves to the case of the binary classification discussed above and cast the multi-label tasks as a set of binary discrimination subtasks.\newline

Denote $p(\lambda)$ as the probability of the measurement result on the designated qubit being $\lambda$ $(\lambda \in\{0,1\}$). The cost function of each training data point $L_i(\btheta)$, as a function of $y_i$ and $ y^{(i)}$ and hence a function of $y_i,x_i,\btheta$ which we denote as $L(x_i,y_i,\btheta)$, is chosen to be the the probability of the measurement result on the designated qubit being identical to the given label \cite{Schuld_2020}, namely: \begin{equation}
  L_i(\btheta)= L(x_i,y_i,\btheta) :=p(y_i).
  \label{definecost}
\end{equation}
Note here the larger the probability is, the more correct the prediction is, so we want to maximize the cost (in this paper, in order to be coherent with the former narrative, we use "cost" instead of commonly used "likelihood" of inferring the correct label for a data sample.) \newline

On the other hand, the quantum state of the system after the state preparation and QNN inference can be written as:
\begin{align}
    \ket{\Psi_i(\btheta)}=U(\btheta)\ket{\varphi(x_i)}
    =\sqrt{p(0)}|0\rangle|u_{\btheta}\rangle+\sqrt{1-p(0)}|1\rangle|v_{\btheta}\rangle
    \\=\begin{cases}
     \sqrt{p(y_i)}|0\rangle|u_{\btheta}\rangle+\sqrt{1-p(y_i)}|1\rangle|v_{\btheta}\rangle,& y_i=0  \\
     \sqrt{p(y_i)}|1\rangle|u_{\btheta}\rangle+\sqrt{1-p(y_i)}|0\rangle|v_{\btheta}\rangle,& y_i=1\label{state}
    \end{cases}
\end{align}
in which $|u_{\btheta}\rangle,|v_{\btheta}\rangle$ are some normalized state that depend on $\btheta$.\newline

\begin{figure}[h!]
\centering
\includegraphics[width=0.9\linewidth]{ 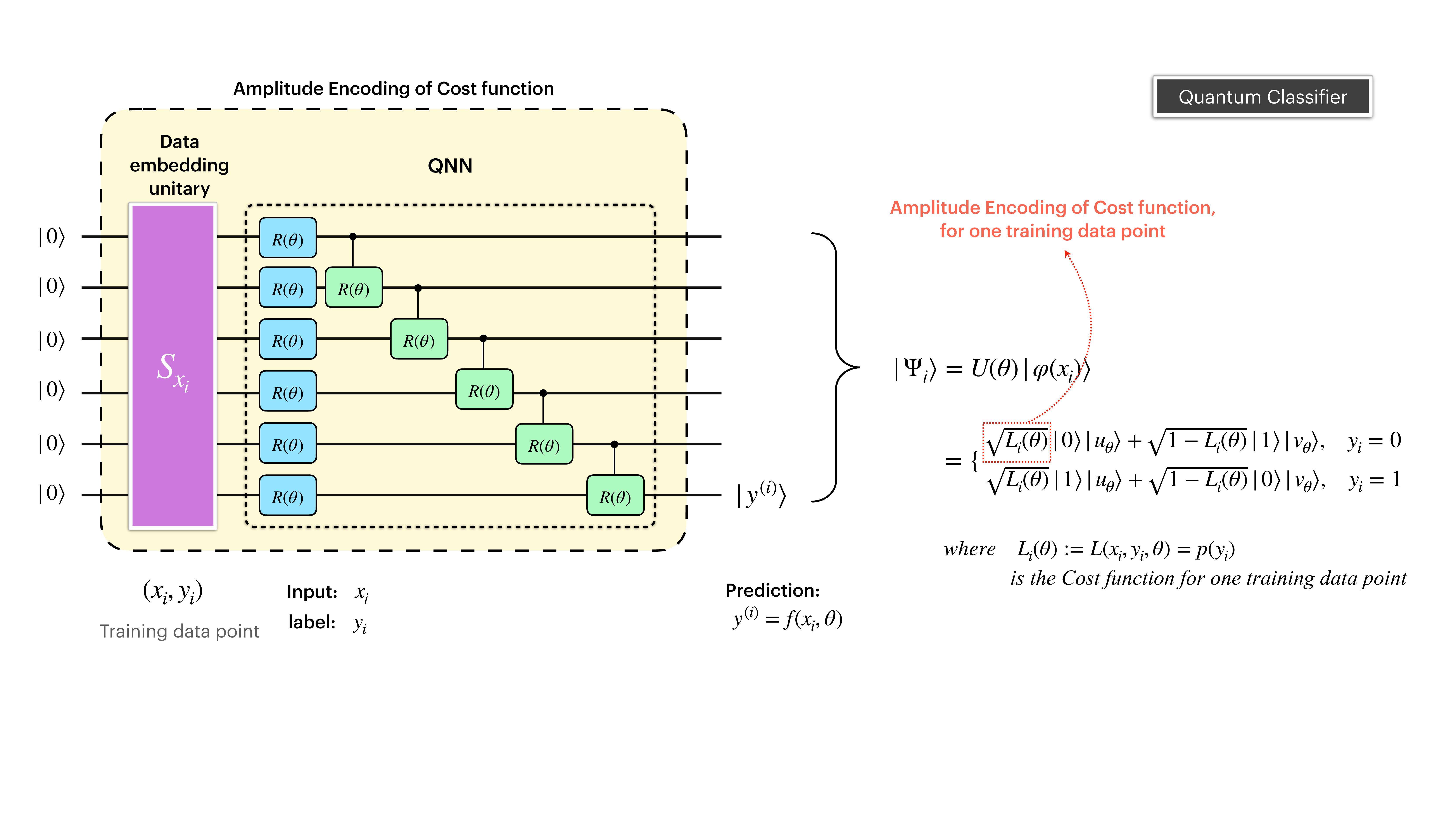}
\caption{\textit{Amplitude encoding of the cost function for quantum classifier.} For a training data point $(x_i,y_i)$, the quantum classifier first embeds $x_i$ into the state of a $n$-qubit quantum system via a data embedding circuit $S_{x_i}$ (purple box) such that $S_{x_i}\ket{0}=\ket{\varphi({x_i})}$, and subsequently uses a learnable quantum circuit $U(\btheta)$ (QNN) as a predictive model to make inference (here for simplicity we use the same symbol $\theta$ for all the angles of different gates). The predicted class label $ y^{(i)} = f (x_i, \btheta)$ is retrieved by measuring a designated qubit in the state $U(\btheta)\ket{\varphi(x)}$. Denote $p(\lambda)$ as the probability of the measurement result on the designated qubit being $\lambda$ $(\lambda \in\{0,1\}$). The cost function of each training data point $L_i(\btheta)$, as a function of $y_i$ and $ y^{(i)}$ and hence a function of $y_i,x_i,\btheta$ which we denote as $L(x_i,y_i,\btheta)$, is chosen to be the the probability of the measurement result on the designated qubit being identical to the given label, namely: $L_i(\btheta)= L(x_i,y_i,\btheta) :=p(y_i)$. We can see that the cost of each data sample is naturally encoded in the amplitude of the output state of QNN.}
\label{ampliclass}
\end{figure}

From Eqs. \ref{definecost} and \ref{state} we can see that the cost of each data sample $L(x_i,y_i,\btheta)$ is naturally encoded in the amplitude of the output state of QNN $\ket{\Psi_i(\btheta)}$. We illustrate the amplitude encoding of the cost function for quantum classifier in Fig.~\ref{ampliclass}. Constructing the "controlled-QNN" will achieve the amplitude encoding for all the parameter configurations in parallel. Based on this amplitude encoding, we can construct $e^{-i \gamma L(x_i,y_i,\btheta)}$ using the methods discussed in section \ref{phaseo}. \footnote{Note that for the training data point with $y_i=0$, $C_1$ in the Grover Operator $G$ and $G^*$ has to be adjusted to $-Z$}

The total cost function of the whole training set can be defined as: (for simplicity we omit $\frac{1}{M}$ here)
 \begin{equation}
C(\btheta)=\sum_i L(x_i,y_i,\btheta).
\end{equation}

It follows immediately
 \begin{equation}
e^{-i\gamma C(\btheta)}= \Pi_i e^{-i\gamma  L(x_i,y_i,\btheta)}.
\end{equation}

\begin{figure}[h!]
\centering
\includegraphics[width=0.7\linewidth]{ 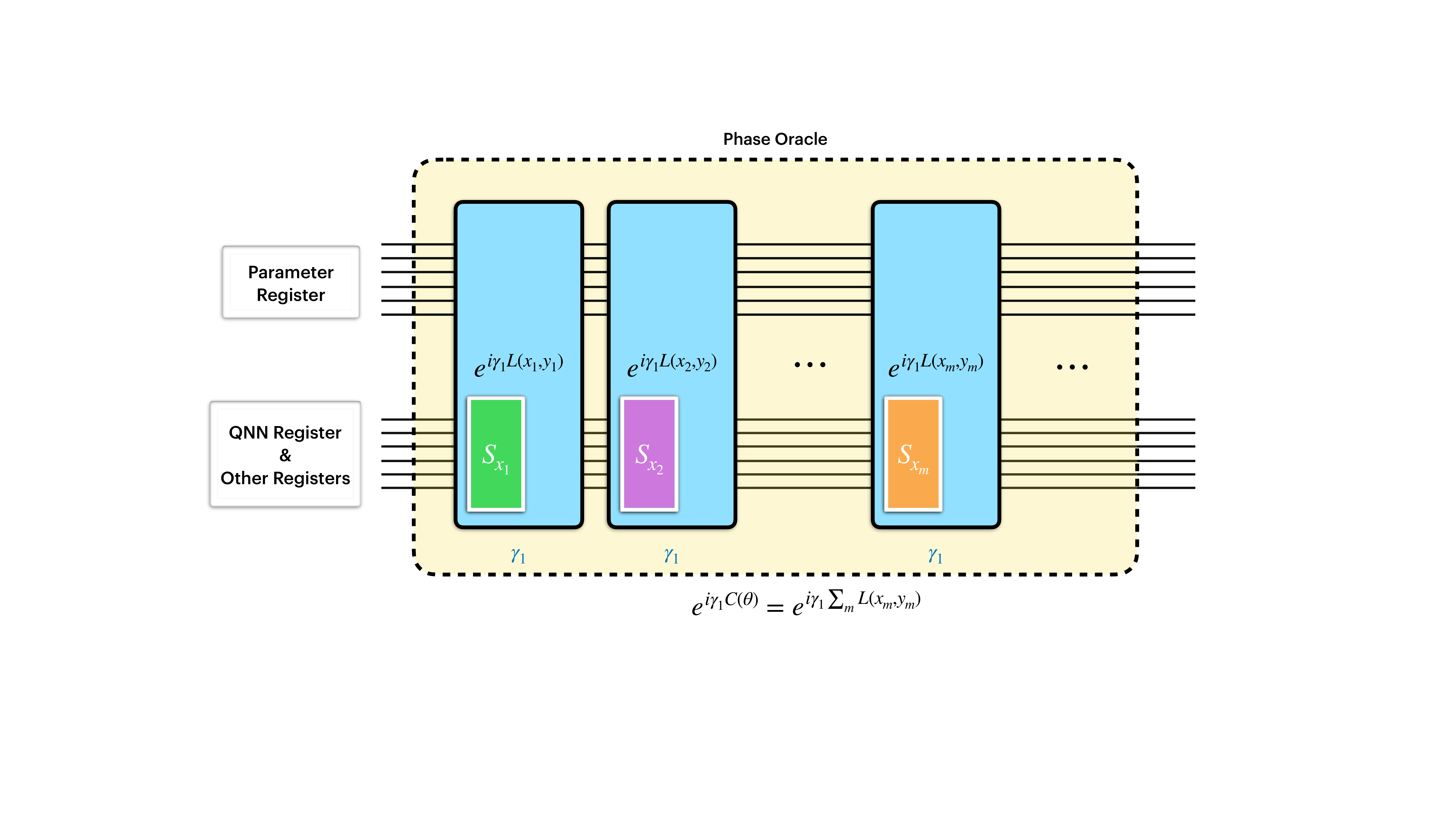}
\caption{\textit{Phase encoding of the total cost function of quantum classifier.} The total cost function of the whole training set can be defined as:  $C(\btheta)=\sum_i L(x_i,y_i,\btheta)$. It follows immediately
 $e^{-i\gamma C(\btheta)}= \Pi_i e^{-i\gamma  L(x_i,y_i,\btheta)}$. Therefore the phase encoding of the total cost function (the overall yellow box) can be implemented by accumulating individual phase encoding for each training sample(blue boxes). In this figure, we omit $\btheta$ in $L(x_i,y_i,\btheta)$ for simplicity. The inner boxes in the blue boxes represent different data embedding unitary for the training data points.}
\label{accumu}
\end{figure}

Therefore the phase encoding of the total cost function can be implemented by accumulating individual phase encoding for each training sample, this process can be illustrated in Fig.~\ref{accumu}.

\begin{figure}[h!]
\centering
\includegraphics[width=\linewidth]{ 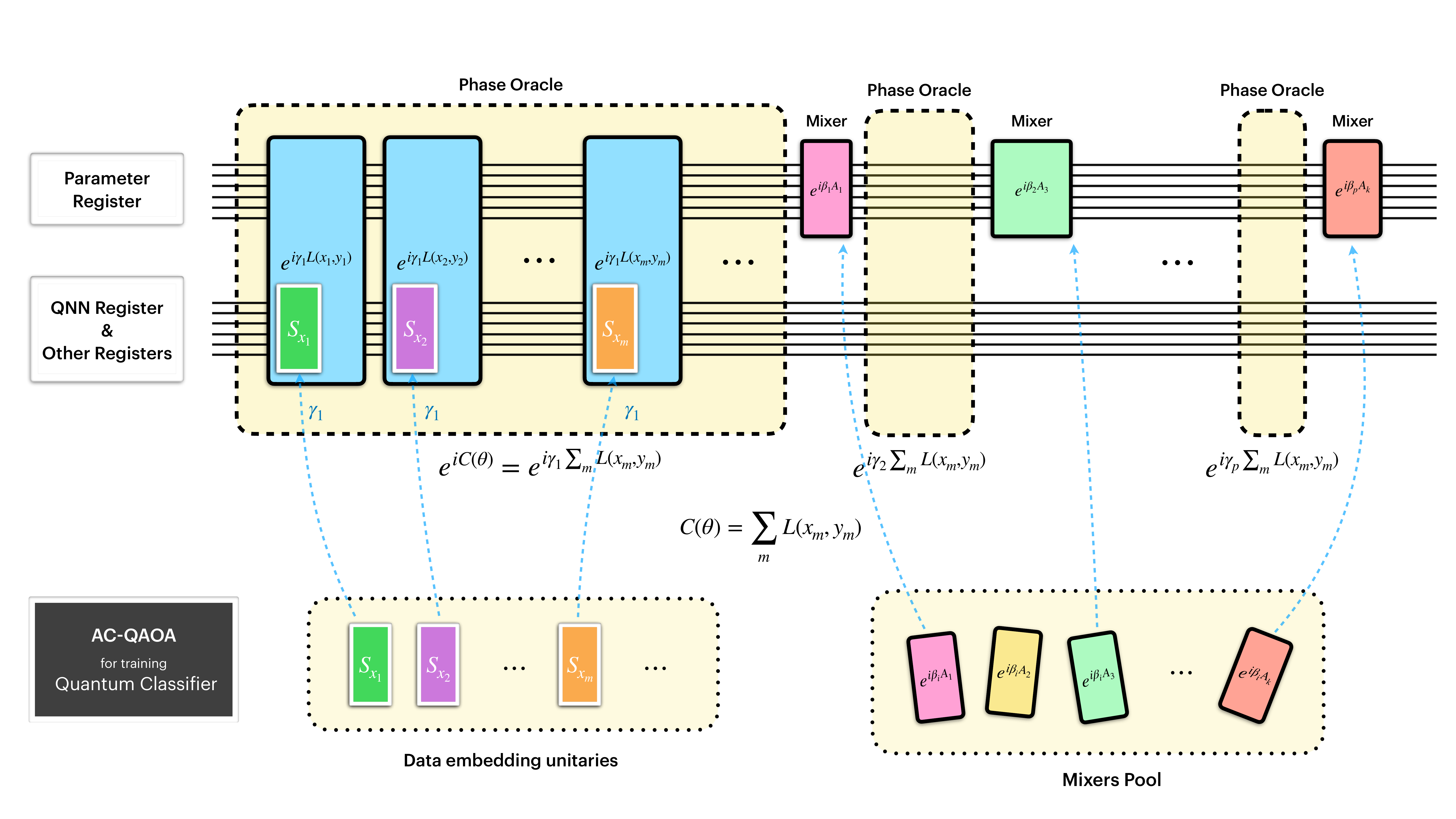}
\caption{\textit{Schematic of our quantum training protocol for quantum classifier.} The full quantum training protocol consists of the alternation of the Phase Oracle that achieve coherent phase encoding of the cost function and the Adaptive Mixers chosen from a Mixers pool. The phase encoding of the total cost function for the quantum classifier are detailed in Fig.~\ref{accumu}. The total cost function of the whole training set can be defined as:  $C(\btheta)=\sum_i L(x_i,y_i,\btheta)$. It follows that
 $e^{-i\gamma C(\btheta)}= \Pi_i e^{-i\gamma  L(x_i,y_i,\btheta)}$. Therefore the Phase Oracle for the total cost function (the yellow boxes in the upper part of this figure) can be implemented by accumulating individual phase encoding for each training sample(blue boxes). In this figure, we omit $\btheta$ in $L(x_i,y_i,\btheta)$ for simplicity. The colorful boxes with white border represent different data embedding unitary for the training data points. The colorful boxes with black border (excluding the blue ones for the Phase encoding) represent different Mixers chosen from a Mixers Pool.}
\label{classfull}
\end{figure}

Armed with the phase encoding of the total cost function, we can now construct the full quantum training protocol as in Fig.~\ref{classfull}.

\section{Discussion}

In this paper we proposed a framework leveraging quantum optimisation routines for training QNNs.  We have designed a variant of QAOA (AC-QAOA) tailored for QNN training problems. Our framework of using AC-QAOA to train QNNs consist of two major components: \textit{1.} A \textit{Phase Oracle} that can achieve coherent phase encoding of the cost function of QNN, and
\textit{2.} \textit{Adaptive Mixers} that can dramatically shorten the depth of QAOA layers while significantly improving the quality of the solution. We adopt RNN and RL to determine mixers sequence and optimize hyper-parameters.
Various applications which our quantum algorithm can apply to were presented.  \newline

QAOA itself and all of its variants are, by construction, heuristics, and therefore their advantages are ultimately determined by testing performance on concrete problems. Heuristically, AC-QAOA is expected to process the advantages of its ancestors, i,e, ADAPT-QAOA and QDD. We leave as future work for demonstrating the advantages by numerical experiments. The estimation of the number of qubits needed in given in Appendix.~\ref{number}. For a small toy example with 5 qubits and 10 rotation gates in the QNN, our protocol requires roughly 60 qubits to implement. Thus, we expect to demonstrate our protocol on near-term devices. \newline

In this paper we have only discussed optimizing the rotation parameters in QNNs (which belongs to a continuous optimisation problem). However, our framework can also be used in \textit{learning circuit structure} --- i.e., to find better acircuit ansatz (which belongs to discrete optimisation problem) --- or even learning the structure and parameters simultaneously. We leave these extensions to future work. Furthermore, we would like to explore the possibility of applying other quantum optimisation algorithms (such as adiabatic quantum evolution, quantum walks, etc.) to QNN training. We hope this work will provide a useful framework for quantum training of future quantum devices.
\bibliographystyle{apsrev4-1}
\bibliography{Ref.bib}

\begin{thebibliography}{66}%
\makeatletter
\providecommand \@ifxundefined [1]{%
 \@ifx{#1\undefined}
}%
\providecommand \@ifnum [1]{%
 \ifnum #1\expandafter \@firstoftwo
 \else \expandafter \@secondoftwo
 \fi
}%
\providecommand \@ifx [1]{%
 \ifx #1\expandafter \@firstoftwo
 \else \expandafter \@secondoftwo
 \fi
}%
\providecommand \natexlab [1]{#1}%
\providecommand \enquote  [1]{``#1''}%
\providecommand \bibnamefont  [1]{#1}%
\providecommand \bibfnamefont [1]{#1}%
\providecommand \citenamefont [1]{#1}%
\providecommand \href@noop [0]{\@secondoftwo}%
\providecommand \href [0]{\begingroup \@sanitize@url \@href}%
\providecommand \@href[1]{\@@startlink{#1}\@@href}%
\providecommand \@@href[1]{\endgroup#1\@@endlink}%
\providecommand \@sanitize@url [0]{\catcode `\\12\catcode `\$12\catcode
  `\&12\catcode `\#12\catcode `\^12\catcode `\_12\catcode `\%12\relax}%
\providecommand \@@startlink[1]{}%
\providecommand \@@endlink[0]{}%
\providecommand \url  [0]{\begingroup\@sanitize@url \@url }%
\providecommand \@url [1]{\endgroup\@href {#1}{\urlprefix }}%
\providecommand \urlprefix  [0]{URL }%
\providecommand \Eprint [0]{\href }%
\providecommand \doibase [0]{http://dx.doi.org/}%
\providecommand \selectlanguage [0]{\@gobble}%
\providecommand \bibinfo  [0]{\@secondoftwo}%
\providecommand \bibfield  [0]{\@secondoftwo}%
\providecommand \translation [1]{[#1]}%
\providecommand \BibitemOpen [0]{}%
\providecommand \bibitemStop [0]{}%
\providecommand \bibitemNoStop [0]{.\EOS\space}%
\providecommand \EOS [0]{\spacefactor3000\relax}%
\providecommand \BibitemShut  [1]{\csname bibitem#1\endcsname}%
\let\auto@bib@innerbib\@empty
\bibitem [{\citenamefont {McClean}\ \emph {et~al.}(2016)\citenamefont
  {McClean}, \citenamefont {Romero}, \citenamefont {Babbush},\ and\
  \citenamefont {Aspuru-Guzik}}]{mcclean2016theory}%
  \BibitemOpen
  \bibfield  {author} {\bibinfo {author} {\bibfnamefont {J.~R.}\ \bibnamefont
  {McClean}}, \bibinfo {author} {\bibfnamefont {J.}~\bibnamefont {Romero}},
  \bibinfo {author} {\bibfnamefont {R.}~\bibnamefont {Babbush}}, \ and\
  \bibinfo {author} {\bibfnamefont {A.}~\bibnamefont {Aspuru-Guzik}},\ }\href
  {https://iopscience.iop.org/article/10.1088/1367-2630/18/2/023023/meta}
  {\bibfield  {journal} {\bibinfo  {journal} {New Journal of Physics}\ }\textbf
  {\bibinfo {volume} {18}},\ \bibinfo {pages} {023023} (\bibinfo {year}
  {2016})}\BibitemShut {NoStop}%
\bibitem [{\citenamefont {{Peruzzo}}\ \emph {et~al.}(2014)\citenamefont
  {{Peruzzo}}, \citenamefont {{McClean}}, \citenamefont {{Shadbolt}},
  \citenamefont {{Yung}}, \citenamefont {{Zhou}}, \citenamefont {{Love}},
  \citenamefont {{Aspuru-Guzik}},\ and\ \citenamefont {{O'Brien}}}]{VQE}%
  \BibitemOpen
  \bibfield  {author} {\bibinfo {author} {\bibfnamefont {A.}~\bibnamefont
  {{Peruzzo}}}, \bibinfo {author} {\bibfnamefont {J.}~\bibnamefont
  {{McClean}}}, \bibinfo {author} {\bibfnamefont {P.}~\bibnamefont
  {{Shadbolt}}}, \bibinfo {author} {\bibfnamefont {M.-H.}\ \bibnamefont
  {{Yung}}}, \bibinfo {author} {\bibfnamefont {X.-Q.}\ \bibnamefont {{Zhou}}},
  \bibinfo {author} {\bibfnamefont {P.~J.}\ \bibnamefont {{Love}}}, \bibinfo
  {author} {\bibfnamefont {A.}~\bibnamefont {{Aspuru-Guzik}}}, \ and\ \bibinfo
  {author} {\bibfnamefont {J.~L.}\ \bibnamefont {{O'Brien}}},\ }\href {\doibase
  10.1038/ncomms5213} {\bibfield  {journal} {\bibinfo  {journal} {Nature
  Communications}\ }\textbf {\bibinfo {volume} {5}},\ \bibinfo {eid} {4213}
  (\bibinfo {year} {2014})}\BibitemShut {NoStop}%
\bibitem [{\citenamefont {Farhi}\ \emph {et~al.}(2014)\citenamefont {Farhi},
  \citenamefont {Goldstone},\ and\ \citenamefont {Gutmann}}]{qaoa2014}%
  \BibitemOpen
  \bibfield  {author} {\bibinfo {author} {\bibfnamefont {E.}~\bibnamefont
  {Farhi}}, \bibinfo {author} {\bibfnamefont {J.}~\bibnamefont {Goldstone}}, \
  and\ \bibinfo {author} {\bibfnamefont {S.}~\bibnamefont {Gutmann}},\
  }\href@noop {} {\enquote {\bibinfo {title} {A quantum approximate
  optimization algorithm},}\ } (\bibinfo {year} {2014}),\ \Eprint
  {http://arxiv.org/abs/1411.4028} {arXiv:1411.4028 [quant-ph]} \BibitemShut
  {NoStop}%
\bibitem [{\citenamefont {Farhi}\ and\ \citenamefont
  {Neven}(2018)}]{farhi2018classification}%
  \BibitemOpen
  \bibfield  {author} {\bibinfo {author} {\bibfnamefont {E.}~\bibnamefont
  {Farhi}}\ and\ \bibinfo {author} {\bibfnamefont {H.}~\bibnamefont {Neven}},\
  }\href {https://arxiv.org/abs/1802.06002} {\bibfield  {journal} {\bibinfo
  {journal} {arXiv preprint arXiv:1802.06002}\ } (\bibinfo {year}
  {2018})}\BibitemShut {NoStop}%
\bibitem [{\citenamefont {Schuld}\ \emph {et~al.}(2020)\citenamefont {Schuld},
  \citenamefont {Bocharov}, \citenamefont {Svore},\ and\ \citenamefont
  {Wiebe}}]{Schuld_2020}%
  \BibitemOpen
  \bibfield  {author} {\bibinfo {author} {\bibfnamefont {M.}~\bibnamefont
  {Schuld}}, \bibinfo {author} {\bibfnamefont {A.}~\bibnamefont {Bocharov}},
  \bibinfo {author} {\bibfnamefont {K.~M.}\ \bibnamefont {Svore}}, \ and\
  \bibinfo {author} {\bibfnamefont {N.}~\bibnamefont {Wiebe}},\ }\href
  {\doibase 10.1103/physreva.101.032308} {\bibfield  {journal} {\bibinfo
  {journal} {Physical Review A}\ }\textbf {\bibinfo {volume} {101}} (\bibinfo
  {year} {2020}),\ 10.1103/physreva.101.032308}\BibitemShut {NoStop}%
\bibitem [{\citenamefont {Du}\ \emph {et~al.}(2018{\natexlab{a}})\citenamefont
  {Du}, \citenamefont {Hsieh}, \citenamefont {Liu},\ and\ \citenamefont
  {Tao}}]{du2018implementable}%
  \BibitemOpen
  \bibfield  {author} {\bibinfo {author} {\bibfnamefont {Y.}~\bibnamefont
  {Du}}, \bibinfo {author} {\bibfnamefont {M.-H.}\ \bibnamefont {Hsieh}},
  \bibinfo {author} {\bibfnamefont {T.}~\bibnamefont {Liu}}, \ and\ \bibinfo
  {author} {\bibfnamefont {D.}~\bibnamefont {Tao}},\ }\href@noop {} {\enquote
  {\bibinfo {title} {Implementable quantum classifier for nonlinear data},}\ }
  (\bibinfo {year} {2018}{\natexlab{a}}),\ \Eprint
  {http://arxiv.org/abs/1809.06056} {arXiv:1809.06056 [quant-ph]} \BibitemShut
  {NoStop}%
\bibitem [{\citenamefont {Benedetti}\ \emph {et~al.}(2019)\citenamefont
  {Benedetti}, \citenamefont {Garcia-Pintos}, \citenamefont {Perdomo},
  \citenamefont {Leyton-Ortega}, \citenamefont {Nam},\ and\ \citenamefont
  {Perdomo-Ortiz}}]{Benedetti_2019}%
  \BibitemOpen
  \bibfield  {author} {\bibinfo {author} {\bibfnamefont {M.}~\bibnamefont
  {Benedetti}}, \bibinfo {author} {\bibfnamefont {D.}~\bibnamefont
  {Garcia-Pintos}}, \bibinfo {author} {\bibfnamefont {O.}~\bibnamefont
  {Perdomo}}, \bibinfo {author} {\bibfnamefont {V.}~\bibnamefont
  {Leyton-Ortega}}, \bibinfo {author} {\bibfnamefont {Y.}~\bibnamefont {Nam}},
  \ and\ \bibinfo {author} {\bibfnamefont {A.}~\bibnamefont {Perdomo-Ortiz}},\
  }\href {\doibase 10.1038/s41534-019-0157-8} {\bibfield  {journal} {\bibinfo
  {journal} {npj Quantum Information}\ }\textbf {\bibinfo {volume} {5}}
  (\bibinfo {year} {2019}),\ 10.1038/s41534-019-0157-8}\BibitemShut {NoStop}%
\bibitem [{\citenamefont {Zeng}\ \emph {et~al.}(2019)\citenamefont {Zeng},
  \citenamefont {Wu}, \citenamefont {Liu}, \citenamefont {Wang},\ and\
  \citenamefont {Hu}}]{Zeng_2019}%
  \BibitemOpen
  \bibfield  {author} {\bibinfo {author} {\bibfnamefont {J.}~\bibnamefont
  {Zeng}}, \bibinfo {author} {\bibfnamefont {Y.}~\bibnamefont {Wu}}, \bibinfo
  {author} {\bibfnamefont {J.-G.}\ \bibnamefont {Liu}}, \bibinfo {author}
  {\bibfnamefont {L.}~\bibnamefont {Wang}}, \ and\ \bibinfo {author}
  {\bibfnamefont {J.}~\bibnamefont {Hu}},\ }\href {\doibase
  10.1103/physreva.99.052306} {\bibfield  {journal} {\bibinfo  {journal}
  {Physical Review A}\ }\textbf {\bibinfo {volume} {99}} (\bibinfo {year}
  {2019}),\ 10.1103/physreva.99.052306}\BibitemShut {NoStop}%
\bibitem [{\citenamefont {Hamilton}\ \emph {et~al.}(2019)\citenamefont
  {Hamilton}, \citenamefont {Dumitrescu},\ and\ \citenamefont
  {Pooser}}]{Hamilton_2019}%
  \BibitemOpen
  \bibfield  {author} {\bibinfo {author} {\bibfnamefont {K.~E.}\ \bibnamefont
  {Hamilton}}, \bibinfo {author} {\bibfnamefont {E.~F.}\ \bibnamefont
  {Dumitrescu}}, \ and\ \bibinfo {author} {\bibfnamefont {R.~C.}\ \bibnamefont
  {Pooser}},\ }\href {\doibase 10.1103/physreva.99.062323} {\bibfield
  {journal} {\bibinfo  {journal} {Physical Review A}\ }\textbf {\bibinfo
  {volume} {99}} (\bibinfo {year} {2019}),\
  10.1103/physreva.99.062323}\BibitemShut {NoStop}%
\bibitem [{\citenamefont {Mitarai}\ \emph {et~al.}(2018)\citenamefont
  {Mitarai}, \citenamefont {Negoro}, \citenamefont {Kitagawa},\ and\
  \citenamefont {Fujii}}]{mitarai2018quantum}%
  \BibitemOpen
  \bibfield  {author} {\bibinfo {author} {\bibfnamefont {K.}~\bibnamefont
  {Mitarai}}, \bibinfo {author} {\bibfnamefont {M.}~\bibnamefont {Negoro}},
  \bibinfo {author} {\bibfnamefont {M.}~\bibnamefont {Kitagawa}}, \ and\
  \bibinfo {author} {\bibfnamefont {K.}~\bibnamefont {Fujii}},\ }\href
  {\doibase 10.1103/PhysRevA.98.032309} {\bibfield  {journal} {\bibinfo
  {journal} {Phys. Rev. A}\ }\textbf {\bibinfo {volume} {98}},\ \bibinfo
  {pages} {032309} (\bibinfo {year} {2018})}\BibitemShut {NoStop}%
\bibitem [{\citenamefont {Huang}\ \emph {et~al.}(2020)\citenamefont {Huang},
  \citenamefont {Du}, \citenamefont {Gong}, \citenamefont {Zhao}, \citenamefont
  {Wu}, \citenamefont {Wang}, \citenamefont {Li}, \citenamefont {Liang},
  \citenamefont {Lin}, \citenamefont {Xu}, \citenamefont {Yang}, \citenamefont
  {Liu}, \citenamefont {Hsieh}, \citenamefont {Deng}, \citenamefont {Rong},
  \citenamefont {Peng}, \citenamefont {Lu}, \citenamefont {Chen}, \citenamefont
  {Tao}, \citenamefont {Zhu},\ and\ \citenamefont
  {Pan}}]{huang2020experimental}%
  \BibitemOpen
  \bibfield  {author} {\bibinfo {author} {\bibfnamefont {H.-L.}\ \bibnamefont
  {Huang}}, \bibinfo {author} {\bibfnamefont {Y.}~\bibnamefont {Du}}, \bibinfo
  {author} {\bibfnamefont {M.}~\bibnamefont {Gong}}, \bibinfo {author}
  {\bibfnamefont {Y.}~\bibnamefont {Zhao}}, \bibinfo {author} {\bibfnamefont
  {Y.}~\bibnamefont {Wu}}, \bibinfo {author} {\bibfnamefont {C.}~\bibnamefont
  {Wang}}, \bibinfo {author} {\bibfnamefont {S.}~\bibnamefont {Li}}, \bibinfo
  {author} {\bibfnamefont {F.}~\bibnamefont {Liang}}, \bibinfo {author}
  {\bibfnamefont {J.}~\bibnamefont {Lin}}, \bibinfo {author} {\bibfnamefont
  {Y.}~\bibnamefont {Xu}}, \bibinfo {author} {\bibfnamefont {R.}~\bibnamefont
  {Yang}}, \bibinfo {author} {\bibfnamefont {T.}~\bibnamefont {Liu}}, \bibinfo
  {author} {\bibfnamefont {M.-H.}\ \bibnamefont {Hsieh}}, \bibinfo {author}
  {\bibfnamefont {H.}~\bibnamefont {Deng}}, \bibinfo {author} {\bibfnamefont
  {H.}~\bibnamefont {Rong}}, \bibinfo {author} {\bibfnamefont {C.-Z.}\
  \bibnamefont {Peng}}, \bibinfo {author} {\bibfnamefont {C.-Y.}\ \bibnamefont
  {Lu}}, \bibinfo {author} {\bibfnamefont {Y.-A.}\ \bibnamefont {Chen}},
  \bibinfo {author} {\bibfnamefont {D.}~\bibnamefont {Tao}}, \bibinfo {author}
  {\bibfnamefont {X.}~\bibnamefont {Zhu}}, \ and\ \bibinfo {author}
  {\bibfnamefont {J.-W.}\ \bibnamefont {Pan}},\ }\href@noop {} {\enquote
  {\bibinfo {title} {Experimental quantum generative adversarial networks for
  image generation},}\ } (\bibinfo {year} {2020}),\ \Eprint
  {http://arxiv.org/abs/2010.06201} {arXiv:2010.06201 [quant-ph]} \BibitemShut
  {NoStop}%
\bibitem [{\citenamefont {Du}\ \emph {et~al.}(2018{\natexlab{b}})\citenamefont
  {Du}, \citenamefont {Hsieh}, \citenamefont {Liu},\ and\ \citenamefont
  {Tao}}]{du2018expressive}%
  \BibitemOpen
  \bibfield  {author} {\bibinfo {author} {\bibfnamefont {Y.}~\bibnamefont
  {Du}}, \bibinfo {author} {\bibfnamefont {M.-H.}\ \bibnamefont {Hsieh}},
  \bibinfo {author} {\bibfnamefont {T.}~\bibnamefont {Liu}}, \ and\ \bibinfo
  {author} {\bibfnamefont {D.}~\bibnamefont {Tao}},\ }\href@noop {} {\bibfield
  {journal} {\bibinfo  {journal} {arXiv preprint arXiv:1810.11922}\ } (\bibinfo
  {year} {2018}{\natexlab{b}})}\BibitemShut {NoStop}%
\bibitem [{\citenamefont {McClean}\ \emph {et~al.}(2018)\citenamefont
  {McClean}, \citenamefont {Boixo}, \citenamefont {Smelyanskiy}, \citenamefont
  {Babbush},\ and\ \citenamefont {Neven}}]{mcclean2018barren}%
  \BibitemOpen
  \bibfield  {author} {\bibinfo {author} {\bibfnamefont {J.~R.}\ \bibnamefont
  {McClean}}, \bibinfo {author} {\bibfnamefont {S.}~\bibnamefont {Boixo}},
  \bibinfo {author} {\bibfnamefont {V.~N.}\ \bibnamefont {Smelyanskiy}},
  \bibinfo {author} {\bibfnamefont {R.}~\bibnamefont {Babbush}}, \ and\
  \bibinfo {author} {\bibfnamefont {H.}~\bibnamefont {Neven}},\ }\href
  {https://www.nature.com/articles/s41467-018-07090-4} {\bibfield  {journal}
  {\bibinfo  {journal} {Nature communications}\ }\textbf {\bibinfo {volume}
  {9}},\ \bibinfo {pages} {4812} (\bibinfo {year} {2018})}\BibitemShut
  {NoStop}%
\bibitem [{\citenamefont {Skolik}\ \emph {et~al.}(2020)\citenamefont {Skolik},
  \citenamefont {McClean}, \citenamefont {Mohseni}, \citenamefont {van~der
  Smagt},\ and\ \citenamefont {Leib}}]{skolik2020layerwise}%
  \BibitemOpen
  \bibfield  {author} {\bibinfo {author} {\bibfnamefont {A.}~\bibnamefont
  {Skolik}}, \bibinfo {author} {\bibfnamefont {J.~R.}\ \bibnamefont {McClean}},
  \bibinfo {author} {\bibfnamefont {M.}~\bibnamefont {Mohseni}}, \bibinfo
  {author} {\bibfnamefont {P.}~\bibnamefont {van~der Smagt}}, \ and\ \bibinfo
  {author} {\bibfnamefont {M.}~\bibnamefont {Leib}},\ }\href
  {https://arxiv.org/abs/2006.14904} {\bibfield  {journal} {\bibinfo  {journal}
  {arXiv preprint arXiv:2006.14904}\ } (\bibinfo {year} {2020})}\BibitemShut
  {NoStop}%
\bibitem [{\citenamefont {Cerezo}\ \emph {et~al.}(2020)\citenamefont {Cerezo},
  \citenamefont {Sone}, \citenamefont {Volkoff}, \citenamefont {Cincio},\ and\
  \citenamefont {Coles}}]{cerezo2020cost}%
  \BibitemOpen
  \bibfield  {author} {\bibinfo {author} {\bibfnamefont {M.}~\bibnamefont
  {Cerezo}}, \bibinfo {author} {\bibfnamefont {A.}~\bibnamefont {Sone}},
  \bibinfo {author} {\bibfnamefont {T.}~\bibnamefont {Volkoff}}, \bibinfo
  {author} {\bibfnamefont {L.}~\bibnamefont {Cincio}}, \ and\ \bibinfo {author}
  {\bibfnamefont {P.~J.}\ \bibnamefont {Coles}},\ }\href
  {https://arxiv.org/abs/2001.00550} {\bibfield  {journal} {\bibinfo  {journal}
  {arXiv preprint arXiv:2001.00550}\ } (\bibinfo {year} {2020})}\BibitemShut
  {NoStop}%
\bibitem [{\citenamefont {Volkoff}\ and\ \citenamefont
  {Coles}(2020)}]{volkoff2020large}%
  \BibitemOpen
  \bibfield  {author} {\bibinfo {author} {\bibfnamefont {T.}~\bibnamefont
  {Volkoff}}\ and\ \bibinfo {author} {\bibfnamefont {P.~J.}\ \bibnamefont
  {Coles}},\ }\href {https://arxiv.org/abs/2005.12200} {\bibfield  {journal}
  {\bibinfo  {journal} {arXiv preprint arXiv:2005.12200}\ } (\bibinfo {year}
  {2020})}\BibitemShut {NoStop}%
\bibitem [{\citenamefont {Verdon}\ \emph
  {et~al.}(2019{\natexlab{a}})\citenamefont {Verdon}, \citenamefont
  {Broughton}, \citenamefont {McClean}, \citenamefont {Sung}, \citenamefont
  {Babbush}, \citenamefont {Jiang}, \citenamefont {Neven},\ and\ \citenamefont
  {Mohseni}}]{verdon2019learning}%
  \BibitemOpen
  \bibfield  {author} {\bibinfo {author} {\bibfnamefont {G.}~\bibnamefont
  {Verdon}}, \bibinfo {author} {\bibfnamefont {M.}~\bibnamefont {Broughton}},
  \bibinfo {author} {\bibfnamefont {J.~R.}\ \bibnamefont {McClean}}, \bibinfo
  {author} {\bibfnamefont {K.~J.}\ \bibnamefont {Sung}}, \bibinfo {author}
  {\bibfnamefont {R.}~\bibnamefont {Babbush}}, \bibinfo {author} {\bibfnamefont
  {Z.}~\bibnamefont {Jiang}}, \bibinfo {author} {\bibfnamefont
  {H.}~\bibnamefont {Neven}}, \ and\ \bibinfo {author} {\bibfnamefont
  {M.}~\bibnamefont {Mohseni}},\ }\href@noop {} {\enquote {\bibinfo {title}
  {Learning to learn with quantum neural networks via classical neural
  networks},}\ } (\bibinfo {year} {2019}{\natexlab{a}}),\ \Eprint
  {http://arxiv.org/abs/1907.05415} {arXiv:1907.05415 [quant-ph]} \BibitemShut
  {NoStop}%
\bibitem [{\citenamefont {Du}\ \emph {et~al.}(2020{\natexlab{a}})\citenamefont
  {Du}, \citenamefont {Hsieh}, \citenamefont {Liu}, \citenamefont {You},\ and\
  \citenamefont {Tao}}]{du2020learnability}%
  \BibitemOpen
  \bibfield  {author} {\bibinfo {author} {\bibfnamefont {Y.}~\bibnamefont
  {Du}}, \bibinfo {author} {\bibfnamefont {M.-H.}\ \bibnamefont {Hsieh}},
  \bibinfo {author} {\bibfnamefont {T.}~\bibnamefont {Liu}}, \bibinfo {author}
  {\bibfnamefont {S.}~\bibnamefont {You}}, \ and\ \bibinfo {author}
  {\bibfnamefont {D.}~\bibnamefont {Tao}},\ }\href@noop {} {\enquote {\bibinfo
  {title} {On the learnability of quantum neural networks},}\ } (\bibinfo
  {year} {2020}{\natexlab{a}}),\ \Eprint {http://arxiv.org/abs/2007.12369}
  {arXiv:2007.12369 [quant-ph]} \BibitemShut {NoStop}%
\bibitem [{\citenamefont {Du}\ \emph {et~al.}(2020{\natexlab{b}})\citenamefont
  {Du}, \citenamefont {Huang}, \citenamefont {You}, \citenamefont {Hsieh},\
  and\ \citenamefont {Tao}}]{du2020quantum}%
  \BibitemOpen
  \bibfield  {author} {\bibinfo {author} {\bibfnamefont {Y.}~\bibnamefont
  {Du}}, \bibinfo {author} {\bibfnamefont {T.}~\bibnamefont {Huang}}, \bibinfo
  {author} {\bibfnamefont {S.}~\bibnamefont {You}}, \bibinfo {author}
  {\bibfnamefont {M.-H.}\ \bibnamefont {Hsieh}}, \ and\ \bibinfo {author}
  {\bibfnamefont {D.}~\bibnamefont {Tao}},\ }\href@noop {} {\enquote {\bibinfo
  {title} {Quantum circuit architecture search: error mitigation and
  trainability enhancement for variational quantum solvers},}\ } (\bibinfo
  {year} {2020}{\natexlab{b}}),\ \Eprint {http://arxiv.org/abs/2010.10217}
  {arXiv:2010.10217 [quant-ph]} \BibitemShut {NoStop}%
\bibitem [{\citenamefont {Zhang}\ \emph {et~al.}(2020)\citenamefont {Zhang},
  \citenamefont {Hsieh}, \citenamefont {Liu},\ and\ \citenamefont
  {Tao}}]{zhang2020trainability}%
  \BibitemOpen
  \bibfield  {author} {\bibinfo {author} {\bibfnamefont {K.}~\bibnamefont
  {Zhang}}, \bibinfo {author} {\bibfnamefont {M.-H.}\ \bibnamefont {Hsieh}},
  \bibinfo {author} {\bibfnamefont {L.}~\bibnamefont {Liu}}, \ and\ \bibinfo
  {author} {\bibfnamefont {D.}~\bibnamefont {Tao}},\ }\href@noop {} {\enquote
  {\bibinfo {title} {Toward trainability of quantum neural networks},}\ }
  (\bibinfo {year} {2020}),\ \Eprint {http://arxiv.org/abs/2011.06258}
  {arXiv:2011.06258 [quant-ph]} \BibitemShut {NoStop}%
\bibitem [{\citenamefont {Wang}\ \emph {et~al.}(2020)\citenamefont {Wang},
  \citenamefont {Fontana}, \citenamefont {Cerezo}, \citenamefont {Sharma},
  \citenamefont {Sone}, \citenamefont {Cincio},\ and\ \citenamefont
  {Coles}}]{wang2020noiseinduced}%
  \BibitemOpen
  \bibfield  {author} {\bibinfo {author} {\bibfnamefont {S.}~\bibnamefont
  {Wang}}, \bibinfo {author} {\bibfnamefont {E.}~\bibnamefont {Fontana}},
  \bibinfo {author} {\bibfnamefont {M.}~\bibnamefont {Cerezo}}, \bibinfo
  {author} {\bibfnamefont {K.}~\bibnamefont {Sharma}}, \bibinfo {author}
  {\bibfnamefont {A.}~\bibnamefont {Sone}}, \bibinfo {author} {\bibfnamefont
  {L.}~\bibnamefont {Cincio}}, \ and\ \bibinfo {author} {\bibfnamefont {P.~J.}\
  \bibnamefont {Coles}},\ }\href@noop {} {\enquote {\bibinfo {title}
  {Noise-induced barren plateaus in variational quantum algorithms},}\ }
  (\bibinfo {year} {2020}),\ \Eprint {http://arxiv.org/abs/2007.14384}
  {arXiv:2007.14384 [quant-ph]} \BibitemShut {NoStop}%
\bibitem [{\citenamefont {Campos}\ \emph {et~al.}(2020)\citenamefont {Campos},
  \citenamefont {Nasrallah},\ and\ \citenamefont
  {Biamonte}}]{campos2020abrupt}%
  \BibitemOpen
  \bibfield  {author} {\bibinfo {author} {\bibfnamefont {E.}~\bibnamefont
  {Campos}}, \bibinfo {author} {\bibfnamefont {A.}~\bibnamefont {Nasrallah}}, \
  and\ \bibinfo {author} {\bibfnamefont {J.}~\bibnamefont {Biamonte}},\
  }\href@noop {} {\enquote {\bibinfo {title} {Abrupt transitions in variational
  quantum circuit training},}\ } (\bibinfo {year} {2020}),\ \Eprint
  {http://arxiv.org/abs/2010.09720} {arXiv:2010.09720 [quant-ph]} \BibitemShut
  {NoStop}%
\bibitem [{\citenamefont {Marrero}\ \emph {et~al.}(2020)\citenamefont
  {Marrero}, \citenamefont {Kieferová},\ and\ \citenamefont
  {Wiebe}}]{marrero2020entanglement}%
  \BibitemOpen
  \bibfield  {author} {\bibinfo {author} {\bibfnamefont {C.~O.}\ \bibnamefont
  {Marrero}}, \bibinfo {author} {\bibfnamefont {M.}~\bibnamefont {Kieferová}},
  \ and\ \bibinfo {author} {\bibfnamefont {N.}~\bibnamefont {Wiebe}},\
  }\href@noop {} {\enquote {\bibinfo {title} {Entanglement induced barren
  plateaus},}\ } (\bibinfo {year} {2020}),\ \Eprint
  {http://arxiv.org/abs/2010.15968} {arXiv:2010.15968 [quant-ph]} \BibitemShut
  {NoStop}%
\bibitem [{\citenamefont {Arrasmith}\ \emph {et~al.}(2020)\citenamefont
  {Arrasmith}, \citenamefont {Cerezo}, \citenamefont {Czarnik}, \citenamefont
  {Cincio},\ and\ \citenamefont {Coles}}]{arrasmith2020effect}%
  \BibitemOpen
  \bibfield  {author} {\bibinfo {author} {\bibfnamefont {A.}~\bibnamefont
  {Arrasmith}}, \bibinfo {author} {\bibfnamefont {M.}~\bibnamefont {Cerezo}},
  \bibinfo {author} {\bibfnamefont {P.}~\bibnamefont {Czarnik}}, \bibinfo
  {author} {\bibfnamefont {L.}~\bibnamefont {Cincio}}, \ and\ \bibinfo {author}
  {\bibfnamefont {P.~J.}\ \bibnamefont {Coles}},\ }\href@noop {} {\enquote
  {\bibinfo {title} {Effect of barren plateaus on gradient-free
  optimization},}\ } (\bibinfo {year} {2020}),\ \Eprint
  {http://arxiv.org/abs/2011.12245} {arXiv:2011.12245 [quant-ph]} \BibitemShut
  {NoStop}%
\bibitem [{\citenamefont {Verdon}\ \emph {et~al.}(2018)\citenamefont {Verdon},
  \citenamefont {Pye},\ and\ \citenamefont {Broughton}}]{verdon2018universal}%
  \BibitemOpen
  \bibfield  {author} {\bibinfo {author} {\bibfnamefont {G.}~\bibnamefont
  {Verdon}}, \bibinfo {author} {\bibfnamefont {J.}~\bibnamefont {Pye}}, \ and\
  \bibinfo {author} {\bibfnamefont {M.}~\bibnamefont {Broughton}},\ }\href
  {https://arxiv.org/abs/1806.09729} {\bibfield  {journal} {\bibinfo  {journal}
  {arXiv preprint arXiv:1806.09729}\ } (\bibinfo {year} {2018})}\BibitemShut
  {NoStop}%
\bibitem [{\citenamefont {Gilyén}\ \emph
  {et~al.}(2019{\natexlab{a}})\citenamefont {Gilyén}, \citenamefont
  {Arunachalam},\ and\ \citenamefont {Wiebe}}]{Gily_n_2019}%
  \BibitemOpen
  \bibfield  {author} {\bibinfo {author} {\bibfnamefont {A.}~\bibnamefont
  {Gilyén}}, \bibinfo {author} {\bibfnamefont {S.}~\bibnamefont
  {Arunachalam}}, \ and\ \bibinfo {author} {\bibfnamefont {N.}~\bibnamefont
  {Wiebe}},\ }\href {\doibase 10.1137/1.9781611975482.87} {\bibfield  {journal}
  {\bibinfo  {journal} {Proceedings of the Thirtieth Annual ACM-SIAM Symposium
  on Discrete Algorithms}\ ,\ \bibinfo {pages} {1425–1444}} (\bibinfo {year}
  {2019}{\natexlab{a}})}\BibitemShut {NoStop}%
\bibitem [{\citenamefont {Zhu}\ \emph {et~al.}(2020)\citenamefont {Zhu},
  \citenamefont {Tang}, \citenamefont {Barron}, \citenamefont {Mayhall},
  \citenamefont {Barnes},\ and\ \citenamefont {Economou}}]{zhu2020adaptive}%
  \BibitemOpen
  \bibfield  {author} {\bibinfo {author} {\bibfnamefont {L.}~\bibnamefont
  {Zhu}}, \bibinfo {author} {\bibfnamefont {H.~L.}\ \bibnamefont {Tang}},
  \bibinfo {author} {\bibfnamefont {G.~S.}\ \bibnamefont {Barron}}, \bibinfo
  {author} {\bibfnamefont {N.~J.}\ \bibnamefont {Mayhall}}, \bibinfo {author}
  {\bibfnamefont {E.}~\bibnamefont {Barnes}}, \ and\ \bibinfo {author}
  {\bibfnamefont {S.~E.}\ \bibnamefont {Economou}},\ }\href@noop {} {\enquote
  {\bibinfo {title} {An adaptive quantum approximate optimization algorithm for
  solving combinatorial problems on a quantum computer},}\ } (\bibinfo {year}
  {2020}),\ \Eprint {http://arxiv.org/abs/2005.10258} {arXiv:2005.10258
  [quant-ph]} \BibitemShut {NoStop}%
\bibitem [{\citenamefont {Wauters}\ \emph {et~al.}(2020)\citenamefont
  {Wauters}, \citenamefont {Panizon}, \citenamefont {Mbeng},\ and\
  \citenamefont {Santoro}}]{PhysRevResearch.2.033446}%
  \BibitemOpen
  \bibfield  {author} {\bibinfo {author} {\bibfnamefont {M.~M.}\ \bibnamefont
  {Wauters}}, \bibinfo {author} {\bibfnamefont {E.}~\bibnamefont {Panizon}},
  \bibinfo {author} {\bibfnamefont {G.~B.}\ \bibnamefont {Mbeng}}, \ and\
  \bibinfo {author} {\bibfnamefont {G.~E.}\ \bibnamefont {Santoro}},\ }\href
  {\doibase 10.1103/PhysRevResearch.2.033446} {\bibfield  {journal} {\bibinfo
  {journal} {Phys. Rev. Research}\ }\textbf {\bibinfo {volume} {2}},\ \bibinfo
  {pages} {033446} (\bibinfo {year} {2020})}\BibitemShut {NoStop}%
\bibitem [{\citenamefont {Yao}\ \emph {et~al.}(2020{\natexlab{a}})\citenamefont
  {Yao}, \citenamefont {Lin},\ and\ \citenamefont
  {Bukov}}]{yao2020reinforcement}%
  \BibitemOpen
  \bibfield  {author} {\bibinfo {author} {\bibfnamefont {J.}~\bibnamefont
  {Yao}}, \bibinfo {author} {\bibfnamefont {L.}~\bibnamefont {Lin}}, \ and\
  \bibinfo {author} {\bibfnamefont {M.}~\bibnamefont {Bukov}},\ }\href@noop {}
  {\enquote {\bibinfo {title} {Reinforcement learning for many-body ground
  state preparation based on counter-diabatic driving},}\ } (\bibinfo {year}
  {2020}{\natexlab{a}}),\ \Eprint {http://arxiv.org/abs/2010.03655}
  {arXiv:2010.03655 [quant-ph]} \BibitemShut {NoStop}%
\bibitem [{\citenamefont {Warren}\ \emph {et~al.}(2020)\citenamefont {Warren},
  \citenamefont {Zhu}, \citenamefont {Tang}, \citenamefont {Najafi},
  \citenamefont {Barnes},\ and\ \citenamefont {Economou}}]{Warren2020RNNVQEAM}%
  \BibitemOpen
  \bibfield  {author} {\bibinfo {author} {\bibfnamefont {A.}~\bibnamefont
  {Warren}}, \bibinfo {author} {\bibfnamefont {L.}~\bibnamefont {Zhu}},
  \bibinfo {author} {\bibfnamefont {H.~L.}\ \bibnamefont {Tang}}, \bibinfo
  {author} {\bibfnamefont {K.}~\bibnamefont {Najafi}}, \bibinfo {author}
  {\bibfnamefont {E.}~\bibnamefont {Barnes}}, \ and\ \bibinfo {author}
  {\bibfnamefont {S.}~\bibnamefont {Economou}},\ }\href@noop {} {\bibfield
  {journal} {\bibinfo  {journal} {Bulletin of the American Physical Society}\ }
  (\bibinfo {year} {2020})}\BibitemShut {NoStop}%
\bibitem [{\citenamefont {Morley}\ \emph {et~al.}(2019)\citenamefont {Morley},
  \citenamefont {Chancellor}, \citenamefont {Bose},\ and\ \citenamefont
  {Kendon}}]{PhysRevA.99.022339}%
  \BibitemOpen
  \bibfield  {author} {\bibinfo {author} {\bibfnamefont {J.~G.}\ \bibnamefont
  {Morley}}, \bibinfo {author} {\bibfnamefont {N.}~\bibnamefont {Chancellor}},
  \bibinfo {author} {\bibfnamefont {S.}~\bibnamefont {Bose}}, \ and\ \bibinfo
  {author} {\bibfnamefont {V.}~\bibnamefont {Kendon}},\ }\href {\doibase
  10.1103/PhysRevA.99.022339} {\bibfield  {journal} {\bibinfo  {journal} {Phys.
  Rev. A}\ }\textbf {\bibinfo {volume} {99}},\ \bibinfo {pages} {022339}
  (\bibinfo {year} {2019})}\BibitemShut {NoStop}%
\bibitem [{\citenamefont {Marsh}\ and\ \citenamefont
  {Wang}(2018)}]{marsh2018quantum}%
  \BibitemOpen
  \bibfield  {author} {\bibinfo {author} {\bibfnamefont {S.}~\bibnamefont
  {Marsh}}\ and\ \bibinfo {author} {\bibfnamefont {J.}~\bibnamefont {Wang}},\
  }\href@noop {} {\enquote {\bibinfo {title} {A quantum walk assisted
  approximate algorithm for bounded np optimisation problems},}\ } (\bibinfo
  {year} {2018}),\ \Eprint {http://arxiv.org/abs/1804.08227} {arXiv:1804.08227
  [quant-ph]} \BibitemShut {NoStop}%
\bibitem [{\citenamefont {Jiang}\ \emph {et~al.}(2017)\citenamefont {Jiang},
  \citenamefont {Rieffel},\ and\ \citenamefont {Wang}}]{Jiang_2017}%
  \BibitemOpen
  \bibfield  {author} {\bibinfo {author} {\bibfnamefont {Z.}~\bibnamefont
  {Jiang}}, \bibinfo {author} {\bibfnamefont {E.~G.}\ \bibnamefont {Rieffel}},
  \ and\ \bibinfo {author} {\bibfnamefont {Z.}~\bibnamefont {Wang}},\ }\href
  {\doibase 10.1103/physreva.95.062317} {\bibfield  {journal} {\bibinfo
  {journal} {Physical Review A}\ }\textbf {\bibinfo {volume} {95}} (\bibinfo
  {year} {2017}),\ 10.1103/physreva.95.062317}\BibitemShut {NoStop}%
\bibitem [{\citenamefont {Mbeng}\ \emph {et~al.}(2019)\citenamefont {Mbeng},
  \citenamefont {Fazio},\ and\ \citenamefont {Santoro}}]{mbeng2019quantum}%
  \BibitemOpen
  \bibfield  {author} {\bibinfo {author} {\bibfnamefont {G.~B.}\ \bibnamefont
  {Mbeng}}, \bibinfo {author} {\bibfnamefont {R.}~\bibnamefont {Fazio}}, \ and\
  \bibinfo {author} {\bibfnamefont {G.}~\bibnamefont {Santoro}},\ }\href@noop
  {} {\enquote {\bibinfo {title} {Quantum annealing: a journey through
  digitalization, control, and hybrid quantum variational schemes},}\ }
  (\bibinfo {year} {2019}),\ \Eprint {http://arxiv.org/abs/1906.08948}
  {arXiv:1906.08948 [quant-ph]} \BibitemShut {NoStop}%
\bibitem [{\citenamefont {Wang}(2017)}]{wang2017quantum}%
  \BibitemOpen
  \bibfield  {author} {\bibinfo {author} {\bibfnamefont {Y.}~\bibnamefont
  {Wang}},\ }\href@noop {} {\enquote {\bibinfo {title} {A quantum walk enhanced
  grover search algorithm for global optimization},}\ } (\bibinfo {year}
  {2017}),\ \Eprint {http://arxiv.org/abs/1711.07825} {arXiv:1711.07825
  [quant-ph]} \BibitemShut {NoStop}%
\bibitem [{\citenamefont {Guéry-Odelin}\ \emph {et~al.}(2019)\citenamefont
  {Guéry-Odelin}, \citenamefont {Ruschhaupt}, \citenamefont {Kiely},
  \citenamefont {Torrontegui}, \citenamefont {Martínez-Garaot},\ and\
  \citenamefont {Muga}}]{Gu_ry_Odelin_2019}%
  \BibitemOpen
  \bibfield  {author} {\bibinfo {author} {\bibfnamefont {D.}~\bibnamefont
  {Guéry-Odelin}}, \bibinfo {author} {\bibfnamefont {A.}~\bibnamefont
  {Ruschhaupt}}, \bibinfo {author} {\bibfnamefont {A.}~\bibnamefont {Kiely}},
  \bibinfo {author} {\bibfnamefont {E.}~\bibnamefont {Torrontegui}}, \bibinfo
  {author} {\bibfnamefont {S.}~\bibnamefont {Martínez-Garaot}}, \ and\
  \bibinfo {author} {\bibfnamefont {J.}~\bibnamefont {Muga}},\ }\href {\doibase
  10.1103/revmodphys.91.045001} {\bibfield  {journal} {\bibinfo  {journal}
  {Reviews of Modern Physics}\ }\textbf {\bibinfo {volume} {91}} (\bibinfo
  {year} {2019}),\ 10.1103/revmodphys.91.045001}\BibitemShut {NoStop}%
\bibitem [{\citenamefont {Hegade}\ \emph {et~al.}(2020)\citenamefont {Hegade},
  \citenamefont {Paul}, \citenamefont {Ding}, \citenamefont {Sanz},
  \citenamefont {Albarrán-Arriagada}, \citenamefont {Solano},\ and\
  \citenamefont {Chen}}]{hegade2020shortcuts}%
  \BibitemOpen
  \bibfield  {author} {\bibinfo {author} {\bibfnamefont {N.~N.}\ \bibnamefont
  {Hegade}}, \bibinfo {author} {\bibfnamefont {K.}~\bibnamefont {Paul}},
  \bibinfo {author} {\bibfnamefont {Y.}~\bibnamefont {Ding}}, \bibinfo {author}
  {\bibfnamefont {M.}~\bibnamefont {Sanz}}, \bibinfo {author} {\bibfnamefont
  {F.}~\bibnamefont {Albarrán-Arriagada}}, \bibinfo {author} {\bibfnamefont
  {E.}~\bibnamefont {Solano}}, \ and\ \bibinfo {author} {\bibfnamefont
  {X.}~\bibnamefont {Chen}},\ }\href@noop {} {\enquote {\bibinfo {title}
  {Shortcuts to adiabaticity in digitized adiabatic quantum computing},}\ }
  (\bibinfo {year} {2020}),\ \Eprint {http://arxiv.org/abs/2009.03539}
  {arXiv:2009.03539 [quant-ph]} \BibitemShut {NoStop}%
\bibitem [{\citenamefont {Whitfield}\ \emph {et~al.}(2010)\citenamefont
  {Whitfield}, \citenamefont {Rodríguez-Rosario},\ and\ \citenamefont
  {Aspuru-Guzik}}]{Whitfield_2010}%
  \BibitemOpen
  \bibfield  {author} {\bibinfo {author} {\bibfnamefont {J.~D.}\ \bibnamefont
  {Whitfield}}, \bibinfo {author} {\bibfnamefont {C.~A.}\ \bibnamefont
  {Rodríguez-Rosario}}, \ and\ \bibinfo {author} {\bibfnamefont
  {A.}~\bibnamefont {Aspuru-Guzik}},\ }\href {\doibase
  10.1103/physreva.81.022323} {\bibfield  {journal} {\bibinfo  {journal}
  {Physical Review A}\ }\textbf {\bibinfo {volume} {81}} (\bibinfo {year}
  {2010}),\ 10.1103/physreva.81.022323}\BibitemShut {NoStop}%
\bibitem [{\citenamefont {Hadfield}\ \emph
  {et~al.}(2019{\natexlab{a}})\citenamefont {Hadfield}, \citenamefont {Wang},
  \citenamefont {O'Gorman}, \citenamefont {Rieffel}, \citenamefont
  {Venturelli},\ and\ \citenamefont {Biswas}}]{nasaQAOA2019}%
  \BibitemOpen
  \bibfield  {author} {\bibinfo {author} {\bibfnamefont {S.}~\bibnamefont
  {Hadfield}}, \bibinfo {author} {\bibfnamefont {Z.}~\bibnamefont {Wang}},
  \bibinfo {author} {\bibfnamefont {B.}~\bibnamefont {O'Gorman}}, \bibinfo
  {author} {\bibfnamefont {E.~G.}\ \bibnamefont {Rieffel}}, \bibinfo {author}
  {\bibfnamefont {D.}~\bibnamefont {Venturelli}}, \ and\ \bibinfo {author}
  {\bibfnamefont {R.}~\bibnamefont {Biswas}},\ }\href {\doibase
  10.3390/a12020034} {\bibfield  {journal} {\bibinfo  {journal} {Algorithms}\
  }\textbf {\bibinfo {volume} {12}},\ \bibinfo {pages} {34} (\bibinfo {year}
  {2019}{\natexlab{a}})}\BibitemShut {NoStop}%
\bibitem [{\citenamefont {{Bulger}}(2005)}]{2005quant.ph..7193B}%
  \BibitemOpen
  \bibfield  {author} {\bibinfo {author} {\bibfnamefont {D.}~\bibnamefont
  {{Bulger}}},\ }\href@noop {} {\bibfield  {journal} {\bibinfo  {journal}
  {arXiv e-prints}\ ,\ \bibinfo {eid} {quant-ph/0507193}} (\bibinfo {year}
  {2005})},\ \Eprint {http://arxiv.org/abs/quant-ph/0507193}
  {arXiv:quant-ph/0507193 [quant-ph]} \BibitemShut {NoStop}%
\bibitem [{\citenamefont {Streif}\ and\ \citenamefont
  {Leib}(2019)}]{streif2019comparison}%
  \BibitemOpen
  \bibfield  {author} {\bibinfo {author} {\bibfnamefont {M.}~\bibnamefont
  {Streif}}\ and\ \bibinfo {author} {\bibfnamefont {M.}~\bibnamefont {Leib}},\
  }\href@noop {} {\enquote {\bibinfo {title} {Comparison of qaoa with quantum
  and simulated annealing},}\ } (\bibinfo {year} {2019}),\ \Eprint
  {http://arxiv.org/abs/1901.01903} {arXiv:1901.01903 [quant-ph]} \BibitemShut
  {NoStop}%
\bibitem [{\citenamefont {Niu}\ \emph {et~al.}(2019)\citenamefont {Niu},
  \citenamefont {Lu},\ and\ \citenamefont {Chuang}}]{niu2019optimizing}%
  \BibitemOpen
  \bibfield  {author} {\bibinfo {author} {\bibfnamefont {M.~Y.}\ \bibnamefont
  {Niu}}, \bibinfo {author} {\bibfnamefont {S.}~\bibnamefont {Lu}}, \ and\
  \bibinfo {author} {\bibfnamefont {I.~L.}\ \bibnamefont {Chuang}},\
  }\href@noop {} {\enquote {\bibinfo {title} {Optimizing qaoa: Success
  probability and runtime dependence on circuit depth},}\ } (\bibinfo {year}
  {2019}),\ \Eprint {http://arxiv.org/abs/1905.12134} {arXiv:1905.12134
  [quant-ph]} \BibitemShut {NoStop}%
\bibitem [{\citenamefont {Barkoutsos}\ \emph {et~al.}(2020)\citenamefont
  {Barkoutsos}, \citenamefont {Nannicini}, \citenamefont {Robert},
  \citenamefont {Tavernelli},\ and\ \citenamefont {Woerner}}]{Barkoutsos_2020}%
  \BibitemOpen
  \bibfield  {author} {\bibinfo {author} {\bibfnamefont {P.~K.}\ \bibnamefont
  {Barkoutsos}}, \bibinfo {author} {\bibfnamefont {G.}~\bibnamefont
  {Nannicini}}, \bibinfo {author} {\bibfnamefont {A.}~\bibnamefont {Robert}},
  \bibinfo {author} {\bibfnamefont {I.}~\bibnamefont {Tavernelli}}, \ and\
  \bibinfo {author} {\bibfnamefont {S.}~\bibnamefont {Woerner}},\ }\href
  {\doibase 10.22331/q-2020-04-20-256} {\bibfield  {journal} {\bibinfo
  {journal} {Quantum}\ }\textbf {\bibinfo {volume} {4}},\ \bibinfo {pages}
  {256} (\bibinfo {year} {2020})}\BibitemShut {NoStop}%
\bibitem [{\citenamefont {Morales}\ \emph {et~al.}(2019)\citenamefont
  {Morales}, \citenamefont {Biamonte},\ and\ \citenamefont
  {Zimborás}}]{morales2019universality}%
  \BibitemOpen
  \bibfield  {author} {\bibinfo {author} {\bibfnamefont {M.~E.~S.}\
  \bibnamefont {Morales}}, \bibinfo {author} {\bibfnamefont {J.}~\bibnamefont
  {Biamonte}}, \ and\ \bibinfo {author} {\bibfnamefont {Z.}~\bibnamefont
  {Zimborás}},\ }\href@noop {} {\enquote {\bibinfo {title} {On the
  universality of the quantum approximate optimization algorithm},}\ }
  (\bibinfo {year} {2019}),\ \Eprint {http://arxiv.org/abs/1909.03123}
  {arXiv:1909.03123 [quant-ph]} \BibitemShut {NoStop}%
\bibitem [{\citenamefont {Lloyd}(2018)}]{lloyd2018quantum}%
  \BibitemOpen
  \bibfield  {author} {\bibinfo {author} {\bibfnamefont {S.}~\bibnamefont
  {Lloyd}},\ }\href@noop {} {\enquote {\bibinfo {title} {Quantum approximate
  optimization is computationally universal},}\ } (\bibinfo {year} {2018}),\
  \Eprint {http://arxiv.org/abs/1812.11075} {arXiv:1812.11075 [quant-ph]}
  \BibitemShut {NoStop}%
\bibitem [{\citenamefont {Bravyi}\ \emph {et~al.}(2019)\citenamefont {Bravyi},
  \citenamefont {Kliesch}, \citenamefont {Koenig},\ and\ \citenamefont
  {Tang}}]{bravyi2019obstacles}%
  \BibitemOpen
  \bibfield  {author} {\bibinfo {author} {\bibfnamefont {S.}~\bibnamefont
  {Bravyi}}, \bibinfo {author} {\bibfnamefont {A.}~\bibnamefont {Kliesch}},
  \bibinfo {author} {\bibfnamefont {R.}~\bibnamefont {Koenig}}, \ and\ \bibinfo
  {author} {\bibfnamefont {E.}~\bibnamefont {Tang}},\ }\href@noop {} {\enquote
  {\bibinfo {title} {Obstacles to state preparation and variational
  optimization from symmetry protection},}\ } (\bibinfo {year} {2019}),\
  \Eprint {http://arxiv.org/abs/1910.08980} {arXiv:1910.08980 [quant-ph]}
  \BibitemShut {NoStop}%
\bibitem [{\citenamefont {Farhi}\ \emph {et~al.}(2017)\citenamefont {Farhi},
  \citenamefont {Goldstone}, \citenamefont {Gutmann},\ and\ \citenamefont
  {Neven}}]{farhi2017quantum}%
  \BibitemOpen
  \bibfield  {author} {\bibinfo {author} {\bibfnamefont {E.}~\bibnamefont
  {Farhi}}, \bibinfo {author} {\bibfnamefont {J.}~\bibnamefont {Goldstone}},
  \bibinfo {author} {\bibfnamefont {S.}~\bibnamefont {Gutmann}}, \ and\
  \bibinfo {author} {\bibfnamefont {H.}~\bibnamefont {Neven}},\ }\href@noop {}
  {\enquote {\bibinfo {title} {Quantum algorithms for fixed qubit
  architectures},}\ } (\bibinfo {year} {2017}),\ \Eprint
  {http://arxiv.org/abs/1703.06199} {arXiv:1703.06199 [quant-ph]} \BibitemShut
  {NoStop}%
\bibitem [{\citenamefont {Hadfield}\ \emph
  {et~al.}(2019{\natexlab{b}})\citenamefont {Hadfield}, \citenamefont {Wang},
  \citenamefont {O'Gorman}, \citenamefont {Rieffel}, \citenamefont
  {Venturelli},\ and\ \citenamefont {Biswas}}]{hadfield2019quantum}%
  \BibitemOpen
  \bibfield  {author} {\bibinfo {author} {\bibfnamefont {S.}~\bibnamefont
  {Hadfield}}, \bibinfo {author} {\bibfnamefont {Z.}~\bibnamefont {Wang}},
  \bibinfo {author} {\bibfnamefont {B.}~\bibnamefont {O'Gorman}}, \bibinfo
  {author} {\bibfnamefont {E.~G.}\ \bibnamefont {Rieffel}}, \bibinfo {author}
  {\bibfnamefont {D.}~\bibnamefont {Venturelli}}, \ and\ \bibinfo {author}
  {\bibfnamefont {R.}~\bibnamefont {Biswas}},\ }\href
  {https://www.mdpi.com/1999-4893/12/2/34} {\bibfield  {journal} {\bibinfo
  {journal} {Algorithms}\ }\textbf {\bibinfo {volume} {12}},\ \bibinfo {pages}
  {34} (\bibinfo {year} {2019}{\natexlab{b}})}\BibitemShut {NoStop}%
\bibitem [{\citenamefont {Verdon}\ \emph
  {et~al.}(2019{\natexlab{b}})\citenamefont {Verdon}, \citenamefont {Arrazola},
  \citenamefont {Brádler},\ and\ \citenamefont
  {Killoran}}]{verdon2019quantum3}%
  \BibitemOpen
  \bibfield  {author} {\bibinfo {author} {\bibfnamefont {G.}~\bibnamefont
  {Verdon}}, \bibinfo {author} {\bibfnamefont {J.~M.}\ \bibnamefont
  {Arrazola}}, \bibinfo {author} {\bibfnamefont {K.}~\bibnamefont {Brádler}},
  \ and\ \bibinfo {author} {\bibfnamefont {N.}~\bibnamefont {Killoran}},\
  }\href@noop {} {\enquote {\bibinfo {title} {A quantum approximate
  optimization algorithm for continuous problems},}\ } (\bibinfo {year}
  {2019}{\natexlab{b}}),\ \Eprint {http://arxiv.org/abs/1902.00409}
  {arXiv:1902.00409 [quant-ph]} \BibitemShut {NoStop}%
\bibitem [{\citenamefont {{Durr}}\ and\ \citenamefont
  {{Hoyer}}(1996)}]{1996quant.ph..7014D}%
  \BibitemOpen
  \bibfield  {author} {\bibinfo {author} {\bibfnamefont {C.}~\bibnamefont
  {{Durr}}}\ and\ \bibinfo {author} {\bibfnamefont {P.}~\bibnamefont
  {{Hoyer}}},\ }\href@noop {} {\bibfield  {journal} {\bibinfo  {journal} {arXiv
  e-prints}\ ,\ \bibinfo {eid} {quant-ph/9607014}} (\bibinfo {year} {1996})},\
  \Eprint {http://arxiv.org/abs/quant-ph/9607014} {arXiv:quant-ph/9607014
  [quant-ph]} \BibitemShut {NoStop}%
\bibitem [{\citenamefont {Baritompa}\ \emph {et~al.}(2005)\citenamefont
  {Baritompa}, \citenamefont {Bulger},\ and\ \citenamefont
  {Wood}}]{Baritompa2005}%
  \BibitemOpen
  \bibfield  {author} {\bibinfo {author} {\bibfnamefont {W.~P.}\ \bibnamefont
  {Baritompa}}, \bibinfo {author} {\bibfnamefont {D.~W.}\ \bibnamefont
  {Bulger}}, \ and\ \bibinfo {author} {\bibfnamefont {G.~R.}\ \bibnamefont
  {Wood}},\ }\href {\doibase 10.1137/040605072} {\bibfield  {journal} {\bibinfo
   {journal} {{SIAM} Journal on Optimization}\ }\textbf {\bibinfo {volume}
  {15}},\ \bibinfo {pages} {1170} (\bibinfo {year} {2005})}\BibitemShut
  {NoStop}%
\bibitem [{\citenamefont {Liao}\ \emph {et~al.}(2020)\citenamefont {Liao},
  \citenamefont {Ebler}, \citenamefont {Liu},\ and\ \citenamefont
  {Dahlsten}}]{10.1088/1367-2630/abc9ef}%
  \BibitemOpen
  \bibfield  {author} {\bibinfo {author} {\bibfnamefont {Y.}~\bibnamefont
  {Liao}}, \bibinfo {author} {\bibfnamefont {D.}~\bibnamefont {Ebler}},
  \bibinfo {author} {\bibfnamefont {F.}~\bibnamefont {Liu}}, \ and\ \bibinfo
  {author} {\bibfnamefont {O.}~\bibnamefont {Dahlsten}},\ }\href
  {http://iopscience.iop.org/article/10.1088/1367-2630/abc9ef} {\bibfield
  {journal} {\bibinfo  {journal} {New Journal of Physics}\ } (\bibinfo {year}
  {2020})}\BibitemShut {NoStop}%
\bibitem [{\citenamefont {Gilliam}\ \emph {et~al.}(2020)\citenamefont
  {Gilliam}, \citenamefont {Woerner},\ and\ \citenamefont
  {Gonciulea}}]{gilliam2020grover}%
  \BibitemOpen
  \bibfield  {author} {\bibinfo {author} {\bibfnamefont {A.}~\bibnamefont
  {Gilliam}}, \bibinfo {author} {\bibfnamefont {S.}~\bibnamefont {Woerner}}, \
  and\ \bibinfo {author} {\bibfnamefont {C.}~\bibnamefont {Gonciulea}},\
  }\href@noop {} {\enquote {\bibinfo {title} {Grover adaptive search for
  constrained polynomial binary optimization},}\ } (\bibinfo {year} {2020}),\
  \Eprint {http://arxiv.org/abs/1912.04088} {arXiv:1912.04088 [quant-ph]}
  \BibitemShut {NoStop}%
\bibitem [{\citenamefont {Buhrman}\ \emph {et~al.}(2001)\citenamefont
  {Buhrman}, \citenamefont {Cleve}, \citenamefont {Watrous},\ and\
  \citenamefont {de~Wolf}}]{Buhrman_2001}%
  \BibitemOpen
  \bibfield  {author} {\bibinfo {author} {\bibfnamefont {H.}~\bibnamefont
  {Buhrman}}, \bibinfo {author} {\bibfnamefont {R.}~\bibnamefont {Cleve}},
  \bibinfo {author} {\bibfnamefont {J.}~\bibnamefont {Watrous}}, \ and\
  \bibinfo {author} {\bibfnamefont {R.}~\bibnamefont {de~Wolf}},\ }\href
  {\doibase 10.1103/physrevlett.87.167902} {\bibfield  {journal} {\bibinfo
  {journal} {Physical Review Letters}\ }\textbf {\bibinfo {volume} {87}}
  (\bibinfo {year} {2001}),\ 10.1103/physrevlett.87.167902}\BibitemShut
  {NoStop}%
\bibitem [{\citenamefont {{Brassard}}\ \emph {et~al.}(2000)\citenamefont
  {{Brassard}}, \citenamefont {{Hoyer}}, \citenamefont {{Mosca}},\ and\
  \citenamefont {{Tapp}}}]{2000quant.ph..5055B}%
  \BibitemOpen
  \bibfield  {author} {\bibinfo {author} {\bibfnamefont {G.}~\bibnamefont
  {{Brassard}}}, \bibinfo {author} {\bibfnamefont {P.}~\bibnamefont {{Hoyer}}},
  \bibinfo {author} {\bibfnamefont {M.}~\bibnamefont {{Mosca}}}, \ and\
  \bibinfo {author} {\bibfnamefont {A.}~\bibnamefont {{Tapp}}},\ }\href@noop {}
  {\bibfield  {journal} {\bibinfo  {journal} {arXiv e-prints}\ ,\ \bibinfo
  {eid} {quant-ph/0005055}} (\bibinfo {year} {2000})},\ \Eprint
  {http://arxiv.org/abs/quant-ph/0005055} {arXiv:quant-ph/0005055 [quant-ph]}
  \BibitemShut {NoStop}%
\bibitem [{\citenamefont {McClean}\ \emph {et~al.}(2020)\citenamefont
  {McClean}, \citenamefont {Harrigan}, \citenamefont {Mohseni}, \citenamefont
  {Rubin}, \citenamefont {Jiang}, \citenamefont {Boixo}, \citenamefont
  {Smelyanskiy}, \citenamefont {Babbush},\ and\ \citenamefont
  {Neven}}]{mcclean2020low}%
  \BibitemOpen
  \bibfield  {author} {\bibinfo {author} {\bibfnamefont {J.~R.}\ \bibnamefont
  {McClean}}, \bibinfo {author} {\bibfnamefont {M.~P.}\ \bibnamefont
  {Harrigan}}, \bibinfo {author} {\bibfnamefont {M.}~\bibnamefont {Mohseni}},
  \bibinfo {author} {\bibfnamefont {N.~C.}\ \bibnamefont {Rubin}}, \bibinfo
  {author} {\bibfnamefont {Z.}~\bibnamefont {Jiang}}, \bibinfo {author}
  {\bibfnamefont {S.}~\bibnamefont {Boixo}}, \bibinfo {author} {\bibfnamefont
  {V.~N.}\ \bibnamefont {Smelyanskiy}}, \bibinfo {author} {\bibfnamefont
  {R.}~\bibnamefont {Babbush}}, \ and\ \bibinfo {author} {\bibfnamefont
  {H.}~\bibnamefont {Neven}},\ }\href@noop {} {\bibfield  {journal} {\bibinfo
  {journal} {arXiv preprint arXiv:2008.08615}\ } (\bibinfo {year}
  {2020})}\BibitemShut {NoStop}%
\bibitem [{\citenamefont {Berry}\ \emph {et~al.}(2015)\citenamefont {Berry},
  \citenamefont {Childs}, \citenamefont {Cleve}, \citenamefont {Kothari},\ and\
  \citenamefont {Somma}}]{PhysRevLett.114.090502}%
  \BibitemOpen
  \bibfield  {author} {\bibinfo {author} {\bibfnamefont {D.~W.}\ \bibnamefont
  {Berry}}, \bibinfo {author} {\bibfnamefont {A.~M.}\ \bibnamefont {Childs}},
  \bibinfo {author} {\bibfnamefont {R.}~\bibnamefont {Cleve}}, \bibinfo
  {author} {\bibfnamefont {R.}~\bibnamefont {Kothari}}, \ and\ \bibinfo
  {author} {\bibfnamefont {R.~D.}\ \bibnamefont {Somma}},\ }\href {\doibase
  10.1103/PhysRevLett.114.090502} {\bibfield  {journal} {\bibinfo  {journal}
  {Phys. Rev. Lett.}\ }\textbf {\bibinfo {volume} {114}},\ \bibinfo {pages}
  {090502} (\bibinfo {year} {2015})}\BibitemShut {NoStop}%
\bibitem [{Note1()}]{Note1}%
  \BibitemOpen
  \bibinfo {note} {Note that here our definition of $G^*$ is slightly different
  from the $G_U$ in Ref.~\cite {Gily_n_2019}: $C_1$ and $C_2$ being negative to
  their counterpart in the definition of $G_U$. However the two negative sign
  cancel, therefore we have $G^*=G_U$.}\BibitemShut {Stop}%
\bibitem [{\citenamefont {Gilyén}\ \emph
  {et~al.}(2019{\natexlab{b}})\citenamefont {Gilyén}, \citenamefont {Su},
  \citenamefont {Low},\ and\ \citenamefont {Wiebe}}]{Gily_n_20192}%
  \BibitemOpen
  \bibfield  {author} {\bibinfo {author} {\bibfnamefont {A.}~\bibnamefont
  {Gilyén}}, \bibinfo {author} {\bibfnamefont {Y.}~\bibnamefont {Su}},
  \bibinfo {author} {\bibfnamefont {G.~H.}\ \bibnamefont {Low}}, \ and\
  \bibinfo {author} {\bibfnamefont {N.}~\bibnamefont {Wiebe}},\ }\href
  {\doibase 10.1145/3313276.3316366} {\bibfield  {journal} {\bibinfo  {journal}
  {Proceedings of the 51st Annual ACM SIGACT Symposium on Theory of Computing -
  STOC 2019}\ } (\bibinfo {year} {2019}{\natexlab{b}}),\
  10.1145/3313276.3316366}\BibitemShut {NoStop}%
\bibitem [{\citenamefont {Weedbrook}\ \emph {et~al.}(2012)\citenamefont
  {Weedbrook}, \citenamefont {Pirandola}, \citenamefont {Garc\'{\i}a-Patr\'on},
  \citenamefont {Cerf}, \citenamefont {Ralph}, \citenamefont {Shapiro},\ and\
  \citenamefont {Lloyd}}]{RevModPhys.84.621}%
  \BibitemOpen
  \bibfield  {author} {\bibinfo {author} {\bibfnamefont {C.}~\bibnamefont
  {Weedbrook}}, \bibinfo {author} {\bibfnamefont {S.}~\bibnamefont
  {Pirandola}}, \bibinfo {author} {\bibfnamefont {R.}~\bibnamefont
  {Garc\'{\i}a-Patr\'on}}, \bibinfo {author} {\bibfnamefont {N.~J.}\
  \bibnamefont {Cerf}}, \bibinfo {author} {\bibfnamefont {T.~C.}\ \bibnamefont
  {Ralph}}, \bibinfo {author} {\bibfnamefont {J.~H.}\ \bibnamefont {Shapiro}},
  \ and\ \bibinfo {author} {\bibfnamefont {S.}~\bibnamefont {Lloyd}},\ }\href
  {\doibase 10.1103/RevModPhys.84.621} {\bibfield  {journal} {\bibinfo
  {journal} {Rev. Mod. Phys.}\ }\textbf {\bibinfo {volume} {84}},\ \bibinfo
  {pages} {621} (\bibinfo {year} {2012})}\BibitemShut {NoStop}%
\bibitem [{\citenamefont {Bartlett}\ \emph {et~al.}(2002)\citenamefont
  {Bartlett}, \citenamefont {Sanders}, \citenamefont {Braunstein},\ and\
  \citenamefont {Nemoto}}]{PhysRevLett.88.097904}%
  \BibitemOpen
  \bibfield  {author} {\bibinfo {author} {\bibfnamefont {S.~D.}\ \bibnamefont
  {Bartlett}}, \bibinfo {author} {\bibfnamefont {B.~C.}\ \bibnamefont
  {Sanders}}, \bibinfo {author} {\bibfnamefont {S.~L.}\ \bibnamefont
  {Braunstein}}, \ and\ \bibinfo {author} {\bibfnamefont {K.}~\bibnamefont
  {Nemoto}},\ }\href {\doibase 10.1103/PhysRevLett.88.097904} {\bibfield
  {journal} {\bibinfo  {journal} {Phys. Rev. Lett.}\ }\textbf {\bibinfo
  {volume} {88}},\ \bibinfo {pages} {097904} (\bibinfo {year}
  {2002})}\BibitemShut {NoStop}%
\bibitem [{\citenamefont {Khairy}\ \emph {et~al.}(2019)\citenamefont {Khairy},
  \citenamefont {Shaydulin}, \citenamefont {Cincio}, \citenamefont {Alexeev},\
  and\ \citenamefont {Balaprakash}}]{khairy2019reinforcementlearningbased}%
  \BibitemOpen
  \bibfield  {author} {\bibinfo {author} {\bibfnamefont {S.}~\bibnamefont
  {Khairy}}, \bibinfo {author} {\bibfnamefont {R.}~\bibnamefont {Shaydulin}},
  \bibinfo {author} {\bibfnamefont {L.}~\bibnamefont {Cincio}}, \bibinfo
  {author} {\bibfnamefont {Y.}~\bibnamefont {Alexeev}}, \ and\ \bibinfo
  {author} {\bibfnamefont {P.}~\bibnamefont {Balaprakash}},\ }\href@noop {}
  {\enquote {\bibinfo {title} {Reinforcement-learning-based variational quantum
  circuits optimization for combinatorial problems},}\ } (\bibinfo {year}
  {2019}),\ \Eprint {http://arxiv.org/abs/1911.04574} {arXiv:1911.04574
  [cs.LG]} \BibitemShut {NoStop}%
\bibitem [{\citenamefont {Yao}\ \emph {et~al.}(2020{\natexlab{b}})\citenamefont
  {Yao}, \citenamefont {Bukov},\ and\ \citenamefont {Lin}}]{yao2020policy}%
  \BibitemOpen
  \bibfield  {author} {\bibinfo {author} {\bibfnamefont {J.}~\bibnamefont
  {Yao}}, \bibinfo {author} {\bibfnamefont {M.}~\bibnamefont {Bukov}}, \ and\
  \bibinfo {author} {\bibfnamefont {L.}~\bibnamefont {Lin}},\ }\href@noop {}
  {\enquote {\bibinfo {title} {Policy gradient based quantum approximate
  optimization algorithm},}\ } (\bibinfo {year} {2020}{\natexlab{b}}),\ \Eprint
  {http://arxiv.org/abs/2002.01068} {arXiv:2002.01068 [quant-ph]} \BibitemShut
  {NoStop}%
\bibitem [{\citenamefont {Childs}\ and\ \citenamefont
  {Wiebe}(2012)}]{childs2012hamiltonian}%
  \BibitemOpen
  \bibfield  {author} {\bibinfo {author} {\bibfnamefont {A.~M.}\ \bibnamefont
  {Childs}}\ and\ \bibinfo {author} {\bibfnamefont {N.}~\bibnamefont {Wiebe}},\
  }\href@noop {} {\enquote {\bibinfo {title} {Hamiltonian simulation using
  linear combinations of unitary operations},}\ } (\bibinfo {year} {2012}),\
  \Eprint {http://arxiv.org/abs/1202.5822} {arXiv:1202.5822 [quant-ph]}
  \BibitemShut {NoStop}%
\bibitem [{Note2()}]{Note2}%
  \BibitemOpen
  \bibinfo {note} {Note that for the training data point with $y_i=0$, $C_1$ in
  the Grover Operator $G$ and $G^*$ has to be adjusted to $-Z$}\BibitemShut
  {NoStop}%
\bibitem [{\citenamefont {Knill}\ \emph {et~al.}(2007)\citenamefont {Knill},
  \citenamefont {Ortiz},\ and\ \citenamefont {Somma}}]{PhysRevA.75.012328}%
  \BibitemOpen
  \bibfield  {author} {\bibinfo {author} {\bibfnamefont {E.}~\bibnamefont
  {Knill}}, \bibinfo {author} {\bibfnamefont {G.}~\bibnamefont {Ortiz}}, \ and\
  \bibinfo {author} {\bibfnamefont {R.~D.}\ \bibnamefont {Somma}},\ }\href
  {\doibase 10.1103/PhysRevA.75.012328} {\bibfield  {journal} {\bibinfo
  {journal} {Phys. Rev. A}\ }\textbf {\bibinfo {volume} {75}},\ \bibinfo
  {pages} {012328} (\bibinfo {year} {2007})}\BibitemShut {NoStop}%
\end{thebibliography}%


%

\appendix

\section{Proofs} \label{appendix}
\label{app1}
Here we provide proofs for the claims mentioned in the main text.\newline

$$U=\sum_{j}\left|j \left>\right<j\right|\otimes U_{j}$$

\begin{proof}

\begin{align*}
\label{ue}
U=[H\otimes I\otimes I\otimes I]\cdot
[\ket{0}\bra{0}\otimes (\sum_{j}\left|j \left>\right<j\right|\otimes P_{j}\otimes T )+\ket{1}\bra{1}\otimes (\sum_{j}\left|j \left>\right<j\right|\otimes T\otimes P_{j} )]\cdot[H\otimes I\otimes I\otimes I]
&=\\
\sum_{j}[H\otimes I\otimes I\otimes I]\cdot[\ket{0}\bra{0}\otimes \left|j \left>\right<j\right|\otimes P_{j}\otimes T+\ket{1}\bra{1}\otimes \left|j \left>\right<j\right|\otimes T\otimes P_{j}]\cdot[H\otimes I\otimes I\otimes I]&=\\
\sum_{j}[I\otimes H\otimes I\otimes I]\cdot[\left|j \left>\right<j\right|\otimes \ket{0}\bra{0}\otimes P_{j}\otimes T+\left|j \left>\right<j\right|\otimes \ket{1}\bra{1}\otimes  T\otimes P_{j}]\cdot[I\otimes H\otimes I\otimes I]&=\\
\sum_{j}[I\otimes H\otimes I\otimes I]\cdot[\left|j \left>\right<j\right|\otimes (\ket{0}\bra{0}\otimes P_{j}\otimes  T+ \ket{1}\bra{1}\otimes  T\otimes P_{j})]\cdot[I\otimes H\otimes I\otimes I]&=\\
\sum_{j}(I \cdot \left|j \left>\right<j\right|\cdot I) \otimes \left([H\otimes I\otimes I]\cdot[\ket{0}\bra{0}\otimes P_{j}\otimes T+\ket{1}\bra{1}\otimes T\otimes P_{j}]\cdot[H\otimes I\otimes I]\right)  &=\\
\sum_{j}\left|j \left>\right<j\right|\otimes U_{j}
\end{align*}

\end{proof}

$$G=\sum_{j}\left|j \left>\right<j\right|\otimes G_{j}$$

\begin{proof}
\begin{align}
G=C_{1}U^{-1}C_{2}U&=\\
C_{1}\left(\sum_{j}\left|j \left>\right<j\right|\otimes U_{j}^{\dagger}\right)C_{2}\left(\sum_{k}\left|k \left>\right<k\right|\otimes U_{k}\right)&=\\
 \sum_{j}\sum_{k}C_{1}\left(\left|j \left>\right<j\right|\otimes U_{j}^{\dagger}\right)C_{2}\left(\left|k \left>\right<k\right|\otimes U_{k}\right)&=\\
 \sum_{j}\left|j \left>\right<j\right|\otimes C_{1}U_{j}^{\dagger}C_{2} U_{j}&=\\\sum_{j}\left|j \left>\right<j\right|\otimes G_{j}
\end{align}
\end{proof}

\section{Performance of quantum training using Grover Adaptive Search} \label{app2}

It has been shown in Ref.~\cite{Baritompa2005} that Global Optimization by Grover Adaptive Search takes $O(\sqrt{N/s})$ calls of Grover Oracle in which $N$ is the dimension of the search space, $s$ is the number of global optima and assuming $s$ is small compared to $N$. A unique optimum will be found after $O(\log{N})$ improvements on the threshold, in expectation. The number of measurements invoked between the improvements in no larger than  $O(\log{\sqrt{N}})$. Therefore the total number of measurements for Grover Adaptive Search to find a global optimum is no larger than $O(\log{N}\log{\sqrt{N}})$.\newline

Here we apply the above results to our QNN training problem. Taking training VQE as an example, we evaluate the number of “controlled-QNN” runs, the number of QNN runs, and the number of measurements respectively as follow.

\subsection{Number of “controlled-QNN” runs}
For our quantum training, each Grover iteration consist of the steps of Amplitude Encoding(AE), Phase estimation(PE), Threshold Oracle, Uncomputation. Taking training VQE as an example, the number of the number of “controlled-QNN” runs of each step can be listed as:\newline
\begin{itemize}
    \item \textit{In Amplitude Encoding(AE):}   $n_{AE}=1$
    \item \textit{In Phase estimation(PE): } $n_{PE}=(2^t-1)2=2^{t+1}-2$
    \item \textit{In Threshold Oracle:} $0$
    \item \textit{In Uncomputation:}   $n_{PE}+n_{AE}$
\end{itemize}

in which $t$ is the number of qubits in the amplitude register for the phase estimation.\newline

The number of “controlled-QNN” runs of each Grover iteration , which we denote as $N_0$, is the sum of the above numbers:
\begin{align}
    N_0=2(n_{AE}+n_{PE})=2(2^{t+1}-1)
\end{align}

To obtain phase accurate to $n'$ bits with probability of success at least $1-\epsilon_1$, $t$ is chosen as
\begin{align}
    t=n'+\lceil  \log (2+\frac{1}{2\epsilon_1})\rceil.
\end{align}

Hence:
\begin{align}
    N_0\approx
    (2^{n'+2}(2+\frac{1}{2\epsilon_1})-2)
\end{align}

For small $\epsilon_1$ we have:
\begin{align}
N_0\approx2^{n'+2}\frac{1}{2\epsilon_1}
\end{align}

$n'$ determines the precision of the QNN cost function evaluated by phase estimation, which we denote as $\epsilon_2$, and we have:
\begin{align}
2^{-n'}=\epsilon_2
\end{align}

therefore:
\begin{align}
N_0\approx\frac{2}{\epsilon_2\epsilon_1}
\end{align}

Therefore the total number of “controlled-QNN” runs for $O(\sqrt{N/s})$ Grover iterations scales as
\begin{equation}\label{nQNN}
N_{\textsf{\textit{controlled-QNN}}}\sim O(\frac{1}{\epsilon_2\epsilon_1}\sqrt{N/s})
\end{equation}

in which $N$ is the dimension of the parameter space of QNN and $s$ is the number of global optima of the QNN cost function. \newline

Let $r$ be the number of parameters(rotation angles)in QNN, $d$ be the number of control qubits for each rotation angle. Therefore
\begin{align}
\label{tQNN}
N=2^{dr}
\end{align}

On the other hand,
\begin{equation}
\label{eee}
2^{-d}=\epsilon_0,
\end{equation}
where $\epsilon_0$ is the precision of each angle value. Hence
\begin{align}
\label{nn}
N= {(\frac{1}{\epsilon_0})^{r}}
\end{align}

Inserting Eq.~\ref{nn} into \ref{nQNN} we get

\begin{equation}
\label{ccqnn}
N_{\textsf{\textit{controlled-QNN}}}\sim O\left(\frac{1}{\epsilon_2\epsilon_1}{\left(\frac{1}{{\epsilon_0}}\right)}^{r/2}s^{-\frac{1}{2}}\right)
\end{equation}

\subsection{Number of Measurements}

As mentioned before, the total number of measurements for Grover Adaptive Search to find a global optimum scales as
\begin{equation}
\label{measuren}
N_{\textsf{\textit{Measurements}}}\sim O(\log{N}\log{\sqrt{N}})
\end{equation}

Inserting Eq.~\ref{tQNN} into \ref{measuren} we get

\begin{equation}
\label{nnmea}
N_{\textsf{\textit{Measurements}}}\sim O\left((dr)^{3/2}\right)
\end{equation}

From Eq.~\ref{eee} we have $d\sim O\left(\log\left(\frac{1}{\epsilon_0}\right)\right)$, therefore
\begin{equation}
\label{nmeasurement}
N_{\textsf{\textit{Measurements}}}\sim O\left(\left(r\log\left(\frac{1}{\epsilon_0}\right)\right)^{1.5}\right)
\end{equation}

\subsection{Number of QNN runs}

After each measurement on the parameter register, we obtain a specific parameter configuration of QNN. We then need to estimate the cost function for this particular parameter configuration. For VQE, the cost function is the expectation value of some Hamiltonian and the number of the estimation sacle as $O(1/\epsilon^\alpha)$ for some small power $\alpha$ \cite{PhysRevA.75.012328}($\alpha$ is a small integer about 1 or 2), where $\epsilon$ is the desired accuracy of the expectation value. For our QNN training we choose the accuracy $\epsilon$ to be $\epsilon_2$ defined above. Taking $\alpha=1$, the number of QNN runs after each measurements scale as $O(1/\epsilon_2)$ and the total number of QNN runs all the measurements during the quantum training scale as
\begin{equation}
\label{nnn}
N_{\textsf{\textit{QNN}}}\sim O\left(\frac{1}{\epsilon_2}\right)N_{\textsf{\textit{Measurements}}}.
\end{equation}

Inserting \ref{nmeasurement} into \ref{nnn} we have
\begin{equation}
\label{nqqqqq}
N_{\textsf{\textit{QNN}}}\sim O\left(\frac{1}{\epsilon_2} \left(r\log\left(\frac{1}{\epsilon_0}\right)\right)^{1.5}\right)
\end{equation}

\section{Number of qubits needed for Quantum training by AC-QAOA}\label{number}

Taking training VQE as example, the number of qubits in each register can be listed as:\newline
\begin{itemize}
    \item \textit{For QNN register}: $n$ qubits
    \item \textit{For Parameter register}: $dr$ qubits ($r$ is the number of parameters in QNN, $d$ is the number of control qubits for each rotation angle.)
    \item \textit{For Hadamard test}: $1$ ancilla qubit
    \item \textit{For LCU register and other registers}:  $O(\log\log(1/\epsilon))$ \cite{Gily_n_2019} qubits ($\epsilon$ is the precision of implememnting the phase oracle by LCU)

\end{itemize}

In total, the number of qubits needed for quantum training is:
\begin{equation}
n_{total} \sim n+dr+O(\log\log(1/\epsilon))
\end{equation}

For instance, when $n=5$, $d=5$, $r=10$, $\epsilon=10^{-8}$, $n_{total}\approx 60 $.

\end{document}